\documentclass[a4paper,prb,twocolumn,showpacs,amsmath,amssymb,groupedaddress]{revtex4-1}
\usepackage[english]{babel}
\usepackage[latin1]{inputenc}
\usepackage{amscd,amsmath,amssymb,verbatim}
\usepackage[dvips]{graphicx,color}
\usepackage{textcomp}
\usepackage{calc}
\usepackage[OT1,T1]{fontenc}
\usepackage{natbib}
\usepackage[verbose,hypertexnames=false,bookmarksopenlevel=1,filecolor=blue,
linkcolor=blue,citecolor=blue,pdfstartview=FitH,bookmarksopen,bookmarksnumbered,
colorlinks,plainpages=false,linktocpage,ps2pdf]{hyperref}
\usepackage[]{psfrag}
\usepackage{esvect}                         % for vectorarrows
\usepackage[hang,raggedright,nooneline]{subfigure}      % for subfigures

\DeclareMathOperator{\sgn}{sgn}

\begin{document}
\title{Quantum Corrections to Thermopower and Conductivity in Graphene}

\author{Aleksander P. Hinz}
\email[]{a.hinz@jacobs-university.de (corresponding author)}
%\homepage[]{www.jacobs-university.de/directory/ahinz}
%\thanks{With special thanks to Hyunyong Lee and Dr. Paul Wenk.}
\affiliation{School of Engineering and Science, Jacobs University Bremen, Bremen 28759, Germany}
\affiliation{Asia Pacific Center for Theoretical
Physics (APCTP) and Division of Advanced Materials Science, Pohang University of Science and Technology (POSTECH), San31, Hyoja-dong, Nam-gu, Pohang 790-784, South Korea}

\author{Stefan Kettemann}
\email[]{s.kettemann@jacobs-university.de}
%\homepage[]{www.jacobs-university.de/ses/skettemann}
\affiliation{School of Engineering and Science, Jacobs University Bremen, Bremen 28759, Germany}
\affiliation{Asia Pacific Center for Theoretical
Physics (APCTP) and Division of Advanced Materials Science, Pohang University of Science and Technology (POSTECH), San31, Hyoja-dong, Nam-gu, Pohang 790-784, South Korea}

\author{Eduardo R. Mucciolo}
\email[]{mucciolo@physics.ucf.edu}
%homepage[]{physics.ucf.edu/~mucciolo/Welcome.html}
\affiliation{Department of Physics, University of Central Florida, Orlando, Florida 32816-2385, USA}

\date{\today}

\begin{abstract}
The quantum corrections to the conductivity and the thermopower in
monolayer graphene are studied.  We use the recursive Green's function
method to calculate numerically the conductivity and the thermopower
of graphene.  We then analyze these weak localization corrections by
fitting  with the analytical theory as function of the impurity
parameters and the gate potential.  As a result of the quantum
corrections to the thermopower, we find large magnetothermopower
  which is shown to provide a very sensitive measure of the size and strength of
the impurities.  We compare these analytical and numerical results
with existing experimental measurements of magnetoconductance of
single layer graphene and find that the average size and strength of
the impurities in these samples can thereby be determined. We suggest favorable parameter ranges for future
measurements of the magnetothermopower.
\end{abstract}
%\keywords{}
\maketitle
%\tableofcontents
\section{Introduction}
Since its synthesis, graphene\cite{Novoselov22102004} has attracted a
lot of attention, both for its novel electronic properties and its
possible applications. One of the most remarkable aspects is that the
weak localization effect, which is a typical low-temperature
phenomenon due to quantum corrections to the conductivity $\delta
\sigma$, can in graphene be observed up to a temperature range of 200
Kelvin.\cite{PhysRevB.78.125409,PhysRevLett.103.226801} It has been
theoretically predicted that the sign of these corrections strongly
depends on the kind of impurities in the graphene
sample,\cite{PhysRevLett.97.146805} so that both positive and negative
magnetoconductivity, the so-called weak antilocalization effect, can
be observed. This effect is well known to occur due to spin-orbit
interaction, even though the latter is typically very weak in
graphene.  Noting that the graphene lattice is composed of two
sublattices, one can formulate the {\it sublattice} degree of freedom
as an {\it isospin} which is strongly coupled to the
momentum. Thereby, any elastic scattering that breaks the graphene
sublattice symmetry is expected to result in weak
antilocalization. This is made more complicated, however, by the fact
that there is another discrete degree of freedom in graphene, the two
degenerate Dirac cones, which can correspondingly be formulated by
introducing a {\it pseudospin} index. Accordingly, any scattering
which mixes these two valleys can result in yet another change of the
sign of the quantum correction to the conductivity, and thereby to the
restoration of weak localization. More recently, thermopower $S$ in
graphene has been measured by several
groups.\cite{PhysRevB.82.245416,PhysRevB.80.081413,PhysRevLett.102.096807,PhysRevLett.102.166808}
The classical thermopower of graphene has been calculated with a range
of different methods. The experiments show good agreement with the
Mott formula, which corresponds to the leading term of the Sommerfeld
expansion at low temperatures and large gate
voltages.\cite{PhysRevB.76.193401} Close to the Dirac point,
thermopower shows an unusual behavior, being linear in gate voltage,
and changing sign. Thus, while usually thermopower is expected to
increase as the Fermi energy, the electron density and thereby the
conductivity are lowered, in graphene the thermopower becomes smaller
as the Dirac point is approached and the electron density and the
conductivity are reduced.

The main aims of this paper are, firstly, to find out if there are
sizable quantum corrections to the thermopower $\delta S$ in graphene
and whether they are sensitive to weak magnetic fields,  secondly,
to present and analyze the combined analytical and numerical theory in
order to provide the basis for a more quantitative analysis of
experimental results on magnetoconductance of graphene. This will
allow one to characterize samples,  in particular the density,
strength, and size of carrier scatterers, more accurately.

We restrict our attention to the diffusion term of the thermopower and
do not consider the phonon drag contribution, which becomes relevant
only at high temperatures. In normal metals, the quantum corrections
to the diffusion thermopower are known to be dominated by the weak
localization corrections to the conductivity, yielding $\delta S/S
\approx - \delta \sigma/\sigma$.
\cite{Afonin,PhysRevB.33.8841,0022-3719-21-9-004} Thus, when the
conductivity is reduced by the weak localization correction, the
thermopower becomes enhanced. This is expected since the thermopower
is known to increase from a metal towards an insulator. Because of the
particular electronic properties of graphene, where the sign of the
quantum correction can change upon varying the gate voltage, it could
be expected that these quantum corrections to thermopower are
particularly large in graphene. We therefore determine the strength
and the sign of the resulting {\it magnetothermopower} and analyze in
detail how it changes with impurity parameters such as range, density,
and strength and with the gate voltage.

This paper is organized as follows. We start in Sec. \ref{graphene}
with a short introduction to the electronic properties of graphene. We
introduce the model for the description of the impurity potential in
Sec. \ref{graphene_Imp} and list the resulting momentum scattering
rates and their dependence on the impurity matrix elements. In
Sec. \ref{loc_in_gra} we give a brief review of transport theory, in
particular the theory of weak localization corrections to the
conductivity and its application to graphene. In Sec. \ref{thermo} we
review the theory of thermopower and address leading quantum
corrections.

In Sec. \ref{program} we present a numerical method to calculate the
conductance which is based on the recursive Green's function technique
(Sec.\ref{green}). We introduce the Hamiltonian that models impurities
in graphene for the numerical calculation in Sec. \ref{KtoV} and
relate it in Sec. \ref{connection} to the impurity model introduced in
Sec. \ref{graphene_Imp}. This connection makes it possible to display
the related transport scattering rates from Sec. \ref{graphene_Imp}
and Sec. \ref{theory} as functions of the sample parameter of the
numerical method, Sec. \ref{program}, which is done in
Sec. \ref{scatteringrates}.

In Sec. \ref{results} we present the numerical results for the
conductance, Sec.\ref{tata}, and thermopower,
Sec.\ref{thermo_section}. In Sec. \ref{intandana} we analyze the
numerical results by fitting them to the analytical results in
Sec. \ref{fitting}.  In Sec. \ref{imanuel} we attempt an ab initio
analysis. To this end, we use the relation of the scattering rates to
the impurity parameters of the numerical calculations and insert it in
the analytical formulas given in Sec. \ref{theory}.  In
Sec. \ref{qccb0} we present the analytical results for the quantum
corrections to the conductance at zero magnetic field.  In
Sec. \ref{kant_thermo} we present the results for the quantum
corrections to the thermopower at zero magnetic field and the
resulting magnetic field dependence of the thermopower.  In
Sec. \ref{compare} we compare the numerical and the analytical results
with experimental results on weak localization corrections to the
conductance.  Finally, we draw the conclusions and summarize our
results in Sec. \ref{conclusion2}.  In appendix \ref{matrix} the
Hamiltonian is given in matrix notation.  In appendix
\ref{appendix_two} we present analytical results for the
magnetoconductance when the warping rate is neglected, $1/\tau_w=0$.

\section{\label{graphene}Electronic Structure of graphene}

Let us start with a brief review of the electronic properties of
graphene, introducing the notation used in this article. We follow the
convention of McCann and coworkers.\cite{PhysRevLett.97.146805} For an
overview and comparison of notations, see also
\onlinecite{PhysRevB.74.235443}, \onlinecite{PhysRevLett.97.236801}, \onlinecite{Guruswamy2000475} and \onlinecite{PhysRevLett.97.036802}

Graphene is a two-dimensional layer of carbon atoms arranged in a
honeycomb lattice. These carbon atoms are connected by strong
$\sigma$-bonds with their three neighboring atoms. The corresponding
energy bands are filled valence bonds, lying deep below the Fermi
energy. The $\pi$-bond leads to the formation of a $\pi-$band which is
exactly half filled in ungated and undoped graphene. Thus, to study
transport properties we can restrict the model to this $\pi$-band. The
basis of the Bravais lattice consists of two atoms, each one of them
forming one sublattice, named A- and B-sublattice.

The electronic band structure shows two degenerate half filled cones,
which in momentum space are located on the sites of the hexagonal
reciprocal lattice. Close to zero energy, the Dirac points, the
dispersion is in good approximation linear. Each Dirac cone is part of
a sublattice, the K- and K'-valleys.

Accordingly, the Hamiltonian can be written as a $4 \times 4$-matrix
which, close to the Dirac points in linear approximation, becomes
\begin{equation}
H_1 = v_F \zeta_3 \otimes \vv{\sigma } \cdot \vv{k},
\label{bigH}
\end{equation}
where
%$\zeta _0$ is the identity matrix and
$\zeta _{3}$ is the diagonal Pauli matrix in K-K' space, $k_x$ and
$k_y$ are components of the momentum $\vv{k}$ in the plane, and
\begin{equation}
\vv{\sigma }=\begin{pmatrix}\sigma_1\\\sigma_2\end{pmatrix},\label{def2}
\end{equation}
with
% $\sigma_0$ bring the identity matrix and
$\sigma_{1,2}$ the non-diagonal Pauli matrices in the sublattice
(A,B)-space. The next correction term to the dispersion is quadratic
and given by
\begin{equation}
H_2 = - \mu \left[\sigma
  _1\left(k_x^2-k_y^2\right)-2\sigma_2\left(k_xk_y\right)\right]
\zeta_0,
\end{equation}
where $\mu$ is the parameter of warping, which is given
by\cite{RevModPhys.81.109,JPSJ.67.2857}
\begin{equation}
\mu=\frac{3 t a_0^2}{8}.
\end{equation}
In this notation the 4-component Bloch-states are given by
\begin{equation}
\vv{\Psi}^T=\begin{pmatrix}\Phi_{AK}, \Phi_{BK}, \Phi_{BK'}, \Phi_{AK'}\end{pmatrix}.
\end{equation}
The Hamiltonian can be more compactly written in the isospin and
pseudospin notation, which is particularly convenient for the purpose
of understanding the sign of weak localization corrections to the
conductivity. Following the definition of McCann and
coauthors,\cite{PhysRevLett.97.146805} we set
\begin{eqnarray}
\Xi _0&=\zeta _0\otimes \sigma _0, \qquad \Xi _1&=\zeta _3\otimes \sigma _1,\\
\Xi _2&=\zeta _3\otimes \sigma _2, \qquad \Xi _3&=\zeta _0\otimes \sigma _3,
\end{eqnarray}
and
\begin{eqnarray}
\Lambda _0&=\zeta _0\otimes \sigma _0, \qquad \Lambda _1&=\zeta _1\otimes \sigma _3,\\
\Lambda _2&=\zeta _2\otimes \sigma _3, \qquad \Lambda _3&=\zeta _3\otimes \sigma _0,
\end{eqnarray}
where $\sigma_{i}$, $i=1,2,3$ are the Pauli matrices and $\sigma_{0}$
the identity matrix in the sublattice (A,B)-space, and $\zeta _{i}$,
$i=1,2,3$ are the Pauli matrices and $\zeta_{0}$ the identity matrix
in K-K' space. The vector of the $4 \times 4$ matrices $\vv{\Xi
}^T=(\Xi _1,\Xi _2,\Xi _3)$ is the {\it isospin} vector referring to
the {\it sublattice} (A,B) degrees of freedom, and $\vv{\Lambda
}^T=(\Lambda _1,\Lambda _2,\Lambda _3)$ is the pseudospin vector
referring to the valley (K, K') degrees of freedom.\\ In this notation
the Hamiltonian $H_1 + H_2$ can be written as
\begin{equation}
\begin{split}
H_0&=H_1+H_2\\ &=\underbrace{v_F\vv{\Xi } \cdot \vv{k}}_{Dirac\ cone}
- \underbrace{\mu \Xi _1\left(\vv{\Xi } \cdot \vv{k}\right)\Lambda
  _3\Xi _1\left(\vv{\Xi} \cdot \vv{k}\right)\Xi
  _1}_{Warping\ correction\ to\ the\ Dirac\ cone}.
\end{split}
\label{h0}
\end{equation}
Note that $H_1$ is independent of the pseudospin $\vv{\Lambda}$, while
the warping term $H_2$ breaks the pseudospin symmetry.

\section{\label{graphene_Imp}Impurities in graphene}

Disorder can have many different origins in graphene such as adatoms
and vacancies, while substitutional defects are rather unlikely due to
the strength of the $\sigma$-bonding.  For the purpose of the
transport calculations, it is rather convenient to classify the
disorder according to whether it breaks the isospin symmetry (A, B
sublattices) or the the pseudospin symmetry (K, K' valleys).  Thus,
the Hamiltonian of nonmagnetic disorder has in the representation of
the pseudospin $\vv{\Lambda}$ and the isospin $\vv{\Xi}$ the general
form,\cite{PhysRevLett.97.146805}
%We will focus on
 %Gaussian scatterers only and
% follow closely McCann et.al\\\
%In our numerical approach the basis of the description are the two operators $\vv{\Xi }$ and $\vv{\Lambda }$.
\begin{equation}
H_{\text{imp}}=I\,V_{0,0}\left(\vv{r}\right)+\sum _{i,j=1}^3 \Xi
_i\Lambda _jV_{i,j}\left(\vv{r}\right),\label{Hnew}
\end{equation}
where $I$ is the identity matrix.  $V_{0,0}\left(\vv{r}\right)$ is the
part of the disorder potential which leaves both the isospin and the
pseudospin symmetry invariant, while the other terms are breaking
either of these symmetries with the amplitudes
$V_{i,j}\left(\vv{r}\right)$.  We note that this disorder Hamiltonian
is invariant under time reversal, $\vv{\Lambda} \rightarrow -
\vv{\Lambda} $ and $\vv{\Xi} \rightarrow - \vv{\Xi} $ and thus indeed
describes nonmagnetic disorder. See Appendix \ref{matrix} for the
explicit matrix representation of the Hamiltonian.

The corresponding scattering rates in Born approximation are given by
\begin{equation} \label{tauij}
\tau _{ij}^{-1}= \left(\pi \nu V_{i,j}^2\right)/\hbar,
\end{equation}
where $\nu$ is the density of states per spin and valley. We can
thus define the total scattering rate as
 \begin{equation}
\tau^{-1}:= \tau_{00}^{-1} + \sum_{i,j =1,2,3}\tau _{ij}^{-1}.
\end{equation}

The intervalley scattering rate is obtained by summing over all matrix
elements which couple to the transverse pseudo spin components
$\Lambda_{1}$ and $\Lambda_{2}$, yielding
\begin{equation}
\tau _i^{-1}:=4\tau_{\bot \bot' }^{-1}+ 2\tau_{3\bot }^{-1}\label{tau_i},
\end{equation}
where we introduced $\bot = {1,2}$ to denote the transverse
components, noting that $\tau_{i1}^{-1}=\tau_{i2}^{-1}=:\tau_{i
  \bot}^{-1}$ and $\tau _{1j}^{-1}=\tau_{2j}^{-1}=:\tau_{\bot
  j}^{-1},$ since $V_{1,j}^2=V_{2,j}^2=:V_{\bot, j}^2$ and
$V_{i,1}^2=V_{i,2}^2=:V_{i, \bot}^2$.

The intravalley scattering rate is accordingly given by
\begin{equation}
\tau _z^{-1}:=2\tau _{\bot 3}^{-1}+\tau _{33}^{-1}\label{tau_z}.
\end{equation}
The trigonal warping term in the kinetic energy of graphene results in
an asymmetry of the dispersion at each valley with respect to momentum
inversion,\cite{Falko200733} while the total Hamiltonian including
both valleys preserves that symmetry.  Therefore, the pseudospin is
expected to precess in the presence of the warping term. In the
presence of elastic scattering the pseudospin channels are expected to
relax accordingly, as in motional narrowing, in proportion to the
scattering time $\tau$,
\begin{equation}
\tau_w^{-1}= 2 \tau\left( \frac{E_{F}^2 \mu}{\hbar v^2} \right)^2. \label{warp}
\end{equation}
%Furthermore we set
%\begin{equation}
%\tau _w^{-1}+\tau _z^{-1}=:\tau _{\ast }^{-1}\label{ast}
%\end{equation}

\section{\label{theory} Transport Theory}

\subsection{\label{loc_in_gra}Weak Localization Corrections to the Conductance in Graphene}

Quantum corrections to the conductance, the so-called weak
localization corrections, originate from the quantum interference of
electrons propagating through the sample.  On time scales $t$
exceeding the elastic mean free time $\tau$, electrons move
diffusively and their return probability can be enhanced due to the
constructive interference of the amplitudes for propagation on closed
diffusion paths, as long as that time $t$ does not exceed the phase
coherence time $\tau_{\phi}(T)$.  In the presence of spin-orbit
interaction, the spin precesses as the electron moves on  closed
paths, and the phase of the amplitudes for clockwise and anticlockwise
propagation no longer match, unless the total spin of these two
propagations adds up to zero.  This spin singlet channel leads to
destructive interference, and thereby an enhancement of the
conductance, the weak antilocalization correction.

Since the spin-orbit interaction in graphene is weak, in the absence
of magnetic impurities, there are two independent spin channels in
graphene and the quantum corrections are doubled due to this spin
degeneracy.  However, since the electron momentum $\vv {p}$ is
directly coupled to the isospin $\vv{\Xi }$ (due to the A-B sublattice
degree of freedom) in the electron Hamiltonian of Eq. (\ref{h0}), any
momentum scattering will result in the breaking of the isospin
symmetry. Therefore, there is a finite contribution from the
interference of two closed electron paths whose total isospin is zero,
leading to weak antilocalization in the isospin singlet channel.

Let us now consider the influence of the pseudospin degree of freedom
of the two valleys in graphene on the weak localization corrections.
Formulating the conductance corrections in the representation of
pseudospin singlet and triplet modes, one has, in two dimensions,
\cite{PhysRevLett.97.146805}
\begin{equation}
\delta \sigma = \frac{2 e^2 D_{e}}{ \pi \hbar} \int \frac{d^2 Q}{(2
  \pi)^2} \left( - C_0^0 + C_0^1 + C_0^2 + C_0^3 \right),
\label{modes}
\end{equation}
where $D_{e}$ is the diffusion constant, which in graphene is related
to the elastic scattering time as $D_{e} = v_{\rm F}^2 \tau$, and the
momentum integral has an upper cutoff $1/l_{e}$, where $l_{e} =v_{\rm
  F} \tau $ is the elastic mean free path. The superscript corresponds
to the pseudospin and the subscript to the isospin.

Without magnetic field ($\vv{ B} =0$) the Cooperon modes $C_0^j$ are
given by
\begin{equation} \label{cl0}
C^j_0 (\vv{ Q}) = \frac{1}{
\left( D_{e} \vv{ Q}^2 + \Gamma_0^j + \tau_{\phi}^{-1}  \right)}.
\end{equation}
Here, $ \tau_{\phi}^{-1}(T)$ is the dephasing rate caused by
electron-electron and electron-phonon interactions, which provides the
low-energy cutoff of the diffusion pole of Eq. (\ref{cl0}).  The
elastic scattering from impurities can break the pseudospin symmetry
and results in the following pseudospin relaxation rates,
\begin{eqnarray}\label{last}
\Gamma _0^0&=&0,\\
\Gamma _0^3&=&2 \tau _i^{-1},\\
\Gamma _0^1=\Gamma _0^2&=&\tau _i^{-1} +
\tau _{\ast }^{-1},
\end{eqnarray}
where
\begin{equation}
\tau _{\ast }^{-1} = \tau _z^{-1} +\tau _w^{-1}.
\end{equation}
  We note that
Thus, the pseudospin triplet Cooperon modes are attenuated. Performing
the integral over momentum one thus obtains logarithmic weak
localization corrections at $B=0$,
  \begin{eqnarray}
  \delta \sigma &=& \frac{ e^2 }{\pi \rm h}
  \theta\left(\frac{\tau_{\phi}} {\tau_{\rm
      tr}},\frac{\tau_{\phi}}{\tau_{i}},\frac{\tau_{\phi}}{\tau_{\ast
  }}\right),
\label{deltasigma}
\end{eqnarray}
where
\begin{eqnarray}
  \theta\left(\frac{\tau_{\phi}} {\tau_{\rm
      tr}},\frac{\tau_{\phi}}{\tau_{i}},\frac{\tau_{\phi}}{\tau_{\ast
  }}\right) &=&\left[2 \ln \frac{\tau_{\phi}} {\tau_{\rm tr}} - \ln
    \left( 1+ 2 \frac{\tau_{\phi}}{\tau_{i}}
    \right)\right.\nonumber\\ &&\left. - 2 \ln \left( 1 +
    \frac{\tau_{\phi}}{\tau_{i}} +\frac{\tau_{\phi}}{\tau_{\ast
    }} \right) \right],\label{theta}
  \end{eqnarray}
and $\tau_{\rm tr} \approx 2 \tau$. The dependence of the various
scattering rates [Eqs.(\ref{tau_i}), (\ref{tau_z}) and (\ref{warp})]
on the impurity potential amplitudes $V_{ij}$ is given by
Eq. (\ref{tauij}). One can see that both the sign and the amplitude of
the weak localization corrections depend strongly on the impurity
type.

As we will study in detail below, impurities with large correlation
lengths mix valleys weakly, $1/\tau_{i} \approx 0$, and therefore only
attenuate two of the pseudospin triplet modes, in a similar way as the
relaxation rate due to the warping term of Eq. (\ref{warp}). This
effect leads to the vanishing of the weak localization effect and a
flat magnetoconductance.

Upon applying an external magnetic field, the weak localization
corrections are suppressed.  Solving the Cooperon equation in the
presence of a magnetic field by summing over the Landau levels, one
finds that $\Delta\sigma(B) = \sigma (B) - \sigma (B = 0)$ is given by
\begin{widetext}
\begin{equation}
\Delta \sigma (B)=\frac{e^2}{\pi h}\underbrace{
  \overbrace{\left[F\left(\frac{\tau _B^{-1}}{\tau _{\phi
        }^{-1}}\right)\right.}^{\text{\parbox{2cm}{\centering
          Pseudospin\ singlet}}} \overbrace{\left.-F\left(\frac{\tau
        _B^{-1}}{\tau_{\phi }^{-1}+2\tau _i^{-1}}\right) - 2 \cdot
      F\left(\frac{\tau _B^{-1}}{\tau _{\phi }^{-1}+\tau _i^{-1}+\tau
        _{\ast
        }^{-1}}\right)\right]}^{\text{Pseudospin\ triplet}}}_{\text{Isospin\ singlet}}
,\label{final}
\end{equation}
\end{widetext}
where the function $F(z)$ in Eq. (\ref{final}) is given by
\begin{equation} \label{ffull}
F_{\text{full}}(z) = -\psi
\left(\frac{1}{2}+\frac{\tau_B}{\tau_{tr}}\right)+\psi
\left(\frac{1}{2}+\frac{1}{z}\right) + \ln [ z \tau_B/\tau_{tr}],
\end{equation}
where $\psi (z)$ is the Digamma function and the magnetic rate is
\begin{equation}
\tau _B^{-1}=\frac{4 e D_{e} B }{\hbar }.
\end{equation}
For weak magnetic fields, one can use the simplified
form\cite{PhysRevLett.97.146805}
\begin{equation}
F(z)=\ln (z)+ \psi \left(1/2+z^{-1}\right).
\end{equation}

\begin{figure}
\begin{center}
\psfrag{z}{$\ast$}
\includegraphics [width=8cm] {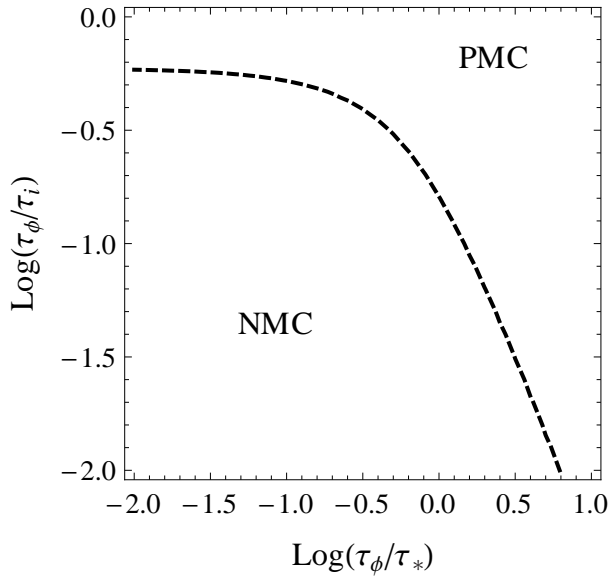}
\caption[]{\label{res2} $\tau _{\phi }/\tau _{\ast }$ - $\tau _{\phi
  }/\tau _{i}$ diagram illustrating the transition from positive
  magneto conductance (PMC) to negative magneto conductance (NMC) as
  obtained from Eq. (\ref{res}). The transition is indicated by the
  dashed black line, corresponding to $\Delta \sigma (B) =0$. This
  behavior is in good agreement with
  experiments.\cite{PhysRevLett.103.226801}}
\end{center}
\end{figure}

Equation (\ref{final}) can be expanded for small magnetic fields,
yielding
\begin{equation}
\Delta \sigma (B)  \thickapprox  \frac{e^2}{24\pi h}{\left(\frac{4
    \text{e D B $\tau$}_{\phi }}{\hbar }\right)}^2\,
\beta\left(\frac{\tau _{\phi }}{\tau _{\ast }}, \frac{\tau _{\phi
}}{\tau _{i}}\right),
\end{equation}
where
\begin{eqnarray}
\beta\left(\frac{\tau _{\phi }}{\tau _{\ast }}, \frac{\tau _{\phi
}}{\tau _{i}}\right) &=& \left[1-\frac{1}{{\left(1+2\frac{\tau_{\phi
      }}{\tau _i}\right)}^2}-\frac{2}{{\left(1+ \frac{\tau_{\phi
      }}{\tau _i} +\frac{\tau _{\phi }}{\tau _{\ast
      }}\right)}^2}\right].\nonumber\\ \phantom{a} \label{res}
\end{eqnarray}
In Fig. \ref{res2} we plot the curve obtained from the solution of
$\Delta \sigma (B) =0$ as function of the parameters $\tau _{\phi
}/\tau _{\ast }$ and $\tau _{\phi }/\tau _{i}$. This curve separates
the region of positive and negative magnetoconductance.  We note that
this separation line does not coincide with the curve $\delta \sigma
(B=0)=0$, which corresponds to the vanishing of the weak localization
correction and separates the parameter space regions of weak
localization and weak antilocalization (corresponding to negative and
positive quantum corrections to the conductivity at $B=0$,
respectively). As seen in Eq. (\ref{deltasigma}) the latter separation
line depends on an additional parameter, the total scattering rate
$1/\tau$.

\subsection{\label{thermo}Thermopower}

Applying a thermal gradient $\nabla T$ to a metallic sample induces
not only a thermal current but also an electrical current density $
\vv{ j} = \sigma (\vv{ E} + \nabla \mu/e) - \eta \nabla T $, where
$\vv{ E}$ is an applied electric field and $\mu$ is the chemical
potential.  Here, $\eta$ denotes the thermoelectric coefficient.
Under open circuit conditions, the thermal gradient results in a
finite voltage $U$ proportional to the temperature difference $\Delta
T$, where the proportionality constant is the thermopower: $S =
\eta/\sigma$.  At low temperatures, the thermopower is dominated by
the diffusion of the electrons in the sample and the phonon drag
contribution becomes small.  Expanding in the ratio of temperature to
Fermi energy and keeping the leading term, one arrives at the Mott
formula, which relates the thermopower $S$ to the derivative of the
conductivity with respect to the Fermi energy,
\begin{equation}
S= \frac{\pi ^2}{3}\frac{k_B^2T}{e}\left[\frac{d \ln (\sigma (E))}{d E
  }\right]_{E=E_F}.
\label{mott}
\end{equation}
This formula is valid for low temperatures, $T \ll E_F$ and
large chemical potential. For $T=0K$, which is the case in this paper,
the chemical potential is equal to the Fermi energy.\\
Here, $e = -|e|$ is the negative electron
charge. Thus, when the carriers have negative charge, thermopower is
expected to be negative, while for holes it becomes positive.  We note
that the {\it elemental} unit of thermopower is given by a ratio of
natural constants $S_0 = k_B/|e| \approx 86 \mu V/K$.

% FORMULA MOVED DOWN
%The thermopower
%can be also written in terms of the back gate voltage $V_{\rm BG}$ as
%\begin{equation}
%S =\frac{\pi ^2}{3} \frac{k_B^2T}{ e}\frac{1}{\sigma}\frac{d \sigma}{d
%  V_{\text{BG}}}{ \left.\frac{d V_{\text{BG}}}{d
%    E}\right|}_{E=E_F}.\label{tatr}
%\end{equation}
%In monolayer graphene, the thermopower is well described by
%Eq. (\ref{tatr}) in a wide range of temperatures and gate voltages not
%too close to the Dirac point.\cite{PhysRevB.82.245416}

Furthermore this formula is valid not only classically but also includes quantum
corrections through the conductivity.  Thus, expanding in the quantum
correction to the conductivity $\delta \sigma$, we can write the
leading quantum corrections to the thermopower, $\delta S$, as
 \begin{equation} \label{quantums}
 \frac{\delta S}{S} = - \frac{\delta \sigma}{\sigma} + \left[\frac{d
     \delta \sigma (E)}{d E }\right]_{E=E_F}/\left[\frac{d \sigma
     (E)}{d E }\right]_{E=E_F}.
 \end{equation}
This relation is also supported by direct diagrammatic calculations
of quantum
corrections.\cite{Afonin,PhysRevB.33.8841,0022-3719-21-9-004} In
standard metals, the last term on the r.h.s of Eq. (\ref{quantums}) is
small and the thermopower is dominated by the quantum correction to
the conductivity.  In this work we revisit this relation to find out
whether this also holds for graphene, or whether the second term in
Eq. (\ref{quantums}) is sizable.

\section{\label{program} The recursive Green's function method }

In this section we introduce the numerical method used to calculate
the electrical conductivity and the thermopower. We employ the
recursive Green's function
method,\cite{PhysRevLett.47.882,PhysRevLett.65.2442}, which has been
previously applied to graphene in
Refs. [\onlinecite{0953-8984-22-27-273201,PhysRevB.79.075407,PhysRevB.77.081410,Mucciolounpublished}]. Below,
we begin by reviewing the essential elements of this method.

The graphene sample is assumed to be connected to two semi-infinite
leads which are modeled by a square lattice. When contacting the leads
at the zigzag edges of the graphene sample, there is no wave function
mismatch between the propagating modes in the square lattice leads and
the graphene sample, provided that a proper energy shift is
used.\cite{PhysRevB.76.045433} The graphene sample is sliced
transversely into $N$ equal cells, with each cell containing $M$
sites. We study the transport as function of $N$ by changing the
length $L$ of the sample, and as function of $M$ by changing the
sample width $W$. When the free edges are of the armchair type, the
sample dimensions are related to to $N$ and $M$ by $\frac{L}{a_0} =
\frac{\sqrt{3}}{2}\left(\frac{N}{2}-1\right)+\frac{\sqrt{3}}{6}$ and
$\frac{W}{a_0} = M-1$, where $a_0=2.46 \mathring{A}$ is the lattice
constant set as the distance between two atoms of the same
sublattice. See Fig. \ref{armchair_walls2} .

\subsection{The conductivity}

In two dimensions, the electrical conductivity $\sigma$ is related to
the conductance $G$ by the standard expression $\sigma = L G/W$ and
thus can be expressed in terms of retarded Green's function through
the Caroli formula
\begin{equation}
\sigma=\frac{L}{W}\frac{2e^2}{h}\text{Tr}_s\left(\Gamma _L G^R\Gamma
_R G^A\right),
\end{equation}
where $G^{A/R}$ denotes the advanced/retarded Green's function
connecting two opposite contact regions (right and left). Here,
$\mbox{Tr}_{s}$ denotes the trace over transverse sites at the
lead-sample interface. The matrix elements of the level width matrix
$\Gamma_p$ can be expressed in term of the transverse wave functions
$\chi_{\nu}(i)$ of the lead propagating modes,
\begin{equation}
\Gamma _p(i,i')=\sum _{\nu} \chi _{\nu}(i)\frac{\hbar
  v_{\nu}}{a_0}\chi _{\nu}(i').
\end{equation}
Here, the sum runs over the propagation modes $\nu$ and $v_{\nu}$ is
the longitudinal propagation velocity. Next we describe the method
used to obtain the Green's functions of the sample.

\subsection{The Green's function\label{green}}

\begin{figure}[b]
\begin{center}
\includegraphics [width=8cm] {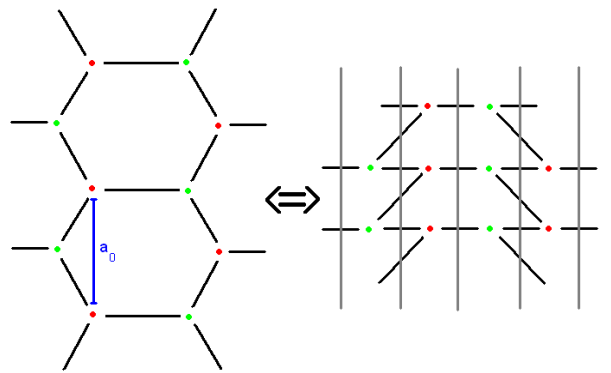}
\caption{\label{armchair_walls2}Slicing of graphene shown for two honeycomb cells when the contacts
  are at zigzag edges. Red and green dots indicate the two
  sublattices A and B, the connecting bonds of the sites are solid lines. The vertical slices, right side, are marked with vertical gray lines. $a_0$ represents the lattice constant, indicated in blue.}
\end{center}
\end{figure}

In order to obtain the Green's function amplitude $G^R_{p q}(i,j)$
between sites $i,j$ at the contacts $p,q$, we start with the surface
Green' function of one of the contacts, which is presumed to be known. Then
we add one slice of transverse sites $i =1,..,M$, and calculate the
Green's function from the contact up to that slice. We repeat the
procedure, adding slice by slice, until the end of the sample is
reached. This procedure is done from left to right, and then in the
opposite direction in order to calculate the full Green's function of
the system. The slicing is displayed in Fig. \ref{armchair_walls2}.

%\begin{figure}[b]
%\begin{center}
%\includegraphics [width=8cm] {armchair_walls2.eps}
%\caption{\label{armchair_walls2}Slicing of graphene when the contacts
%  are at zigzag edges. Filled and empty dots indicate the two
%  sublattices A and B. The vertical slices are marked with dashed
%  lines, the connecting bonds of the sites are solid lines. $a_0$
%  represents the lattice constant.}
%\end{center}
%\end{figure}

Each line in Fig. \ref{armchair_walls2} represents a hopping amplitude
from one sublattice to another. This is described by the hopping
matrix $U$. This matrix, together with the Green's function of every
new single slice $g_{n}$ and the Green's function that includes the
contacts and the sample up to slice $n$, defining $G^{(0)}$, are
inserted in the Dyson equation $G=G^{(0)}+G^{(0)} V G $ to obtain the
new $G$.
One can easily understand that the speed of the procedure strongly
depends on the slicing scheme adopted. In general, more slices mean
more steps to calculate, but each slice consists of fewer atom sites
and so its calculation is done faster since an inversion at each slice
is required. The complexity of the calculation scales as $O(N M^3)$.

%In the Appendix (\ref{appendix_two}) we show a more detailed view of
%the method for the case of graphene including a figure of the
%slicing.\\
The Green's function of the contacts has been derived previously and
is given by \cite{ferry1997transport}
\begin{equation}
G^{\text{semi}}(\nu
)=\frac{2p_{\nu}}{{\left(2\left|t^x\right|\right)}^2}\left(1-\sqrt{1-\left(\frac{2\left|t^x\right|}{p_{\nu}}\right)^2}\right),\label{leadsgreen}
\end{equation}
with
\begin{equation}
p_{\nu}=-V_{\text{gate}}+2\left|t^y\right|\cos \left(\frac{\pi \nu
}{M+1}\right)+\text{i0}^+,
\end{equation}
where $\nu = 1,...,M$ represents the modes of the transverse wave
function and $t^i$ is the hopping rate in $i$-direction. We can
transform $G^{\text{semi}}(\nu )$ from the channel representation into
$G^{\text{semi}}(j,j')$ in the site representation by
\begin{equation}
  G^{\text{semi}}(j,j')=\sum_{\nu=1}^M \chi^*_{\nu}(j)
  G^{\text{semi}}(\nu ) \chi_{\nu}(j'),
\end{equation}
with the transverse wave functions $\chi_{\nu}(j)$ given by
\begin{equation}
\chi_{\nu}(j) = \sqrt{\frac{2}{M+1}} \text{sin}\left(\frac{\pi \nu j}{M+1}\right).
\end{equation}

\subsection{\label{KtoV}Inclusion of impurities}

We now look at the model disorder used in the numerical
calculations. The sites of the graphene sample are occupied by
uniformly randomly distributed Gaussian scatterers with a random
potential $V_n \in [-\delta V,\delta V]$. We can write for the overall
potential resulting from all $N_{\text{imp}}$ scatterers at the points
$\vv{R_n}$ as
\begin{equation}
V\left(\vv{r_j}\right) = \sum_{n=1}^{N_{\text{imp}}} V_n \left(\vv{r_j}\right) =\sum_{n=1}^{N_{\text{imp}}} V_n e^{- \frac{\left|\vv{r_j}-\vv{R_n}\right|^2}{2 \xi^2}}.\label{distwo}
\end{equation}
Here, $\xi$ is the range of the potential and $\vv{r_j}$ are the
lattice sites. The concentration of scatterers is given by
$n_{\text{imp}} = N_{\text{imp}} / N_{\text{tot}}$, where
$N_{\text{tot}}$ is the total number of lattice sites. We focus on
pairwise uncorrelated impurities, $\langle V_n V_{n'}\rangle=\langle
V_n^2 \rangle \delta_{n,n'}$, each vanishing on average, $\langle V_n
\rangle=0$.

We use as an input parameter the dimensionless correlation strength
$K_0$, which is defined by the equation for the impurity potential
correlation function as
\begin{equation}
\langle V\left(\vv{r_i}\right)V\left(\vv{r_j}\right)
\rangle=\frac{K_0{\left(\hbar v_F\right)}^2}{2 \pi \xi^2}
e^{\frac{-{\left|\vv{r_i}-\vv{r_j}\right|}^2}{2\xi^2}}.
\end{equation}
In the limit of dilute impurities, summing over all sites $i,j$ yields
\begin{equation}
K_0=\frac{L W}{{(\hbar v_0)}^2
  N_{\text{tot}}^{2}}\sum_{i=1}^{N_{\text{tot}}}\sum_{j=1}^{N_{\text{tot}}}
\langle
V\left(\vv{r_{i}}\right)V\left(\vv{r_{j}}\right)\rangle.\label{disone}
\end{equation}
Inserting Eq. (\ref{distwo}) into Eq. (\ref{disone}) and using the
relations
\begin{equation}
N_{\text{tot}}=\frac{4\sqrt{3}}{3} \frac{L W}{a_0}, \quad \text{ and }
\quad v_F=\frac{\sqrt{3}}{2} \frac{t a_0^2}{\hbar},
\end{equation}
we get\cite{0295-5075-79-5-57003}
\begin{equation} \label{k0}
K_0={\left(\delta V\right)}^2 \varrho,
\end{equation}
where $\delta V$ is the half width of the box distribution
 of the amplitude of the Gaussian imputities $V_n$, defined in
  Eq. (\ref{distwo}), and
\begin{equation}
\varrho =
\frac{\sqrt{3}}{9}\left(\frac{1}{t}\right)^2\frac{1}{N_{\text{tot}}}\sum
_{n=1}^{N_{\text{imp}}}{\left[ \sum _{i=1}^{N_{\text{tot}}}
    e^{\left(-\frac{{\left(\left|\vv{r_i}-\vv{R_n}\right|\right)}^2}{2
        \xi ^2}\right)}\right]}^2.
\label{varrho}
\end{equation}
Thus, the parameter $\varrho$ depends on $L$, $W$, and $\xi$
 and is proportional to  concentration of impurities $n_{\rm imp}=N_{\text{imp}}/(LW) $. It is
useful to introduce the typical impurity strength amplitude $V_{ta}$,
which is related to $K_{0}$ and $\varrho$ by
\begin{equation}
V_{ta} \equiv \sqrt{\langle V_n^2 \rangle} = \sqrt{\frac{1}{3}
  {\left(\delta V \right)}^2} = \sqrt{\frac{1}{3} \frac{K_0} {
    \varrho}}.\label{lastK0}
\end{equation}

\section{\label{connection}Connection between the two impurity descriptions from section (\ref{graphene_Imp}) and (\ref{KtoV})}

\subsection{Impurity potential}
In this subsection, we establish the link between the description of
the impurities in graphene in the pseudospin and isospin
representation, as introduced in Sec. \ref{graphene_Imp}, and the
Gaussian impurities introduced in the previous
section.\cite{PhysRevB.74.235443,0295-5075-79-5-57003} McCann and
coworkers assume that the different components of the impurity
potential Eq. (\ref{Hnew}) are uncorrelated,
\begin{equation}
\langle V_{i,j}(\vv{r}) V_{i',j'}(\vv{r'}) \rangle = V_{i,j}^2 \delta(\vv{r} - \vv{r'}) \delta_{i,i} \delta_{j,j}.
\end{equation}
where $i,j$ denote the pseudospin and isospin indices as
introduced in Eq. (\ref{Hnew}). We can decompose the potential due to
one impurity at a given site of the lattice, $V_n(\vv{r})$, in Fourier
components. Defining the vector connecting sublattices as $\vv{m}$, we
find for the Fourier component of $V_n(\vv{r})$ on sublattice A
\cite{PhysRevB.74.235443,JPSJ.74.777}
\begin{equation}
V_{\vv{q},n}=\frac{\sqrt{3}{a_0}^2}{2}\sum_{\vv{r}}V_n(\vv{r})e^{-i/\hbar\vv{q}\vv{r}}\label{pot3}
\end{equation}
and for sublattice B
\begin{equation}
V'_{\vv{q},n}=\frac{\sqrt{3}{a_0}^2}{2}\sum_{\vv{r}}V_n(\vv{r}-\vv{m})e^{-i/\hbar\vv{q}\vv{r}},\label{pot4}
\end{equation}
where the sum is over all elementary cells. We assume $V_{\vv{q},n}$
and $V'_{\vv{q},n}$ to be slow functions of the momentum
$\vv{q}$. Thus, the quantity $V_{\vv{q},n}$ is proportional to the
first-order scattering amplitude for electrons on the same sublattice
where the impurity resides, while $V'_{\vv{q},n}$ is the one for
electrons on the other sublattice. We will explicitly consider two
values: the intravalley scattering, $q=0$, and the intervalley
scattering, $q=k_0$, where $k_0$ connects the two different valleys in
the reciprocal space and has the amplitude $k_0=2h/3a_0$. We thus have
$V_{0,n}$ and $V'_{0,n}$ for intravalley scattering in K-K' space and
$V_{k_0,n}$ and $V'_{k_0,n}$ for intervalley scattering. Intravalley
scattering means that the electron that is scattered does not leave
the original cone-shaped valley in k-space, while in the intervalley
 scattering process the
electron starts in the K valley and ends after scattering in the K'
valley or viceversa.

Equations (\ref{pot3}) (\ref{pot4}) can now be displayed in the form
of Eq. (\ref{Hnew}) using a $4 \times 4$ matrix notation. For
impurities $V_n(\vv{r})$ located on sublattice A in the elementary
cell $\vv{r_n}$ and small $\vv{q}$, we can approximate
\begin{eqnarray}
V^A_{\vv{q},n}&&=\nonumber\\
&&\begin{pmatrix}
 V_{0,n} & 0 & 0 & V_{\vv{k_0},n} e^{-2i\vv{k_0}\vv{r_n}} \\
 0 & V'_{0,n} & 0 & 0 \\
 0 & 0 & V'_{0,n} & 0 \\
 V_{\vv{k_0},n} e^{2i\vv{k_0}\vv{r_n}} & 0 & 0 & V_{0,n}
\end{pmatrix}e^{-i\vv{q}\vv{r_n}},\nonumber\\
\phantom{a}\label{ma1}
\end{eqnarray}
and, equivalently, for impurities in sublattice B,
\begin{eqnarray}
V^B_{\vv{q},n}&&=\nonumber\\
&&\begin{pmatrix}
 V'_{0,n} & 0 & 0 & 0 \\
 0 & V_{0,n} & V_{\vv{k_0},n} e^{-2i\vv{k_0}\vv{r_n}} & 0 \\
 0 & V_{\vv{k_0},n} e^{2i\vv{k_0}\vv{r_n}} & V_0 & 0 \\
 0 & 0 & 0 & V'_{0,n}
\end{pmatrix}e^{-i\vv{q}\vv{r_n}},\nonumber\\
\phantom{a}\label{ma2}
\end{eqnarray}
where we used the fact that $V'_{\vv{k_0},n}$ vanishes due to the
symmetry of graphene (otherwise we would not have any zeros in the
secondary diagonal, but terms involving $V'_{\vv{k_0},n}
e^{-2i\vv{k_0}\vv{r_n}}$).

As the next step, we convert Eq. (\ref{ma1}) and Eq. (\ref{ma2}) into
one single impurity potential matrix that can be compared with
Eq. (\ref{Hnew}). This is done by calculating the autocorrelation
function $\langle\!\langle V_{\vv{q}}\otimes
V_{-\vv{q}}\rangle\!\rangle$, with $V_{\vv{q}}=1/2
\sum_{n=1}^{N_{imp}} \left( V^A_{\vv{q},n} + V^B_{\vv{q},n}\right)$,
set by
\begin{equation}
\left\langle\!\langle V_q \otimes V_{-q}^T \right\rangle\!\rangle =\frac{n_{\text{imp}}}{2}\left\langle V_{\vv{q},n}^A\otimes V_{-\vv{q},n}^A+V_{\vv{q},n}^B\otimes V_{-\vv{q},n}^B\right\rangle,\label{autocorrelation}
\end{equation}
where the averaging $\langle\!\langle \cdot \rangle\!\rangle$ is with
respect to the positions of the impurities and the impurity strength
$V_n$, whereas the averaging $\langle \cdot \rangle$ is only with
respect to $V_n$. The normalization factors are already included in
the prefactor of the r.h.s. of Eq. (\ref{autocorrelation}).

\begin{figure*}[htb]
\begin{center}
\subfigure[Effective potentials $V_0$, $V_{k_0}$ and $V'_0$ in
  dependence on $\xi$.]{ \includegraphics [width=0.45\textwidth]
  {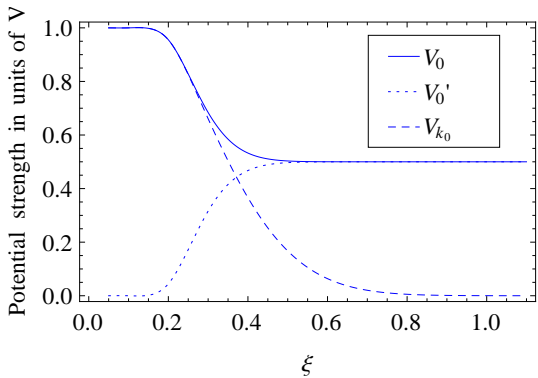}} \subfigure[Potential terms $V_0+V'_0$, $V_0-V'_0$
  and $V_{k_0}$ in dependence on $\xi$.]{ \includegraphics
  [width=0.45\textwidth] {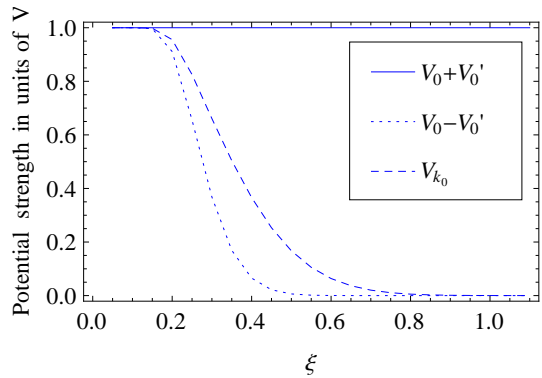}}
\caption[Ando potential]{\label{ando1} $V_0$, $V_{\vv{k_0}}$ and
  $V'_0$ as set by Eq. (\ref{ma1}) and Eq. (\ref{ma2}). For
  short-range impurities, $V_0$ and $V_{\vv{k_0}}$  approach $2V$
  because the potential is localized only at sites of one of the
  sublattices. Wider impurities also cover  the neighboring sublattice and
  $V'_0$ start to increase and $V_0$, $V_{k_0}$ decrease. Also visible
  is the faster decrease of $V_0-V_0'$ than $V_{\vv{k_0}}$. $\xi$ in
  units of the lattice constant $a_0$.}
\end{center}
\end{figure*}

Comparison of Eq. (\ref{autocorrelation}) with the matrix notation of
the impurity as introduced in Eq. (\ref{Hnew}) leads, with the help of
Eq. (\ref{lastK0}) and the use of the impurity position averaged
parameter $\langle \varrho \rangle$ from Eq. (\ref{varrho}), to the
following set of equations
\begin{eqnarray}
{V_{0,0}}^2 & = & 1/4 \enspace (V_0+V_0')^2,\label{vone}\\
{V_{3,3}}^2 & = & 1/4 \enspace (V_0-V_0')^2,\label{vthree}\\
{V_{\bot \bot}}^2 & = & 1/8 \enspace V_{\vv{k_0}}^2, \label{vtwo}
\end{eqnarray}
where
\begin{eqnarray}
V_0=&\sqrt{\langle V_{0,n}^2 \rangle}=&V/\tilde{\varrho}\sum_{\vv{r}}e^{-\frac{\vv{r}^2}{2 \xi^2}},\label{vfour}\\
V_0'=&\sqrt{\langle V_{0,n}'^2 \rangle}=&V/\tilde{\varrho}\sum_{\vv{r}}e^{-\frac{(\vv{r}+\vv{m})^2}{2 \xi^2}},\label{vfive}\\
V_{\vv{k_0}}=&\sqrt{\langle V_{\vv{k_0},n}^2 \rangle}=&V/\tilde{\varrho}\sum_{\vv{r}}e^{-i\vv{k_0}\vv{r}}e^{-\frac{\vv{r}^2}{2 \xi^2}},\label{vsix}
\end{eqnarray}
\begin{eqnarray}
V&=&\sqrt[4]{3}\sqrt{K_0} t,\label{defV}\\
\tilde{\varrho}&=&\sum_{\vv{r}}\left(e^{-\frac{\vv{r}^2}{2\xi^2}}+e^{-\frac{(\vv{r}+\vv{m} )^2}{2\xi^2}}\right).
\end{eqnarray}
  Note that $V$ is a measure of the total impurity strength
   averaged over all impurities. Since $K_0 $ is proportional
    to their concentration $n_{\rm imp}$, $V$ increases with
     $n_{\rm imp}$ as
     $$
    V\sim \sqrt{n_{\rm imp}}$$.
In the limit of short-range impurities, $\xi \rightarrow 0$, we find
\begin{eqnarray}
V_0=V_{\vv{k_0}}&=& V,\\
V'_0=V'_{\vv{k_0}}&=&0
\end{eqnarray}
while for long-range impurities, $\xi \rightarrow \infty$, we find
\begin{eqnarray}
V_0=V'_0&=& V/2,\\
V_{\vv{k_0}}=V'_{\vv{k_0}}&=&0.
\label{eq:70}
\end{eqnarray}
 In that limit,
Eq. (\ref{eq:70}), only ${V_{0,0}}$ is not zero. We can see here
that, due to symmetry, $V'_{\vv{k_0}}$ is  always $0$ and we are left
with only one parameter $V$. This is shown in Fig. \ref{ando1}. Also
displayed are the potential terms $V_0+V_0'$, $V_0-V_0'$ and
$V_{\vv{k_0}}$ which, as we have seen above, are proportional to the
impurity scattering matrix elements ${V_{0,0}}$, ${V_{3,3}}$, and
${V_{\bot \bot}}$.

Within the simplified picture of this subsection, the last result is
again consistent with short-range scatterers mixing valleys and
sublattices while long-range scatterers mixing only sublattices.

\subsection{Scattering rates\label{scatteringrates}}

We now calculate the scattering rates of Eq. (\ref{tau_i}) and
Eq. (\ref{tau_z}) and combine them with the potential Fourier
components of Eqs. (\ref{vone}), (\ref{vthree}), and (\ref{vtwo}).
( Note: We restrict us here to  the Born approximation. Going beyond that
 approximation, one needs to include all multiple impurity scattering, which yields corrections which depend logarithmically on energy\cite{PhysRevLett.97.236801}, and can yield according to Mott's law additional contributions to thermopower.) The
density of state per spin and per valley $\nu$, which is given by
\begin{equation}
\nu =\frac{k_FA_{\text{PC}}}{2\pi \hbar ^2v_0}=\frac{A_{\text{PC}}
  E_F}{2 \pi \hbar ^2 v_0^2},\label{gamma}
\end{equation}
is inserted into the general expression for the scattering rate,
Eq. (\ref{tauij}), where
\begin{equation}
k_F=\frac{E_F}{v_0}
\end{equation}
and $A_{\text{PC}}=a_0^2 \sqrt{3}/2 $ is the area of the primitive
unit cell.\\
%Here we only consider low energies and so we assume that
%$E_F\sim V_{\text{BG}}$.

In case of Gaussian scatterers of Eqs. (\ref{vone}), (\ref{vthree}),
(\ref{vtwo}), and Eqs. (\ref{ma1}) and Eq. (\ref{ma2}), only the main-
and off-diagonal matrix elements of Eq. (\ref{Hnew}) are non
zero. This helps us simplify Eqs. (\ref{tau_i}) and (\ref{tau_z}), so
that we find
\begin{equation}
1/\tau _{\text{3 $\bot$}}^{-1}=1/\tau _{\text{$\bot$ 3}}^{-1}=0
\end{equation}
resulting in
\begin{eqnarray}
1/\tau _i&=&\frac{4}{\tau _{\bot\bot}}\\
1/\tau _z&=&1/\tau _{33}.
\end{eqnarray}
Inserting Eqs. (\ref{vone}), (\ref{vthree}), and (\ref{vtwo}) and by
using Eqs. (\ref{tauij}) and (\ref{warp}) leads to
\begin{eqnarray}
\frac{1}{\tau _i}&=&\frac{(1/2)\pi \nu {V_{k_0}^2}}{\hbar }\\
\frac{1}{\tau _z}&=&\frac{(1/4)\pi \nu (V_0-V_0')^2}{\hbar }\\
\frac{1}{\tau _{00}}&=&\frac{(1/4)\pi \nu (V_0+V_0')^2}{\hbar }.
\end{eqnarray}
Now, using Eq. (\ref{gamma}) we find
\begin{eqnarray}
\frac{1}{\tau _i}&=&\frac{V_{k_0}^2 A_{\text{PC}}
  E_F}{4 \hbar ^3 v_0^2}\label{90}\\
  \frac{1}{\tau_z}&=&\frac{(V_0-V_0')^2 A_{\text{PC}}
  E_F}{8 \hbar ^3 v_0^2}\label{91}\\
  \frac{1}{\tau_{00}}&=&\frac{(V_0+V_0')^2 A_{\text{PC}}
  E_F}{8 \hbar ^3 v_0^2}\label{92}\\
  \frac{1}{\tau_w}&=&\frac{16 \hbar^2 \mu^2
  E_F^3}{(V_0+V_0')^2 A_{\text{PC}}}.\label{92.1}
\end{eqnarray}
Since the dephasing rate $1/\tau _{\phi}=0$ in the numerical
calculations, the low-energy cutoff is provided by the Thouless
energy\cite{0022-3719-5-8-007}
\begin{equation}
E_T=\frac{D}{\Lambda ^2}=\frac{v_0^2\tau _{TR}}{\Lambda ^2}=\frac{4
  \hbar^3 v_0^4}{(V_0+V_0')^2 \Lambda ^2 A_{\text{PC}} E_F},\label{93}
\end{equation}
where $\Lambda$ is the length or width of the sample, depending which
one is smaller, $\tau _{TR}=2\tau _0 \approx 2\tau _{00}$, if we have
$1/\tau _{00}\gg1/\tau_{ij}$, for $i,j\neq0$, and D is the diffusion
constant given by\cite{RevModPhys.57.287}
\begin{equation}
D=v_0^2\tau _{TR}/2.\label{Diffusion}
\end{equation}
Thus, we get the following ratios:
\begin{eqnarray}
\frac{\tau _{\phi }}{\tau _i}&=&\frac{V_{k_0}^2 (V_0+V_0')^2 \Lambda ^2 A_{\text{PC}}^2 E_F^2}{32 \hbar ^6 v_0^6}\label{preprop}\\
\frac{\tau _{\phi }}{\tau _z}&=&\frac{(V_0-V_0')^2 (V_0+V_0')^2 \Lambda ^2 A_{\text{PC}}^2 E_F^2}{64 \hbar ^6 v_0^6}\label{prop}\\
\frac{\tau _{\phi }}{\tau _w}&=&\frac{4 \mu^2 \Lambda ^2 E_F^4}{\hbar v_0^4}\label{prop2}
\end{eqnarray}
Since the factors $(V_0+V_0')^2$, ($V_0-V_0')^2$, and
$V_{\vv{k_0}}^2$, depend on the impurity parameters $V$ and $\xi$, as
shown in Fig. \ref{ando1}, by tuning one of these parameters and the
system size $\Lambda$, and Fermi energy $E_F$, one cam move
in the $\tau _{\phi }/\tau _i$ - $\tau _{\phi }/\tau _{\ast}$ diagram
of Fig. (\ref{res2}). This phase diagram is equivalent to what has
been shown in the experimental paper,\cite{PhysRevLett.103.226801}
where the temperature has been varied to reach the different regions
of the diagram. By tuning these parameters, a change from weak
localization to weak antilocalization can be observed. We note that a
change of the impurity concentration changes $\tau _{\phi }/\tau _i$
and $\tau _{\phi }/\tau _{z}$, while the ratio originating from the
warping rate $\tau _{\phi }/\tau _{w}$ is independent of the impurity
concentration $n_{\rm imp}$.

For the magnetic rate $1/\tau _B$ we find accordingly,
\begin{equation}
\frac{1}{\tau _B}=\frac{64 B e \hbar ^2 v_0^4}{(V_0+V_0')^2 A_{\text{PC}} E_F}.
\end{equation}

\section{\label{results}Numerical Results}

\begin{figure*}
\subfigure[$K_0 = 4$, $\xi$ = 1.] {\label{Weak localization}
\includegraphics[width=0.32\textwidth]{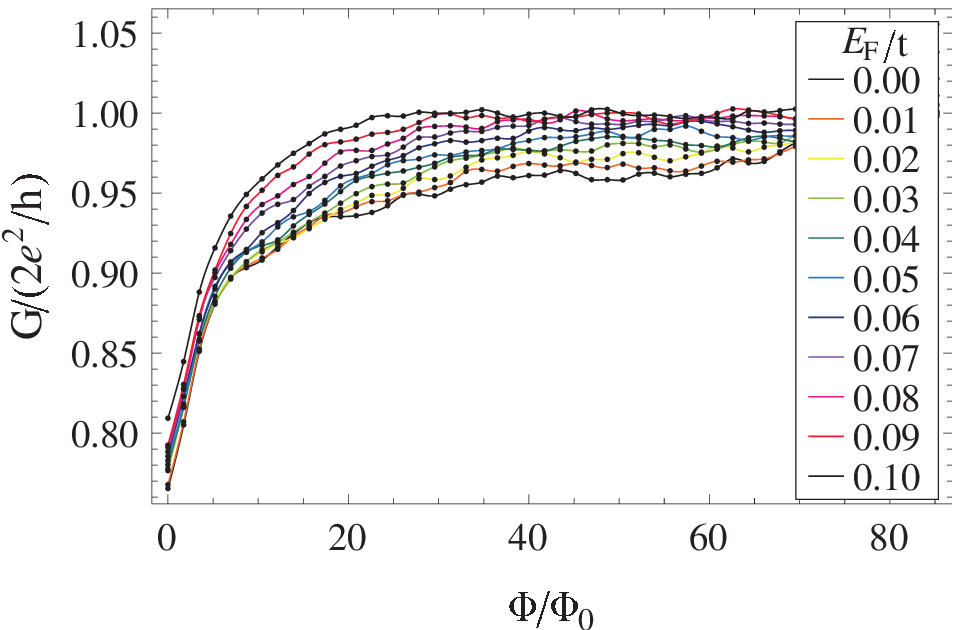}}
\subfigure[$K_0 = 4$, $\xi$ = 3.] {\label{weak antilocalization}
\includegraphics[width=0.32\textwidth]{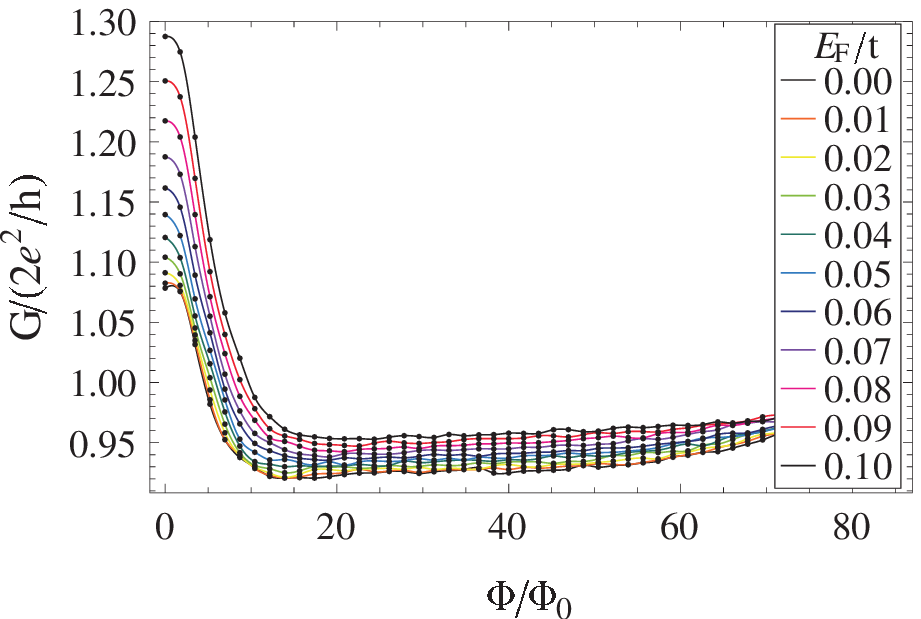}}
\subfigure[$\xi = 0.5$, $K_0 = 0.5$, $n_{\text{imp}}=0.3$.] {\label{weak-anti-imp2}
\includegraphics[width=0.32\textwidth]{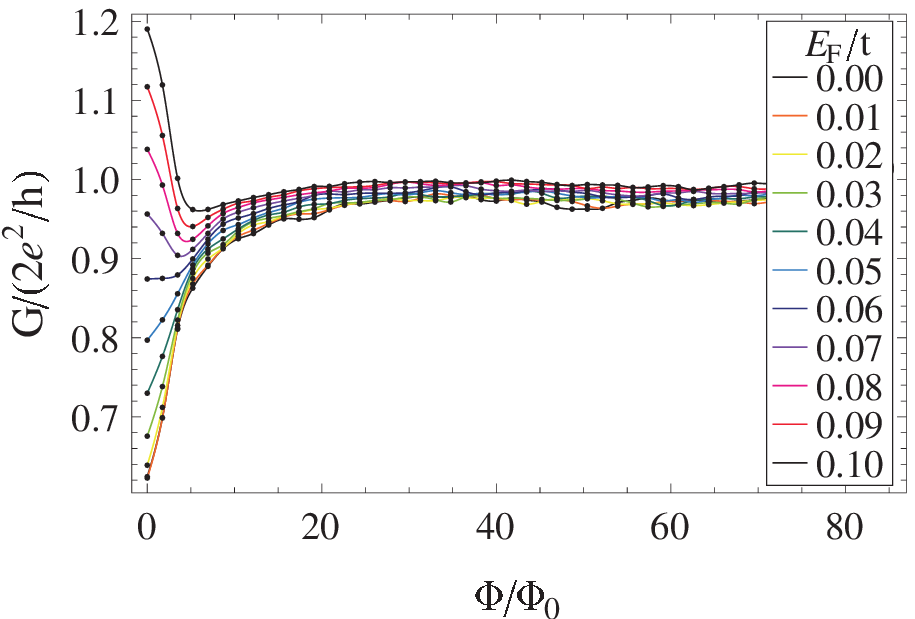}}
\caption[]{\label{crossings} Average conductance $G$, as calculated
  numerically with the recursive Green's function method for
  dimensions $N=30$ and $M=72$. The Fermi energy $E_F$
  is varied as indicated in the figure. (a) Positive
  magnetoconductance, weak localization effect at small correlation
  length $\xi =1$. (b) Negative magnetoconductance, weak
  antilocalization effect at larger correlation length $\xi =3$. (c)
  Change between positive and negative magnetoconductance by tuning
  $E_F$.  The lines are merely a guide for the eye.}
\end{figure*}

All results presented in the following are for samples with armchair
edges. Most calculations have also been done with zig zag edges, especially in the case of small systems, but they do not show any significant difference and are consequently not displayed here.
We consider always samples with an aspect ratio of 1, $L\approx
W$, ranging from $N=20$ and $M=48$; $M=30$, $N=72$, up to $M=80$,
$N=192$. The correlation length $\xi$ is given in units of $a_0$, the
lattice constant. The correlated disorder strength $K_0$ is the
dimensionless parameter, defined in Eq. (\ref{k0}). For small systems,
we use $\xi/a_0 =0.5,..., 3 $ and $K_0 =0.5,...,4$, while for the larger
system, we consider only $\xi/a_0=0.5,...,2$, $K_0 =0.5,...,2$. Unless
explicitly mentioned otherwise, the impurity density is set to
$n_{\text{imp}}=0.03$, meaning that $3\%$ of the atomic sites are
occupied by impurities. The experiments on weak
antilocalization\cite{PhysRevLett.103.226801} indicate indeed that
this is a realistic concentration of impurities in these graphene
samples. We also performed the calculations for higher densities up to
$n_{\text{imp}}=0.3$.

Impurities are uniformly distributed across the sample. For each
impurity realization we calculate the electrical conductivity as
delineated above. We run the numerical calculations for $N_c = 5000$
different realizations in order to average the results for the
conductivity and thermopower. For the larger systems, $N_c=1000$ turns
out to be sufficient, since self-averaging improves with increasing
system size. The Fermi energy, $E_F$ is displayed in
units of the hopping parameter $t$, which is set to $2.7$ eV. We
consider the range $E_F/t = 0,..,0.1$, where
$E_F=0$ corresponds to the Dirac point. The magnetic field
is displayed as the magnetic flux $\Phi$ through the whole sample in
units of the magnetic flux quantum $\Phi_0$.

\begin{figure}
\subfigure[Size 10 $\times$ 24, $\xi = 1.5$, K = 4 ] {\label{diffone}
\includegraphics[width=6cm]{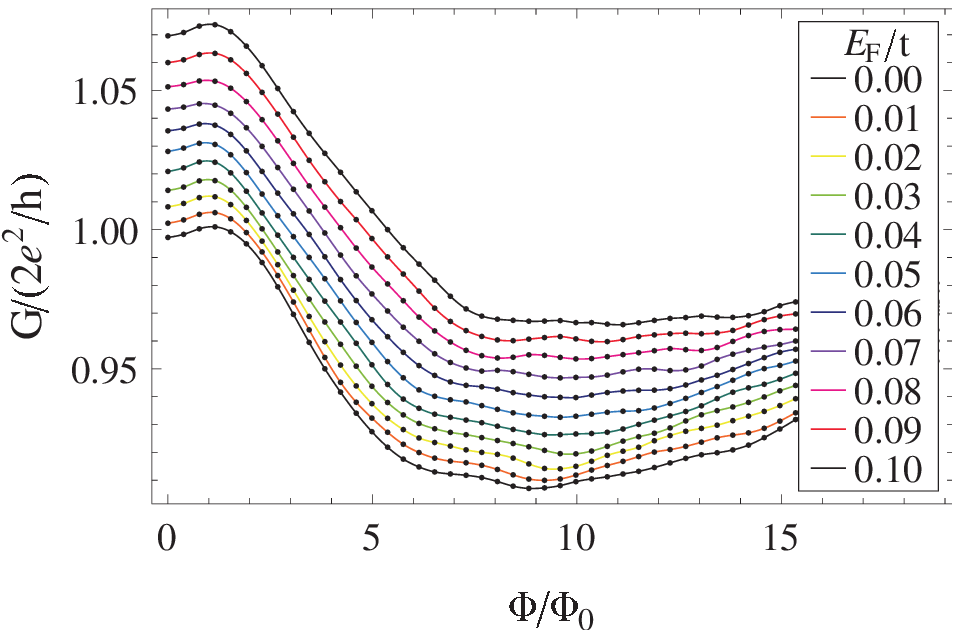}}
\subfigure[Size 20 $\times$ 48, $\xi = 1$, K = 2] {\label{difftwo}
\includegraphics[width=6cm]{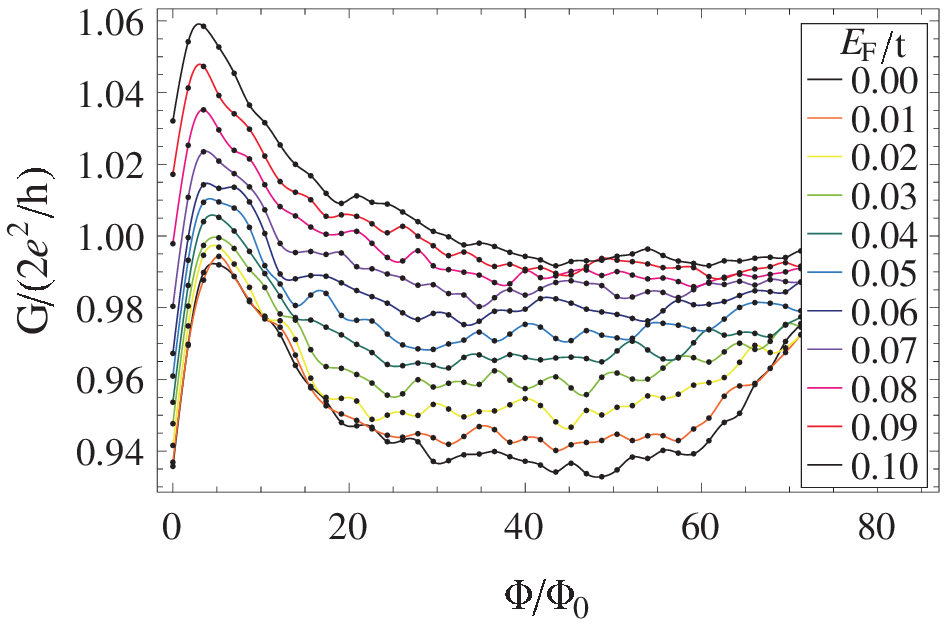}}
\caption[short caption]{\label{diffplot} Samples of numerical  results
 showing strongly nonmonotonic  magnetoconductance.}
\end{figure}

\subsection{\label{tata}Conductivity}

\subsubsection{\label{spsy} Magnetoconductance sign}

Numerical results for the magnetoconductivity are displayed in
Fig. \ref{crossings}. We can distinguish the weak localization from
the weak antilocalization by the sign of the magnetoconductance:
\begin{equation}
\Delta\sigma_{\text{sgn}} (B ) = \sgn {\left( \frac{d \Delta\sigma
    (B)}{dB} \right)}. \label{zeroone}
\end{equation}
Negative magneto conductance $\Delta\sigma_{\text{sgn}} (B) =-1$ at
weak magnetic fields $B$ corresponds to weak antilocalization, since
the magnetic field reduces the quantum correction and thereby the
conductance. Positive magneto conductance $\Delta\sigma_{\text{sgn}}
(B) =1$ corresponds to weak localization.  As seen in
Figs. \ref{diffplot}, this sign can change at large magnetic
fields. This occurs as soon as the magnetic rate  $1/\tau_B$ exceeds
all symmetry breaking rates $1/\tau_{ij}$ defined in
Eq. (\ref{tauij}). We note that the two-dimensional samples considered
in the numerical calculations are rather small, with length $L$ and
width $W$ that are much smaller than the magnetic length $l_B$ for
moderate magnetic fields $B$ ($l_B =0.026 \mu m
\sqrt{T/B}$). Therefore, the sensitivity of the conductivity to an
external magnetic field $B$ is reduced and the magnetic rate
$1/\tau_B$ becomes suppressed for $l_B \gg L, W$
to\cite{PhysRevLett.87.256801}
\begin{equation}
1/\tau_{B} = c D \frac{L W}{l_B^4},\label{new_tau_b}
\end{equation}
where $c$ is a geometrical factor of order unity. On the other hand,
the symmetry breaking rates $1/\tau_{ij}$, Eq. (\ref{tauij}) do not
depend on the system dimensions $L$ and $W$. Only the relaxation rate
originating from the warping term, $1/\tau_w$, acquires a small sample
size dependence and becomes smaller when $D \tau_w \gg LW$. Therefore,
as seen in Fig. \ref{crossings}, the change of the weak localization
corrections occurs only when  the magnetic fluxes through the samples
corresponds to huge magnetic fields. This is due to the small samples
considered in the numerical calculations.

As seen in Fig. \ref{Weak localization}, for small correlation length
one observes positive magnetoconductance, while at a larger
correlation lengths negative magnetoconductance occurs for the same
impurity strength $K_0$, see Fig. \ref{weak antilocalization}. As we
will analyze in detail below, this can be explained by the reduction
of the intervalley scattering amplitude $V_{k_0}$ with the increase of
the correlation length $\xi$, as shown in Fig. \ref{ando1}.

A transition between positive and negative magnetoconductance is
observed as function of the Fermi energy $E_F$, as seen in
Fig. \ref{weak-anti-imp2}. Similar transitions can be observed when
changing $K_0$ or $n_{imp}$, which we do not show here.

\subsubsection{\label{spsy2}Magnetoconductance amplitude}

\begin{figure}[t]
\begin{center}
\includegraphics [width=6cm] {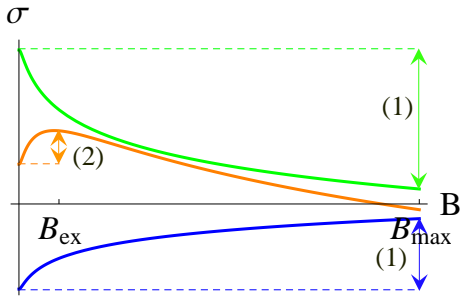}
\caption{\label{draft} Schematic plots of different types of
  magnetoconductance $\sigma(B)$.  At $B_{max}$ all weak localization
  corrections are suppressed. If the magnetoconductance is monotonous
  as in cases (1), the amplitude is $\Delta\sigma (B_{max})$.
  Otherwise, the magnetoconductance has a maximum at $B = B_{ex}$, as
  in case (2).}
\end{center}
\end{figure}

In addition to the sign of the magnetoconductance, the amplitude of
the magnetoconductance $\Delta\sigma$ is important and can reveal more
information on the nature of the impurities in the sample. As a
measure of the amplitude of the magnetoconductance, we take the
difference between the conductance at the first extremum $B_{ex}$ and
the one at zero magnetic field (B=0),
\begin{equation}
\Delta\sigma (B_{ex}) := \sigma (B_{ex}) - \sigma (B=0)\label{continuous}.
\end{equation}
Positive $\Delta\sigma (B_{ex}) $ thus means that there is a weak
localization dip at weak magnetic fields, while negative values
correspond to the amplitude of the weak antilocalization peak.  We
summarize the different types of magnetoconductivity schematically in
Fig. \ref{draft}. The amplitude $\Delta\sigma (B_{ex})$ is indicated
by arrows, both in the case when there is no sign change
$\Delta\sigma_{\text{sgn}} (B)$ as the magnetic field is increased
(1), as well as when the sign changes, (2). The quantum conductance
corrections vanish at large magnetic fields, when $l_B$ is of the
order of the elastic mean free path $l_e$. This magnetic field we
denote as $B_{max}$, as defined by $l_{B_{max}}=l_e$.

As seen in Fig. \ref{crossings}, there are still statistical
fluctuations on smaller magnetic field scales, despite the large
number of realizations $N_c=5000$, which we used in the averaging of
the conductance.

\subsubsection{\label{phase_dia}Phase diagrams}

\paragraph{Magnetoconductance sign}$\;$\\

\begin{figure}
\subfigure[$E_F = 0.01 t$] {\label{01.2}
\includegraphics[width=0.23\textwidth]{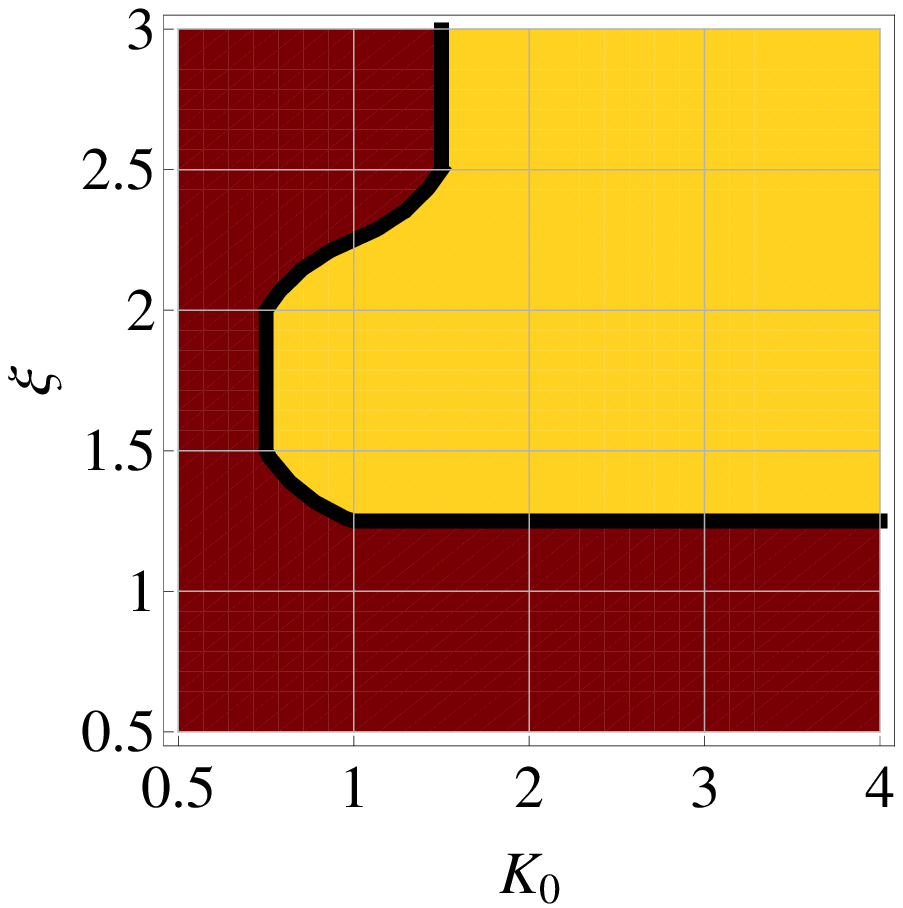}}
\subfigure[$E_F = 0.02 t$] {\label{01.3}
\includegraphics[width=0.23\textwidth]{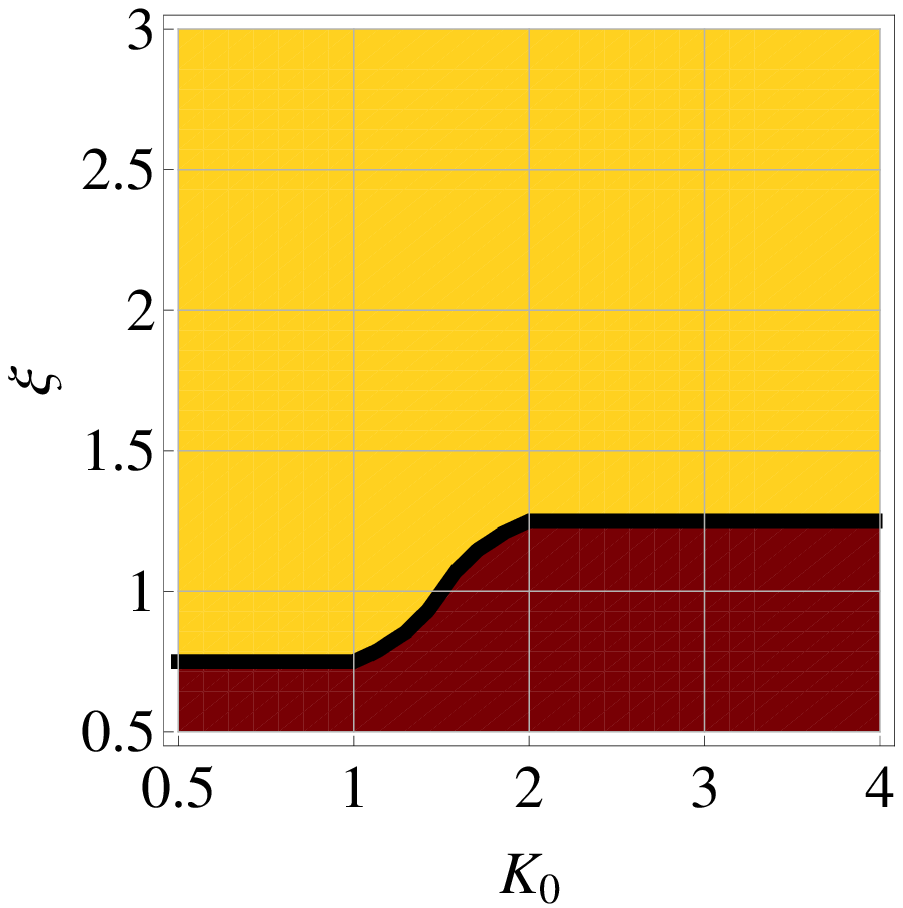}}
\subfigure[$\xi = 2 a_0$] {\label{02.2}
\includegraphics[width=0.23\textwidth]{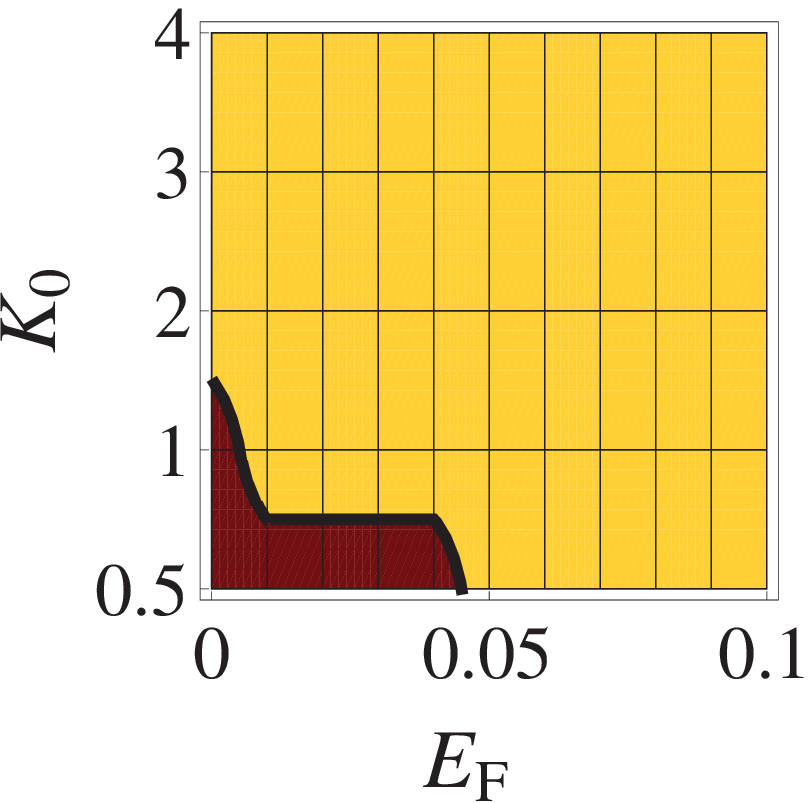}}
\subfigure[$K_0 = 0.5$] {\label{03.2}
\includegraphics[width=0.23\textwidth]{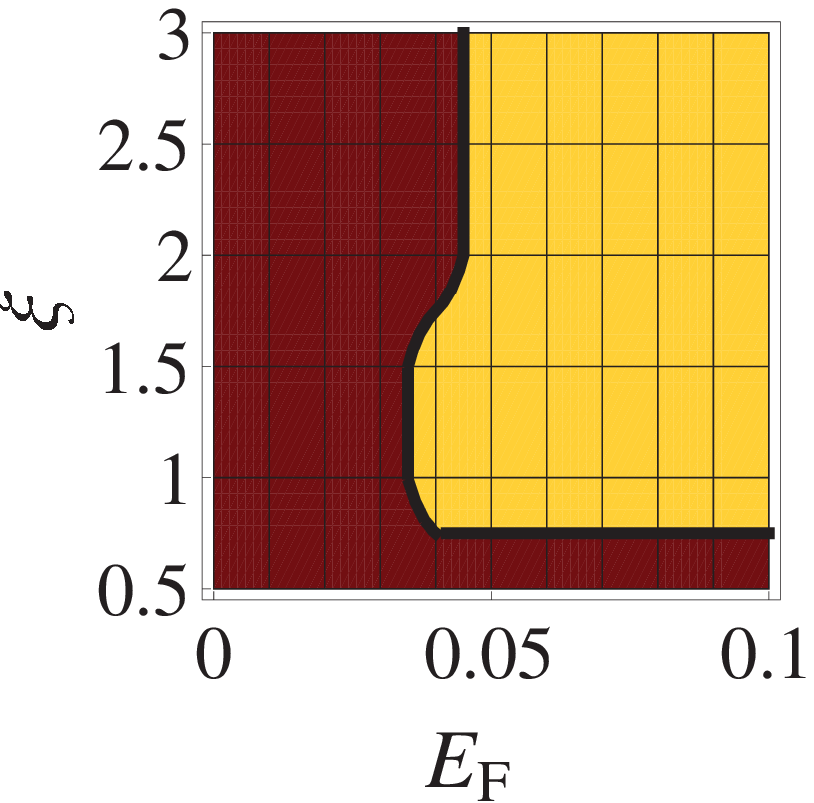}}
\caption[]{\label{xi-k} Numerical magnetoconductance sign phase
  diagrams calculated with Eq. (\ref{zeroone}). (a), (b) as function of
  correlation length $\xi$ and impurity strength parameter $K_0$ at
  fixed Fermi energy, (c) as function of $K_0$ and $E_F$ for fixed
  $\xi$, and (d) as function of $\xi$ and $E_F$ for fixed
  $K_0$. Positive magnetoconductance is indicated by the red area,
  while negative magnetoconductance is yellow. The system size is
  fixed to $M=30$ and $N=72$.}
\end{figure}

In order to study the crossover between positive and negative
magnetoconductance as function of the three impurity parameters $\xi$,
$K_0$, and $E_F$, we use Eq. (\ref{zeroone}) at weak
magnetic fields to assign the value $\Delta\sigma_{\text{sgn}} (B
\rightarrow 0 ) = -1$ for negative magnetoconductance (NMC) and
$\Delta\sigma_{\text{sgn}} (B \rightarrow 0 ) =1$ for the positive
magnetoconductance (PMC). With this information we create three such
sign phase diagrams, varying two parameters while the third parameter
stays fixed.  For better visualization, the diagrams are colored
yellow for negative and red for positive magnetoconductance. The
crossing line is highlighted by the black line. We note that the
numerical data is given only at the crossings of grid lines. We used
an interpolation method to get the continuous crossing line. As seen
in the $\xi - K_0$ diagram, PMC occurs for short-range impurities with
$ \xi < a_0$ in the whole range of impurity strengths $K_0$. This is
expected since short-range impurities cause intervalley scattering and
thereby suppress the pseudospin triplet Cooperons.

For impurities with larger correlation lengths ($\xi >a_0$), NMC is
observed for strong $K_0$ and large $E_F$. Surprisingly, there is a
regime at weak $K_0$ and small $E_F$ where PMC occurs even for
large $\xi$.  We observed similar behavior for other sizes M and N
which we do not display here.  In the $E_F - K_0$ diagram we see a
change from PMC to NMC when moving away from the Dirac point. At the
Dirac point we observe a change at $K_0 \thickapprox 1$. For
$E_F > 0.05$ t only NMC is observed for $\xi =2 a_0$.  The
$\xi - E_F$ phase diagram shows that for $\xi \ll a_0$ PMC
occurs independent of the Fermi energy. At larger $\xi$ a change
to NMC occurs, as expected.

\begin{figure}
\begin{center}
\includegraphics [width=6cm] {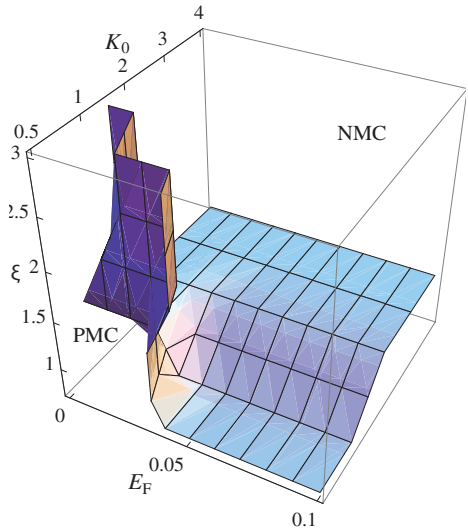}
\caption[legend-back-gate]{\label{sign-3d} $E_F - K_0 - \xi$
  phase diagram for $M=30$ and $N=72$, where the magnetoconductance
  sign is obtained from the numerical results with
  Eq. (\ref{zeroone}). The surface with vanishing magnetoconductance
  is displayed. }
\end{center}
\end{figure}

In order to get a simpler representation of the results, we merge all
three two-dimensional phase diagrams into one three-dimensional phase
diagram, in which each of the parameters $\xi$, $K_0$ and
$E_F$ is represented by one axis, Fig.(\ref{sign-3d}).

\paragraph{Magnetoconductance Amplitude}

\begin{figure}
\subfigure[$E_F = 0.01 t$] {\label{04.2}
\includegraphics[width=0.23\textwidth]{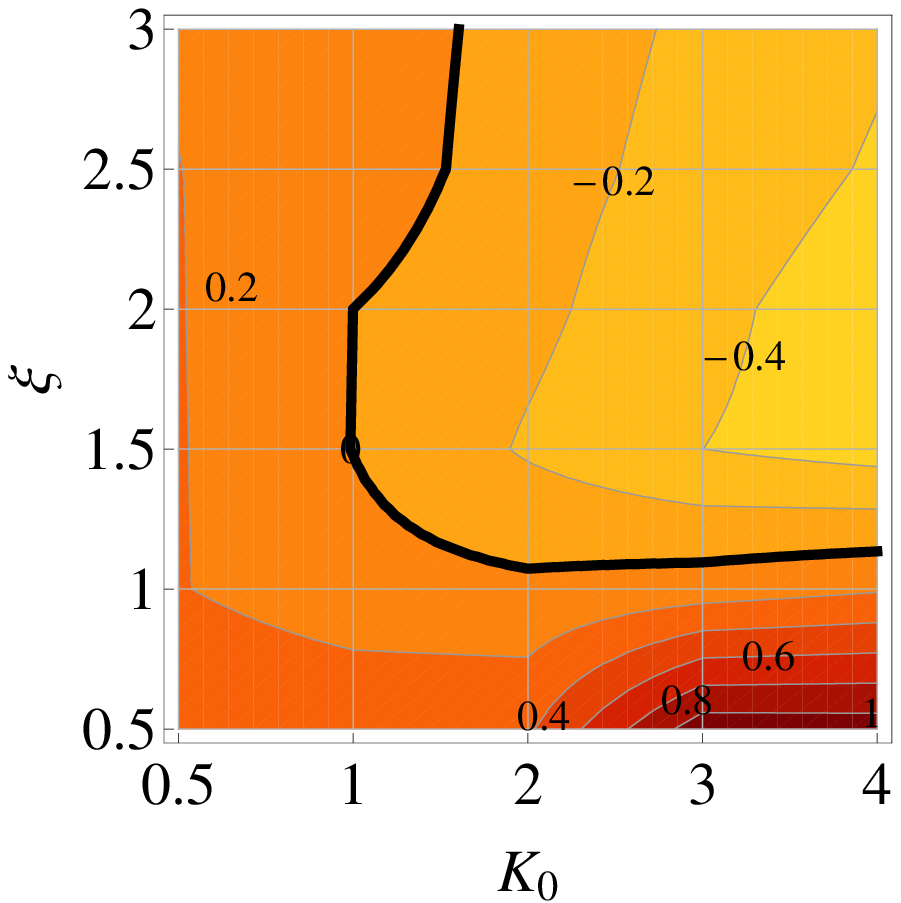}}
\subfigure[$E_F = 0.1 t$] {\label{04.3}
\includegraphics[width=0.23\textwidth]{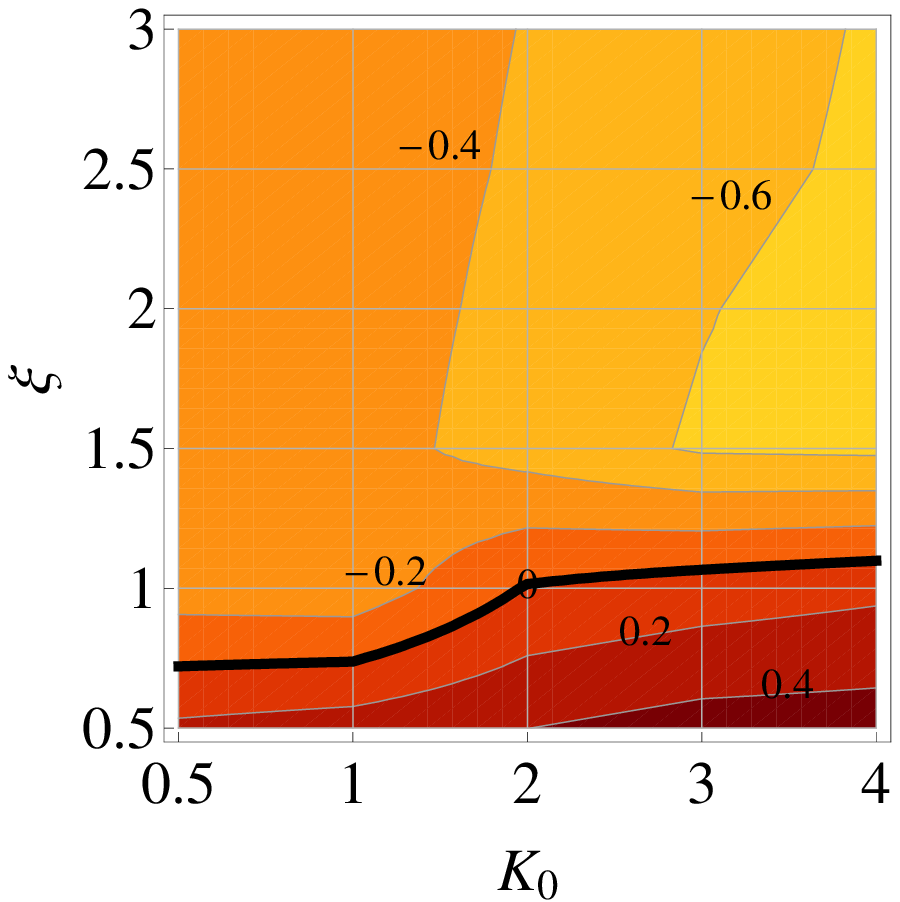}}
\subfigure[$\xi = 2$] {\label{05.2}
\includegraphics[width=0.23\textwidth]{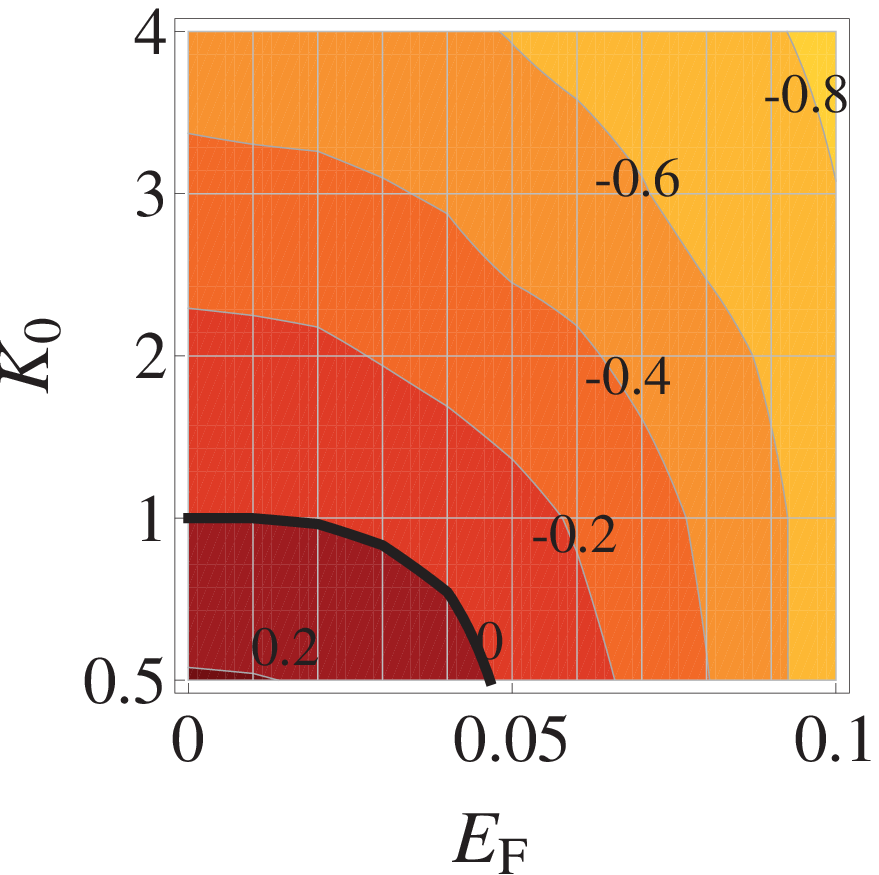}}
\subfigure[$K_0$ = 0.5] {\label{06.2}
\includegraphics[width=0.23\textwidth]{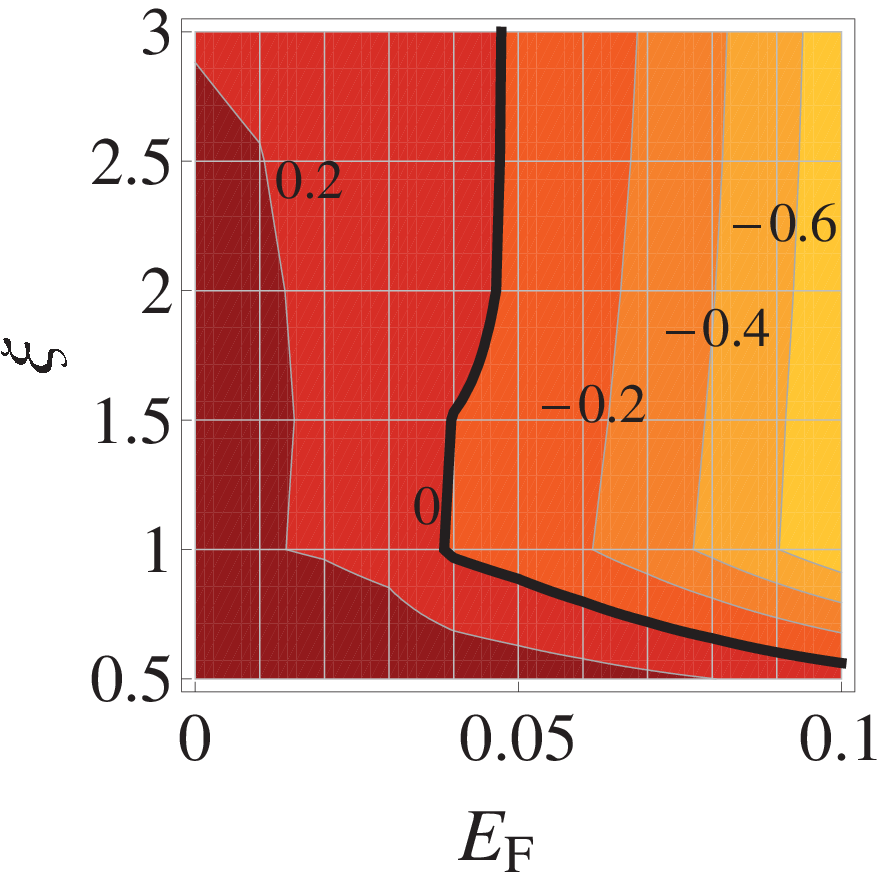}}
\caption[]{\label{xi-k3} Magnetoconductance amplitude obtained from
  the numerical results with Eq. (\ref{continuous}) in (a), (b) the $\xi -
  K_0$ for two $E_F$, (c) the $K_0 - E_F $ and (d) the $\xi -
  E_F$ plane, respectively. The black line indicates
  vanishing magnetoconductance. Contour lines are in units of $2
  e^2/h$. $M=30$ and $N=72$.}
\end{figure}

\begin{figure}
\begin{center}
\includegraphics [width=6cm] {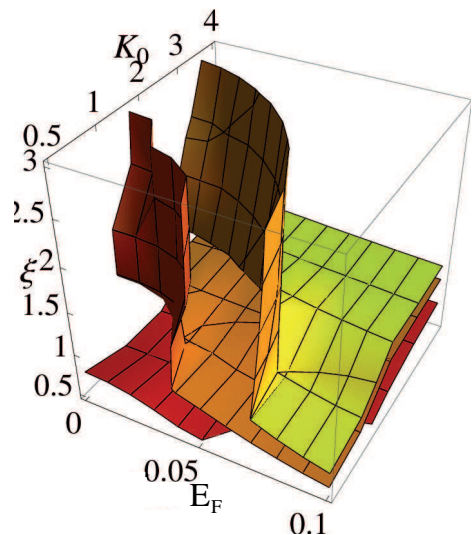}
\caption[legend-back-gate]{\label{cont-3d} Numerical results as
  function of the parameters $E_F, K_0, \xi$ for the
  magnetoconductance amplitude, Eq. (\ref{continuous}), for $M=30$ and
  $N=72$. Surfaces of value $0.3$ (red), $0$ (orange), $-0.3$ (yellow)
  are displayed, in units of $2 e^2/h$. Each grid line crossing
  corresponds to a numerically calculated value.}
\end{center}
\end{figure}

Using Eq. (\ref{continuous}) we determine the amplitude of the
magnetoconductance and plot it in contour plots in the $\xi - K_0$,
the $E_F - K_0$ and the $\xi - E_F$ plane,
respectively, in Fig. \ref{xi-k3}. In these figures, the contour lines
are tagged with the corresponding numbers calculated from
Eq. (\ref{continuous}). The main features already observed in the
magnetoconductance sign phase diagrams can be seen in
Fig. \ref{xi-k3}: PMC occurs for short-range impurities with $ \xi \ll
a_0$ irrespective of $K_0$ and $E_F$. For impurities with larger
correlation lengths ($\xi >a_0$), there is NMC for strong $K_0$ and
large $E_F$, as expected. At weak $K_0$ and small $E_F$, PMC
occurs even for large $\xi$. Its amplitude remains weak, however,
corresponding to only a small weak localization dip at weak magnetic
fields. This is displayed in Fig. \ref{cont-3d}.

\subsection{Thermopower\label{thermo_section}}

\begin{figure}
\subfigure[$\xi = 2$] {\label{thermoverhalten1}
\includegraphics[width=6cm]{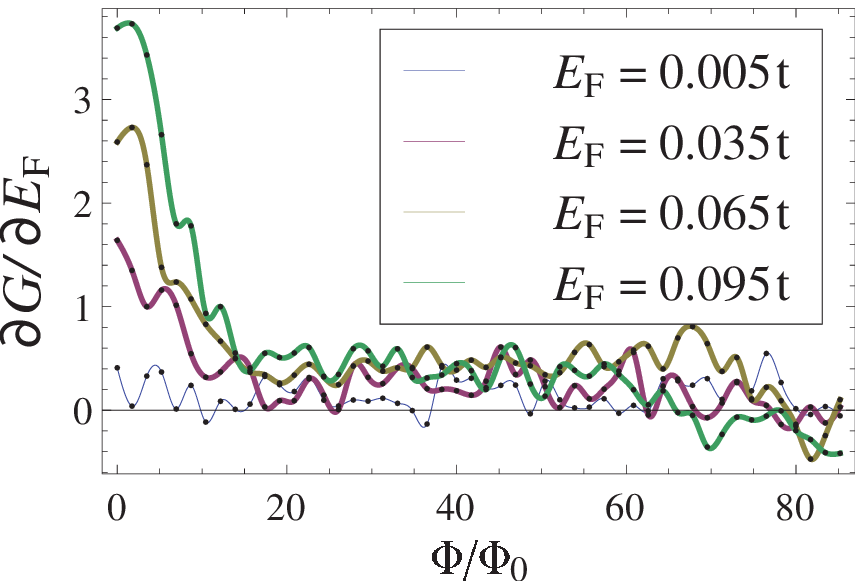}}
\subfigure[$\xi = 2$] {\label{thermoverhalten2}
\includegraphics[width=6cm]{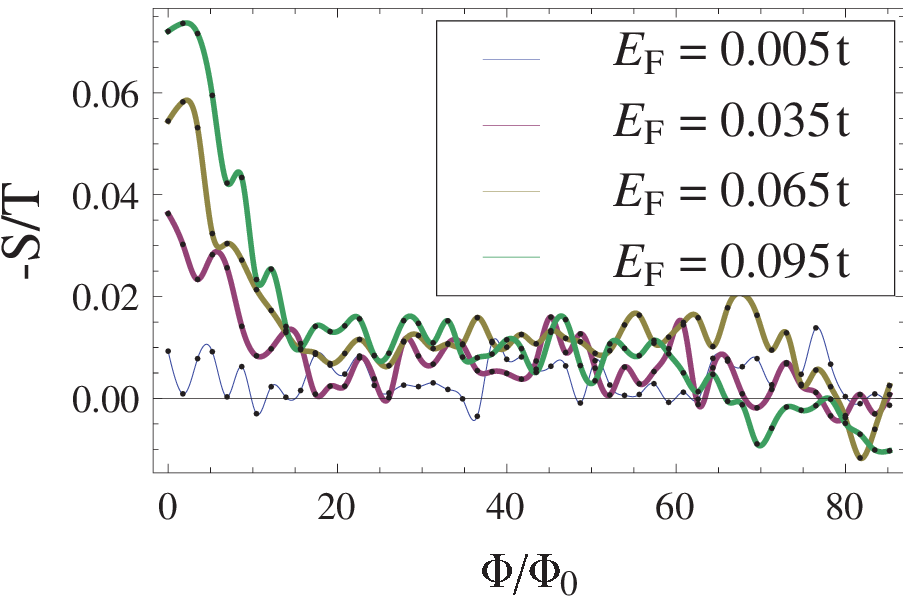}}
\caption[]{\label{xi-v3-thermopower} Numerical results for (a) $d G/d
  E_F$ in units of $(2 e^2/h)/eV$ as function of magnetic
  flux $\phi$ ($M=30$, $N=72$, $K_0=2$). (b) Magnetothermopower in
  units of $\mu V/K^2$ for same system as (a). The corresponding
  conductivity is shown in Fig. \ref{weak antilocalization}.}
\end{figure}

\begin{figure*}
\subfigure[$M=80$, $N=192$] {\label{delle1}
\includegraphics[width=0.30\textwidth]{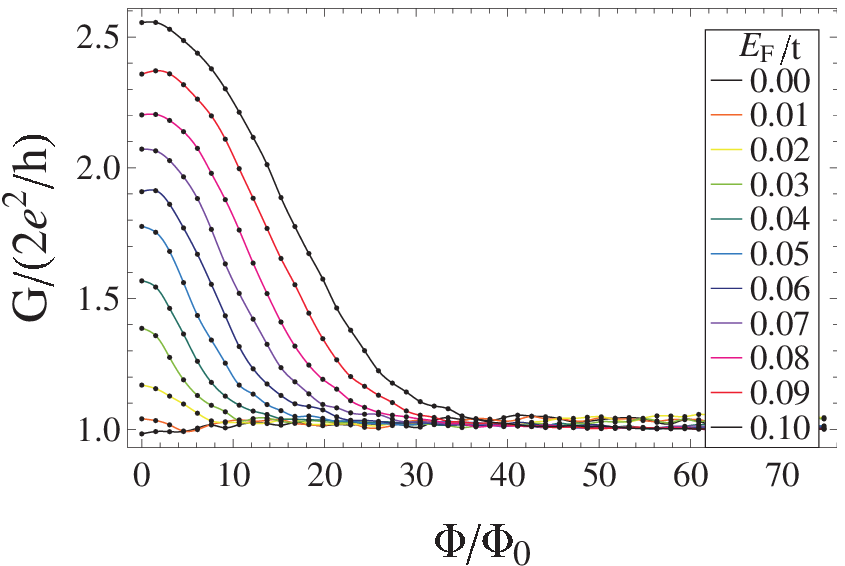}}
\subfigure[$M=80$, $N=192$] {\label{delle2}
\includegraphics[width=0.30\textwidth]{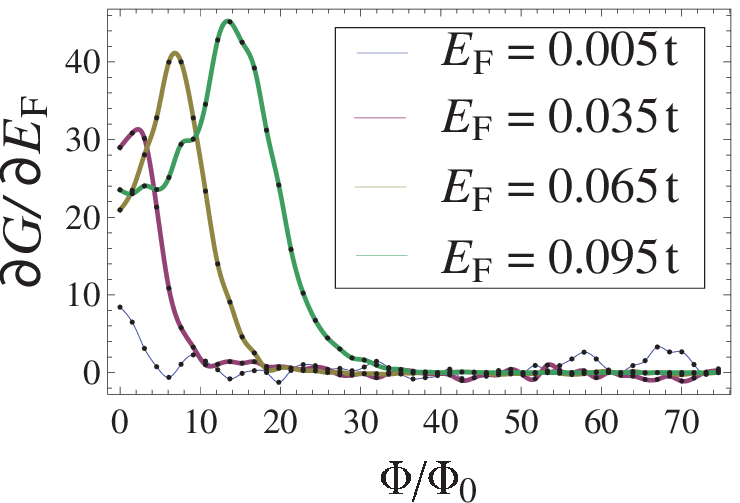}}
\subfigure[$M=80$, $N=192$] {\label{delle3}
\includegraphics[width=0.30\textwidth]{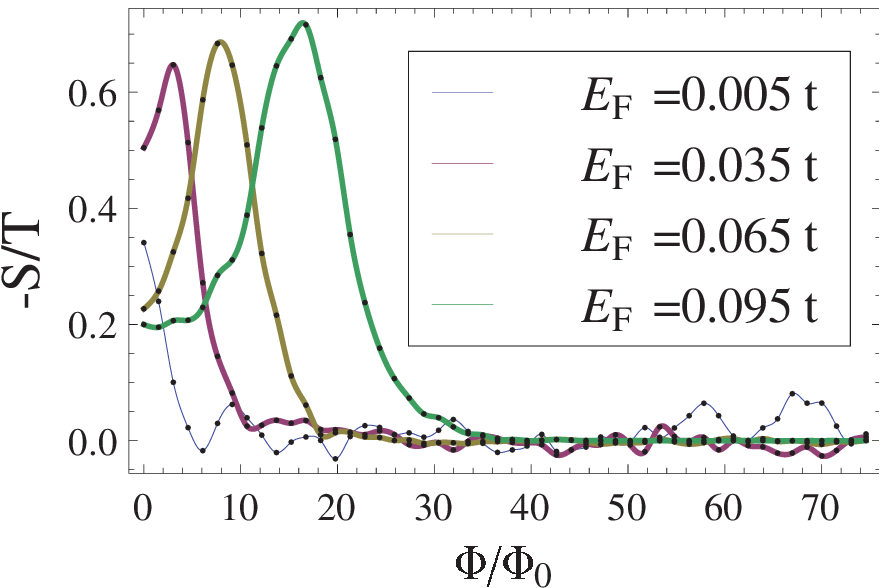}}
\caption[]{\label{delle} Numerical results for (a) conductance (b) $d
  G/d E_F$ in units of $(2 e^2/h)/eV$. (c)
  Magnetothermopower in units of $\mu V/K^2$ ((a)-(c): $K_0=0.5$,
  $\xi=2$).}
\end{figure*}

\begin{figure}
\subfigure[$M=60$ and $N=144$.]{\label{05.3d}
\psfrag{V}{\small{E}}
\psfrag{BG}{\tiny{F}}
\includegraphics[width=0.23\textwidth]{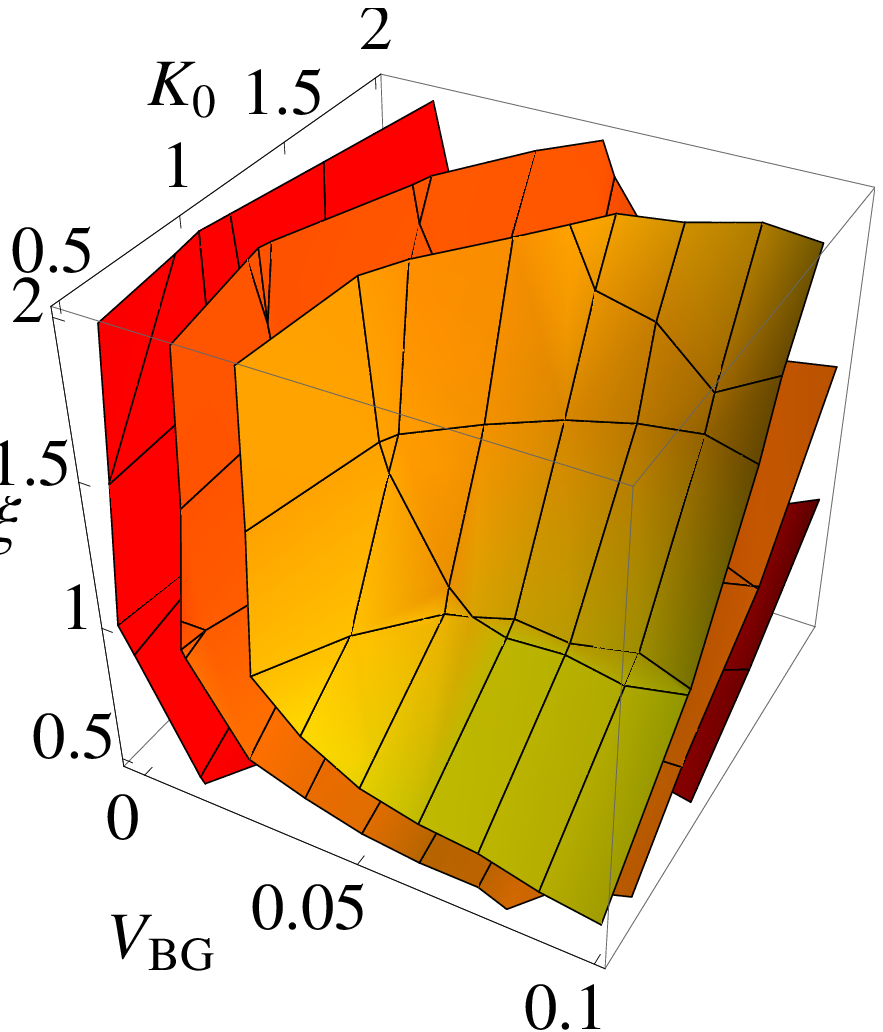}}
\subfigure[$M=60$ and $N=144$.]{\label{06.3d}
\psfrag{V}{\small{E}}
\psfrag{BG}{\tiny{F}}
\includegraphics[width=0.23\textwidth]{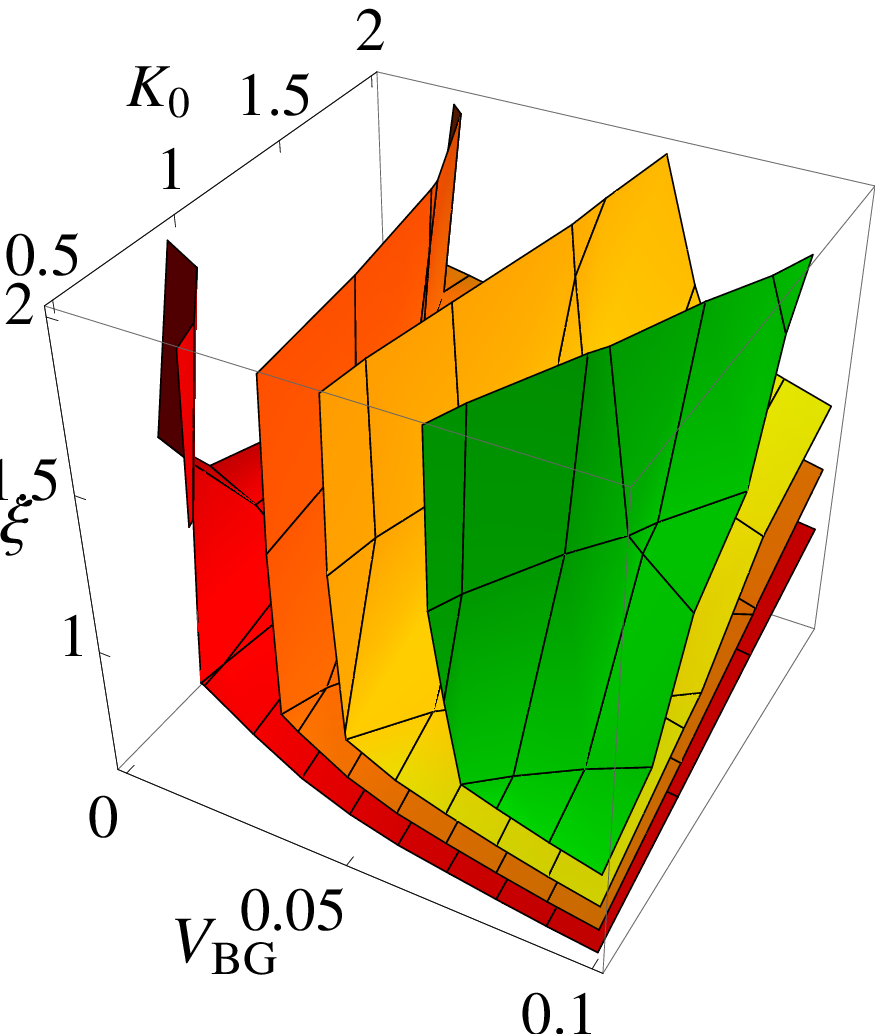}}
\caption[]{\label{crossings2} Numerical results for (a)
  magnetothermopower amplitude, $\Delta S/T$, Eq. (\ref{thermo-sign}),
  as function of $E_F, K_0, \xi$. Surfaces with value
  $-0.05$ (red), $-0.1$ (orange), $-0.2$ (yellow) in units of $\mu
  V/K^2$ are displayed. (b) magnetoconductance amplitude
  $\Delta\sigma$, Eq. (\ref{continuous}), as function of the
  parameters $E_F, K_0, \xi$. Surfaces of value $0$ (red),
  $-0.3$ (orange), $-0.6$ (yellow), $-1$ (green) in units of $2 e^2/h$
  are displayed. (The grid corresponds to the numerically calculated
  values.)}
\end{figure}

Next, we calculate the thermopower $S$, as introduced in
Sec. \ref{thermo}, using the Mott formula of Eq. (\ref{mott}) in order
to see if there are quantum corrections and, as a result,
magnetothermopower. To this end, we use the results for the
conductance $G$ and its dependence on Fermi energy $E_F$
from the previous section to calculate and analyze $\frac{d G}{d
  E_F}$ and the thermopower $S$. We note that by expanding
in the quantum corrections of the conductivity to first order, we can
write the quantum correction to the thermopower $\delta S$ as
\begin{equation} \label{squantum}
\frac{\delta S}{T} = - \frac{S_{cl}}{T} \frac{\delta G}{G_{cl}}  - \frac{\pi^2}{3} \frac{k_B^2}{|e|}
\frac{\delta d G/dE}{G_{cl}},
\end{equation}
where $\frac{\pi^2}{3} \frac{k_B^2}{|e| eV} = 0.024 \mu V/K^2$. Since
the conductance change with magnetic field is only of the order of
$e^2/ h$, the first term in Eq. (\ref{squantum}) is of order
$\frac{S_{cl}}{T}$. Moreover, since $S_{cl}$ vanishes at the Dirac
point and increases linearly with the gate voltage, we expect the last
term to become more important in the vicinity of the Dirac
point. This is confirmed by the numerical results shown in
Fig. \ref{thermoverhalten2} where $\frac{d G}{d E_F}$ is
plotted as function of magnetic field. This quantity decays with
magnetic field; the larger the Fermi energy $E_F$, the larger is
the sensitivity to the magnetic field. As shown in
Fig. \ref{thermoverhalten1}, the thermopower shows the same
qualitative behavior, namely, a reduction of the amplitude of the
thermopower when a magnetic field is applied. For positive gate
voltage the thermopower is negative, so that we obtain {\it positive
  magnetothermopower}.

For the largest samples considered, we also observe an initial
increase of $\frac{d G}{d E_F}$ at weak magnetic fields, and
a peak which becomes enhanced with increasing Fermi energy (see
Fig. \ref{delle}), followed by a decay to its classical value at
larger magnetic fields. Correspondingly, in these samples the
amplitude of the thermopower (see Fig. \ref{delle3}) first is enhanced
when the magnetic field is applied  and then it decreases toward
its classical value at larger magnetic fields.

Since the magnetic field suppresses quantum corrections, we can obtain
the classical value numerically from the high-magnetic field
limit. Since the thermopower is fluctuating even in the range where it
should saturate towards the classical limit, we needed to average over
the fluctuations of the data points at different magnetic fields.  For
every set of parameters we thereby calculate $S_{mean}$ and use as a
definition for the quantum corrections the difference to the
thermopower without magnetic field,
\begin{equation}
\Delta S := S(B = 0) - S_{mean}.\label{thermo-sign}
\end{equation}
In the following, we explore the size and sign of these quantum
corrections of the thermopower, and how they depend on the impurity
parameters, such as the strength $K_0$ and the correlation length
$\xi$, and on the Fermi energy. In Fig. \ref{05.3d} we plot the values
of $\Delta S$ as function of these three parameters. As we are
particularly interested on how these quantum corrections to the
thermopower are related to the weak localization corrections to the
conductivity, we also plot the magnetoconductance amplitude (MCA) in
Fig. \ref{06.3d} in units of $2 e^2/ h$. We notice that the set of
parameters giving the most pronounced weak antilocalization seems also
to yield the strongest thermopower enhancement. We will explain and
analyze this effect in detail in Sec.  \ref{kant_thermo}. This
behaviour can also be seen when looking at the MC amplitude phase
diagrams as function of $E_F$ in Fig. \ref{xi-k3} since the
derivative $\frac{d G}{d E_F}$ is the highest when there is
a strong increase of values in the $E_F$ direction. This is
the case in the regime of weak antilocalization, while a smaller
change is observable in the weak localization regime, see
Figs. \ref{05.2} and \ref{06.2}.

\section{\label{intandana}Analysis of Numerical Results}

In order to get a better understanding of these numerical results for
the magnetoconductance and the magnetothermopower, we compare them
with the analytical expression of Eq. (\ref{final}) and use the
scattering rates $1/\tau _{z }$ and $1/\tau _{i }$ and the effective
dephasing rate $1/\tau _{\phi }$ as fitting parameters.  We also
attempt an ab initio calculation where we use the input parameters of
the numerical calculations to calculate directly the different
scattering rates $1/\tau_{ij}$, insert these in the analytical
expression, and compare the result with the numerical ones.

\subsection{\label{fitting}Fitting of the numerical results}

For the fitting procedure, we picked the sample size $N=30$ and
$M=72$, where we observed a clear transition between weak
antilocalization and weak localization when changing the parameters
$K_0$, $\xi$ and the Fermi energy, see Fig. \ref{xi-k3}.  As
mentioned before, since the magnetic length exceeds the system size
$L$ in the magnetic flux range considered, we have to use the
appropriate magnetic scattering rate $1/\tau _{B}$,
Eq. (\ref{new_tau_b}), which is strongly reduced by a geometrical
factor compared to the two-dimension limit.  As discussed above, since
the numerical results are done at zero temperature, the dephasing rate
$1/\tau _{\phi }$ can be substituted by the Thouless energy, see
Eq. (\ref{93}).

\subsubsection{\label{fitting1}Fitting with a fixed $1/\tau _{\phi }$}

In this subsection, the two fitting parameters in Eq. (\ref{final})
are taken to be $1/\tau _{z }$ and $1/\tau _{i }$, while $1/\tau
_{\phi }$ is calculated directly from Eq. (\ref{93}).  For the
parameter range $\xi/a_0 = 1, 1.5, 2$, $K_0 = 1, 1.5, 2$ and
$E_F/t = 0.01, 0.05$, we find that Eq. (\ref{final}) fits
well.  Within this parameter region, we find the transition line
between weak antilocalization and weak localization.  We concentrate
our fitting procedure to the validity range of the analytical formula
at small magnetic flux, where $l_B > L$, Eq. (\ref{new_tau_b}).

\begin{figure}
\begin{center}
\subfigure[$\enspace$ $E_F = 0.01$, $K_0 = 1$, calculated
  $\tau _{\phi }$][ $\tau _{\phi }$ calculated with
  Eq. (\ref{93})]{\label{fitting_2.1_result_1}
  \includegraphics[width=6cm]{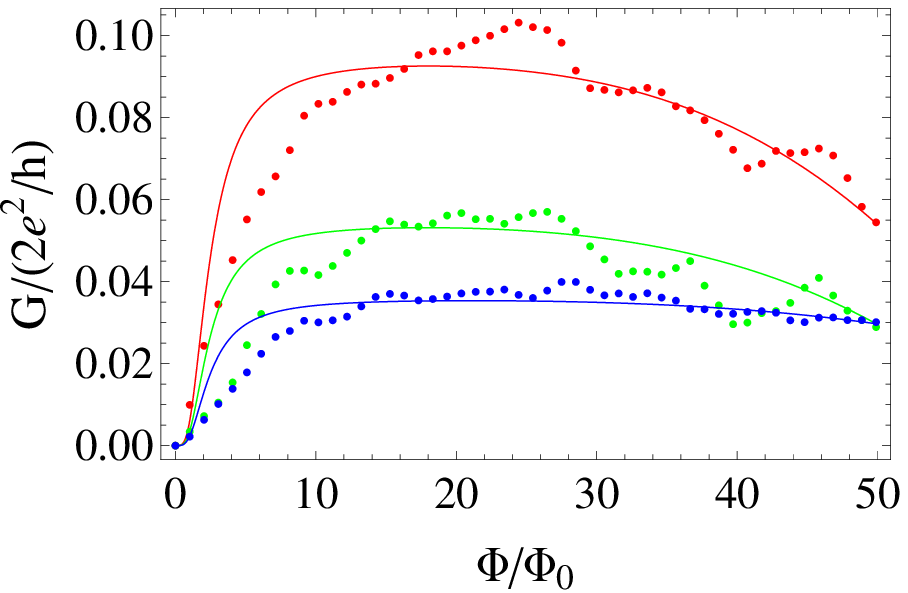}}
\subfigure[$\tau _{\phi }$ fitted][$\tau _{\phi }$ fitted]
          {\label{fitting_2.1_result_2}
            \includegraphics[width=6cm]{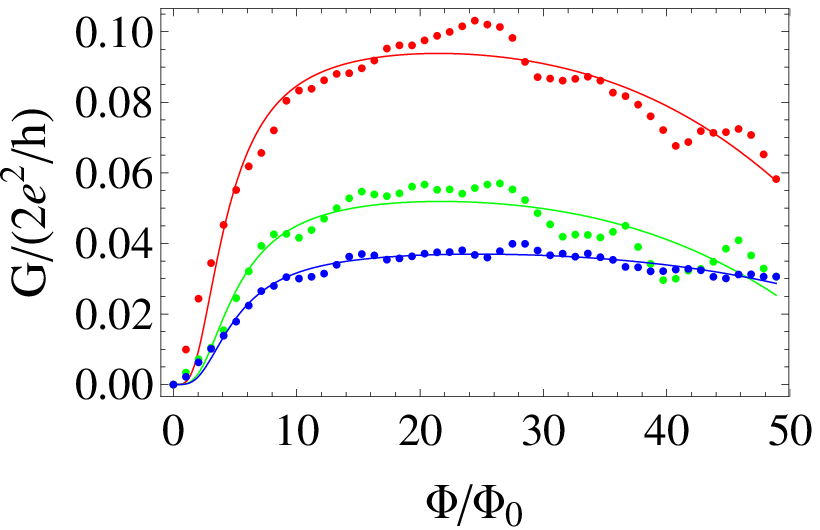}}
\caption[]{\label{fit_big} The numerical results for the
  magnetoconductance amplitude $\Delta G(\phi) = G(\phi) - G(0)$ for
  $M=30$, $N=72$ together with the fitting in terms of $1/\tau_i$,
  $1/\tau_z$ in Eq. (\ref{final}).  Red $\xi = 1$, green $\xi = 1.5$,
  blue $\xi = 2$, $E_F = 0.01$, $K_0 = 1$.}
\end{center}
\end{figure}

We show the results of the fitting exemplary for some parameters in
Fig. \ref{fitting_2.1_result_1}, where the magnetoconductance is
positive.  In that case of pronounced weak localization, the fitting
becomes worse, and the uncertainty in the set of $1/\tau_i$ and
$1/\tau_z$ is larger.

\begin{figure*}
\begin{center}
\subfigure[$\enspace$ With calculated $\tau _{\phi }$][$\tau _{\phi }$ from Eq. (\ref{93}), $E_F = 0.01 t$.] {\label{fitting_2.1_result_3.1}
\includegraphics[width=0.37\textwidth]{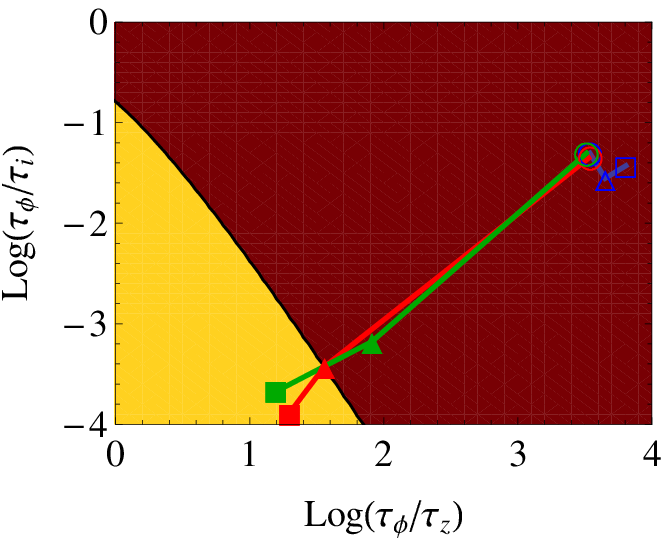}}
\subfigure[$\enspace$ With calculated $\tau _{\phi }$][$\tau _{\phi }$ from Eq. (\ref{93}), $E_F = 0.05 t$] {\label{fitting_2.1_result_3.2}
\includegraphics[width=0.45\textwidth]{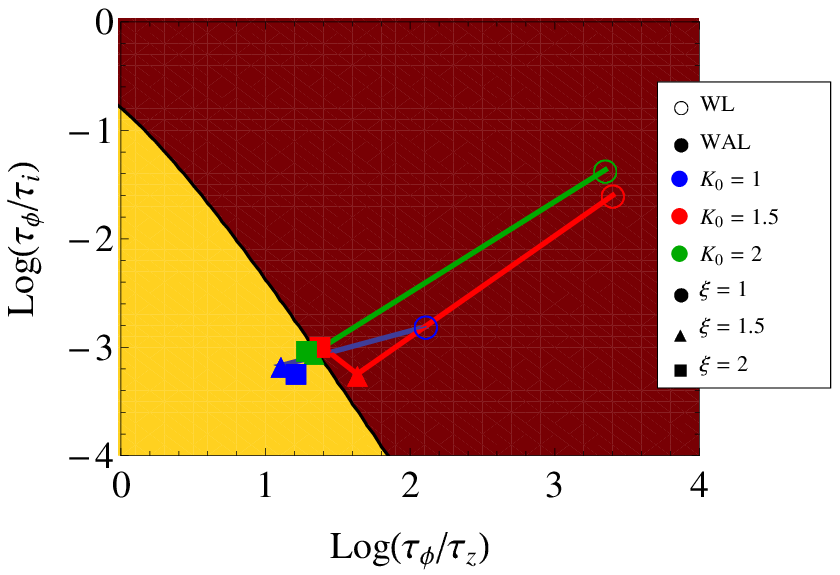}}
\subfigure[$\enspace$ With fitted $\tau _{\phi }$][$\tau _{\phi }$ fitted, $E_F = 0.01 t$] {\label{fitting_2.1_result_4.1}
\includegraphics[width=0.375\textwidth]{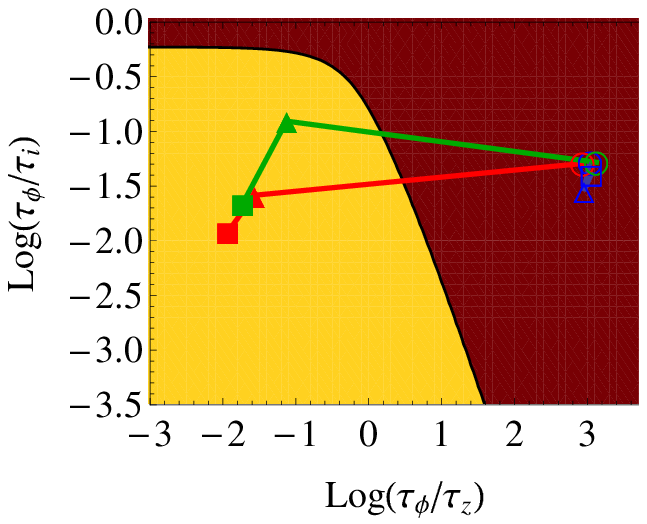}}
\subfigure[$\enspace$ With fitted $\tau _{\phi }$][$\tau _{\phi }$ fitted, $E_F = 0.05 t$] {\label{fitting_2.1_result_4.2}
\includegraphics[width=0.39\textwidth]{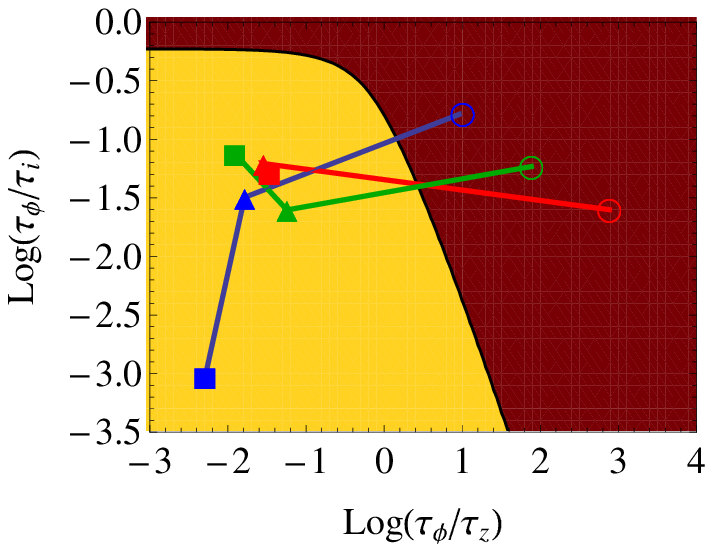}}
\caption[$\enspace$ $\tau _{\phi }/\tau _i$ - $\tau _{\phi }/\tau
  _z$]{\label{fit_big_2.1} $\tau _{\phi }/\tau _i$ - $\tau _{\phi
  }/\tau _z$ diagram obtained from fitting of numerical results. Empty
  symbols: weak localization. Filled symbols: weak
  antilocalization. Black line: obtained from $\Delta \sigma (B) =0$
  in Eq. (\ref{res}).}
\end{center}
\end{figure*}

For each parameter set used in the numerical calculations, we get a
set of fitted values of the rates $\tau _{\phi }/\tau _i$ - $\tau
_{\phi }/\tau _z$ which we plot in the diagrams of
Figs. \ref{fitting_2.1_result_3.1} and \ref{fitting_2.1_result_3.2} as
a symbol. The color of the symbols indicates the value of $K_0$, while
the symbol itself depends on the correlation length as shown in the
legend.  Weak localization corresponds to empty symbols and weak
antilocalization to filled symbols.  The range of scattering rates is
$\log\left(\tau _{\phi }/\tau _i\right) \approx -4 ... 0$ and
$\log\left(\tau _{\phi }/\tau _z\right) \approx 1... 4$. Thus, the
intervalley scattering rates are typically several magnitudes smaller
than the intravalley scattering rates for the range of $\xi$
considered.  The black line is obtained from $\Delta \sigma (B) =0$ in
Eq. (\ref{res}), indicating the transition between weak
antilocalization and weak localization.  With the exception of a few
data points, we find good agreement.  We can improve the agreement by
fitting also $\tau_{\phi}$ as shown in the next section.

\subsubsection{\label{fitting2}Fitting  $1/\tau _{\phi }$}

We repeat the fitting procedure for the same sample but now taking
$1/\tau _{\phi }$ as a third fitting parameter.  We focus on the range
of parameters $\xi/a_0 = 1, 1.5, 2$, $K = 1, 1.5, 2$ and
$E_F/t = 0.01, 0.05$.  Close to the transition the extra
fitting parameter does not change the result significantly.  The
result of the fitting procedure for some parameter values is displayed
in Fig. \ref{fitting_2.1_result_2} . As in Sec. \ref{fitting1}, we
display the resulting $1/\tau_i$, $1/\tau_z$, and $1/\tau_{\phi}$ in a
$\tau _{\phi }/\tau _i$ - $\tau _{\phi }/\tau _z$ diagram in
Figs.\ref{fitting_2.1_result_4.1} and \ref{fitting_2.1_result_4.2}.

We find that the results fall in the range $\log\left(\tau _{\phi
}/\tau _i\right) \approx -3 \ldots 0$ and $\log\left(\tau _{\phi
}/\tau _z\right) \approx -3 \ldots 3$.  The agreement with the
analytical results, the black solid line in Fig. \ref{fit_big_2.1}, is
improved. All numerical results fall on the expected side of the weak
antilocalization to weak localization transition line.

\subsection{\label{imanuel}Ab initio calculation}

In this section we will attempt an ab initio calculation in the sense
that we use the input parameters of the numerical calculations to
calculate directly the different scattering rates $1/\tau_{ij}$,
insert them into the analytical formula of Eq. (\ref{res}) for the
weak localization corrections, and compare the latter to the numerical
results.  For this calculation we choose the following parameter
ranges. The impurity width $\xi$ is varied from $0.05\, a_0$ to $1.5\,
a_0$ in steps of $0.05\, a_0$, the impurity strength V from $0.5 eV$
to $5 eV$ in steps of $0.5$ eV (corresponding to $K_0$ between
approximately 0.05 and 2) and $E_F$ from $0.01\, t$ to
$0.1\, t$ in steps of $0.01\, t$ (note that the analytical result is
not valid too close to the Dirac point $E_F = 0$). The
impurity concentration is set to $3\%$ as before, if not explicitly
mentioned otherwise.

\subsubsection{Conductance}

\paragraph{Magnetoconductance sign transition \label{para}}$\;$

\begin{figure}
\begin{center}
\includegraphics [width=8cm] {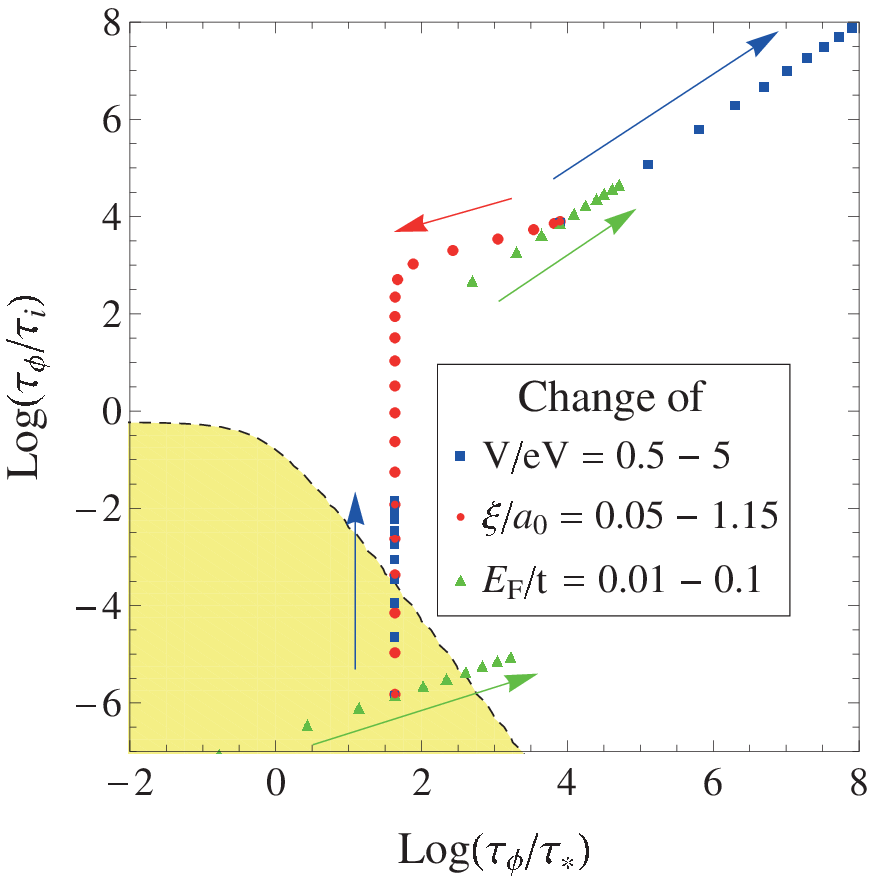}
\caption[]{\label{a_priori1} Magnetoconductance in the $\tau _{\phi
  }/\tau _i$ - $\tau _{\phi }/\tau _\ast $ diagram.  The ratios of
  scattering rates and dephasing rates $\tau_{\phi}/\tau_i$,
  $\tau_{\phi}/\tau _\ast $ are calculated from the parameters $\xi$
  (red dots), $V$ (blue squares) and $E_F$ (green triangles)
  which are varied in the range given in the inset.  Arrows indicate
  the direction of increasing parameter.  The variation with $V$ and
  $E_F$ is shown for $\xi/a_0 =0.05$ and $\xi/a_0 =1.15$,
  $\xi$ is varied at fixed $V/eV =0.5$ and $E_F/t=0.04$.
  Dashed line: localization transition. White area: PMC. Yellow area:
  NMC.  }
\end{center}
\end{figure}

Using the explicit relations between the scattering rates and the
parameters $\xi$, $V$, and $E_F$, as derived in
Sec. \ref{scatteringrates}, in Fig. \ref{a_priori1} we show a $\tau
_{\phi }/\tau _i$ - $\tau _{\phi }/\tau _\ast $ diagram.

In Fig. \ref{a_priori1} we display the magnetoconductance trajectories
in the $\tau _{\phi }/\tau _i$ - $\tau _{\phi }/\tau _{\ast}$ plane by
changing the impurity parameters and $E_F$. For instance,
this is done by changing $\xi$ for fixed $E_F=0.01\,t$ and
$V=5$ eV, red dots, and by changing $V$ and $E_F$ blue/green
dots, respectively, for fixed $\xi = 0.05\, a_0$ and $\xi = 1.5\,
a_0$.  The impurity concentration is kept fixed, its dependence
inferred from Eqs. (\ref{preprop}) and (\ref{prop}).

\begin{figure}
\subfigure[$E_F = 0.01 t$] {\label{a_priori3a}
  \includegraphics[width=0.23\textwidth]{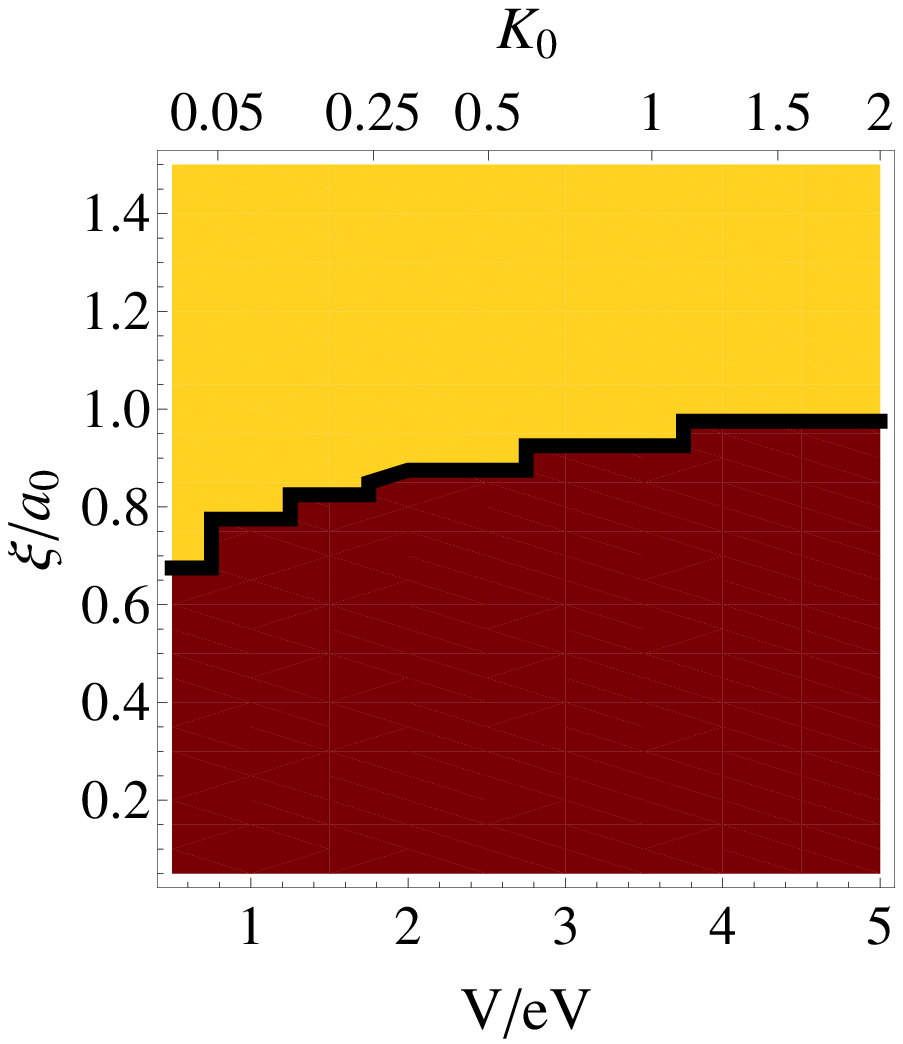}}
\subfigure[$E_F = 0.1 t$] {\label{a_priori3b}
  \includegraphics[width=0.23\textwidth]{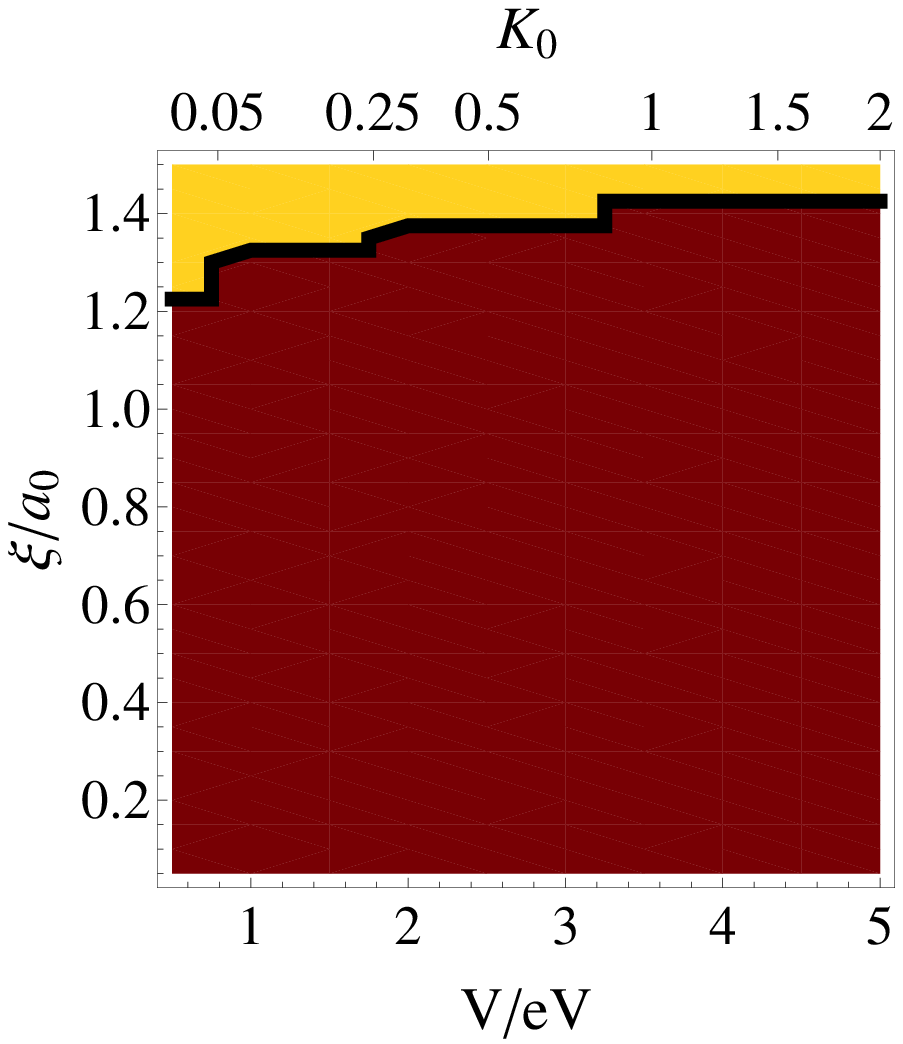}}
\caption[short caption]{\label{a_priori3} Sign of MC as function of
  $V$ and $\xi$, corresponding to Fig. \ref{a_priori1} for two
  different Fermi energies. Red is PMC, yellow is NMC. Compare
  with the numerical results Fig. \ref{01.2}. Black line: transition
  between positive and negative MC.}
\end{figure}

As expected, increasing the range $\xi$ of the impurities induces a
change from weak localization to weak antilocalization, see the red
arrow. When increasing the impurity strength $V$ one moves toward weak
localization, see the blue arrow. Increasing the Fermi energy one
also moves towards weak localization, see the green arrow.  The
diagram shows that $\tau _{\phi }/\tau _{\ast}$ saturates at fixed
Fermi energy $E_F$ toward a lower limit since the ratio of
the warping rate $1/\tau_w$ and the dephasing rate, $\tau _{\phi
}/\tau _{w}$ converges to a finite value, Eq. (\ref{prop2}),
independently of $V$ and $\xi$. At small correlation lengths ($\xi \ll
a_0$) we find that the $1/\tau_z$ term dominates; a $V^4$ dependence
on the strength of the impurities and a $E_F$ dependence
on the Fermi energy emerge, in agreement with Eq. (\ref{prop}).  In
the regime of large correlation lengths ($\xi \gg a_0$), we observe
instead a $E_F^4$ dependence, which is in agreement with the
dependence of the warping rate of Eq. (\ref{prop2}).  For comparison,
we plot the $\tau _{\phi }/\tau _i$ - $\tau _{\phi }/\tau _{z}$
diagram without the warping term in Appendix \ref{appendix_two} (
Fig. \ref{old_a_priori1}).  Next, we transform Fig. \ref{a_priori1}
into the form of the phase diagram of Fig. \ref{01.2}. The resulting
$\xi - V$ magnetoconductance sign diagram is shown in
Fig. \ref{a_priori3}  for two different Fermi energies.

\paragraph{Magnetoconductance Amplitude \label{cond_cont_dia}}$\;$

\begin{figure}
\begin{center}
\subfigure[$E_F = 0.01 t$] {\label{a_priori4a}
\includegraphics[width=0.23\textwidth]{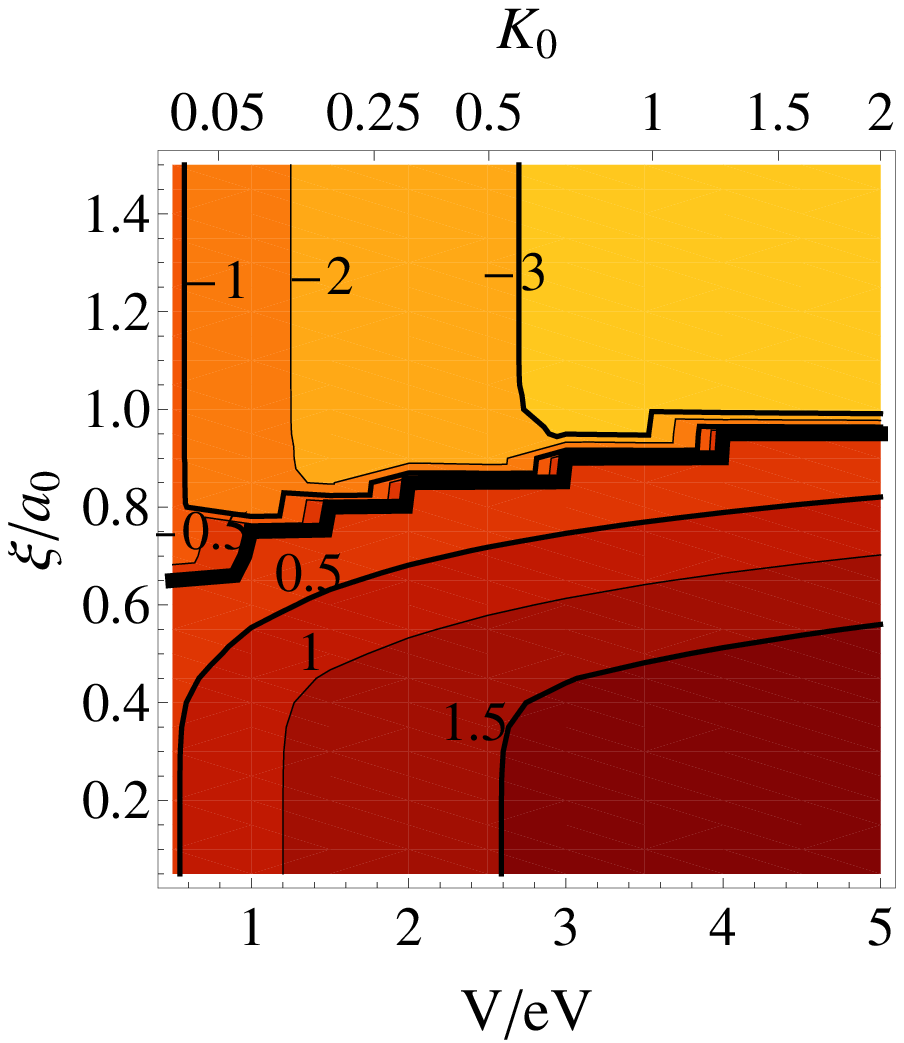}}
%\subfigure[$E_F = 0.1 eV$] {\label{a_priori4b}
%\includegraphics[width=0.23\textwidth]{a_priori4b.eps}}
\subfigure[$E_F = 0.01 t$] {\label{a_priori4b}
\includegraphics[width=0.23\textwidth]{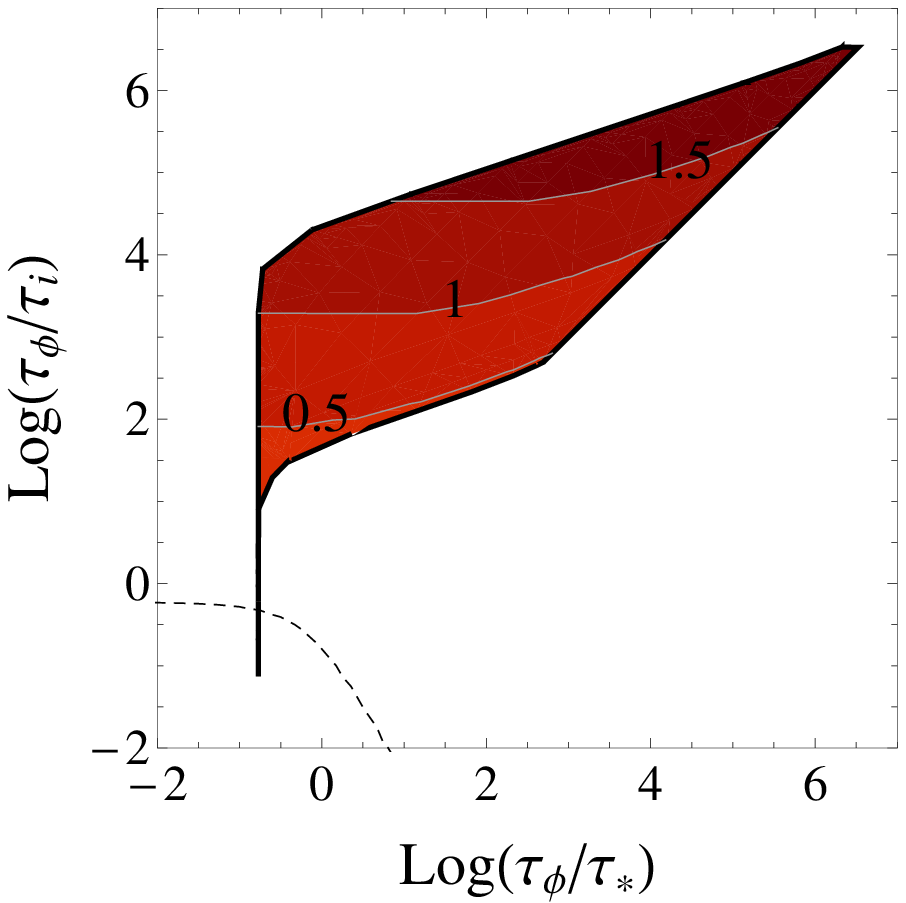}}
%\subfigure[$E_F = 0.1 eV$] {\label{a_priori7b}
%\includegraphics[width=0.23\textwidth]{a_priori7b.eps}}
\caption[short caption]{\label{a_priori4ab} (a): MC amplitude
  analytically calculated, Eq. (\ref{continuous}), in units of $2 e^2/
  h$.  Black line: transition between PMC and NMC.  Positive numbers
  and red color: weak localization.  (b) MC amplitude as function of
  $\tau _{\phi }/\tau _i$ and $\tau _{\phi }/\tau _\ast$.}
\end{center}
\end{figure}

Having investigated the dependence of the sign of the
magnetoconductivity, let us now study its amplitude using
Eq. (\ref{continuous}) to evaluate Eq. (\ref{final}) with
$F_{\text{full}}$ given by Eq. (\ref{ffull}).

\begin{figure}
\begin{center}
\includegraphics [width=8cm] {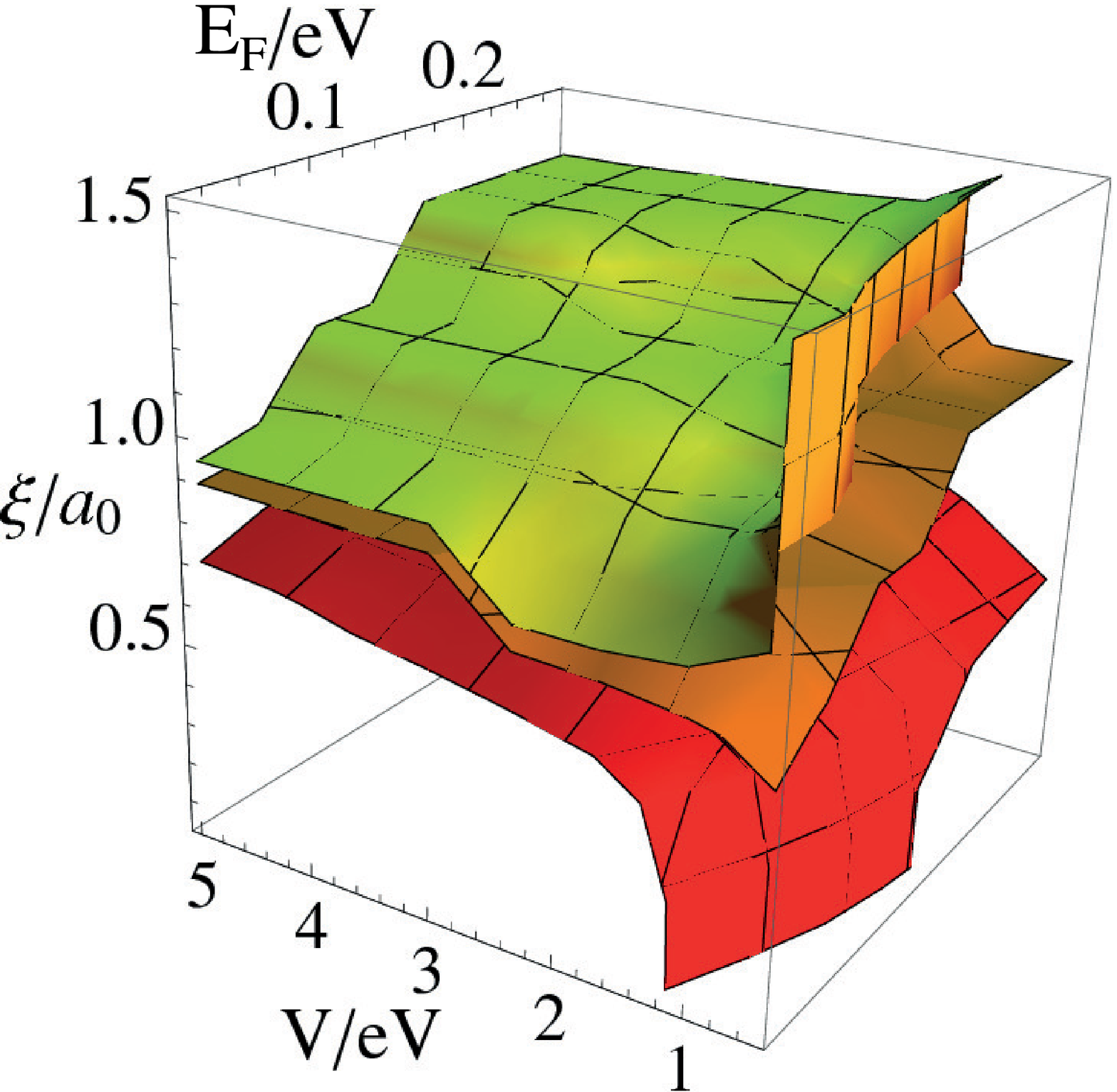}
\caption[short]{\label{a_priori6} Analytically calculated
  magnetoconductance amplitude, Eq. (\ref{continuous}), as function of
  $\xi,V$ and $E_F$. Surfaces with value $-1$ (red), $0$
  (orange) and $1$ (yellow) in units of $2 e^2/ h$ are displayed. Grid
  lines are guides to the eye.}
\end{center}
\end{figure}

The resulting analytical magnetoconductance amplitude is shown in
Figs. \ref{a_priori4ab} and \ref{a_priori6}. In Fig. \ref{a_priori4a}
the amplitude shown in Fig. \ref{a_priori4b} is displayed as function
of $\tau _{\phi }/\tau _i$ and $\tau _{\phi }/\tau _{\ast}$.  Since
the warping term dominates for large correlation lengths ($\xi>a_0$)
and $\tau _{\phi }/\tau _{W}$ does not depend on $\xi$ and $V$,
regions I and II collapse onto a line in Fig.  \ref{a_priori4b}.  We
can recognize the localization transition that we observed in the sign
diagrams of Fig. \ref{a_priori3}. In addition, we see that in the
range of the localization transition obtained by the numerical
calculation in Fig. \ref{sign-3d}, we observe weak antilocalization
with a small amplitude when close to the Dirac point for small back
Fermi energies and for weak scatterers (small $V,K_0$).
A small weak localization amplitude is
suppressed already by a weak magnetic field. Thus, in
 the numerical calculations   it is difficult
to identify a region with a weak localization amplitude.  We typically
 find by comparison
 that the low-energy cutoff in the numerical calculation,
is larger than the one obtained from the Thouless energy,
 Eq. (\ref{93}).

\subsection{Quantum Correction to Conductance at B=0}\label{qccb0}

\begin{figure*}
\begin{center}
\subfigure[$E_F = 0.01 t$] {\label{a_priori8a}
\includegraphics[width=0.23\textwidth]{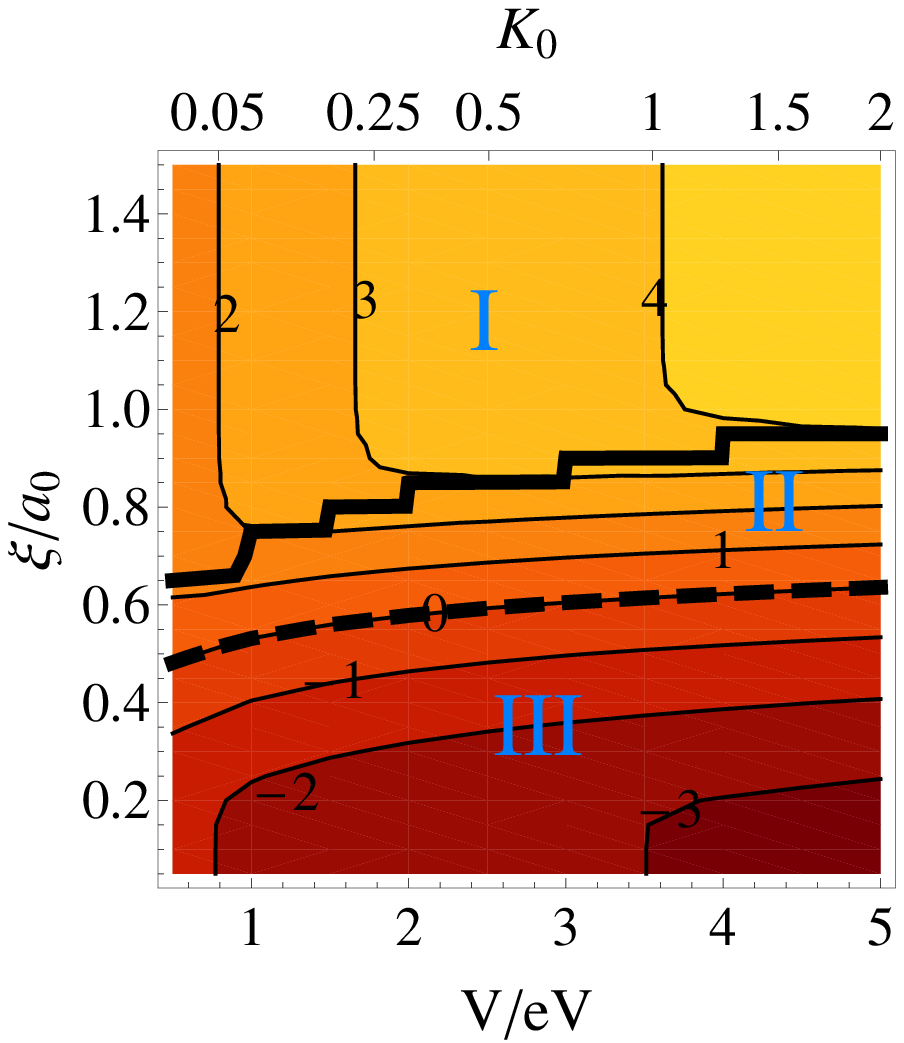}}
\subfigure[$E_F = 0.1 t$] {\label{a_priori8b}
\includegraphics[width=0.23\textwidth]{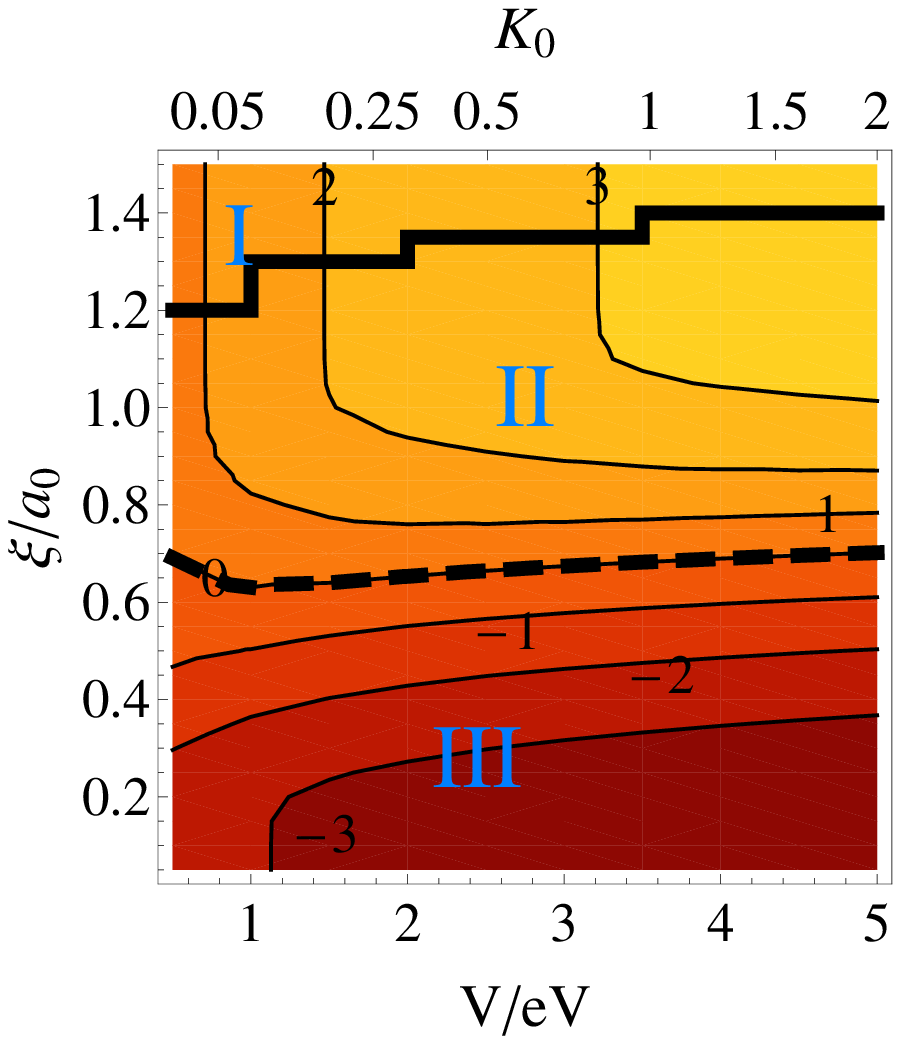}}
\subfigure[$E_F = 0.01 t$] {\label{a_priori9a}
\includegraphics[width=0.23\textwidth]{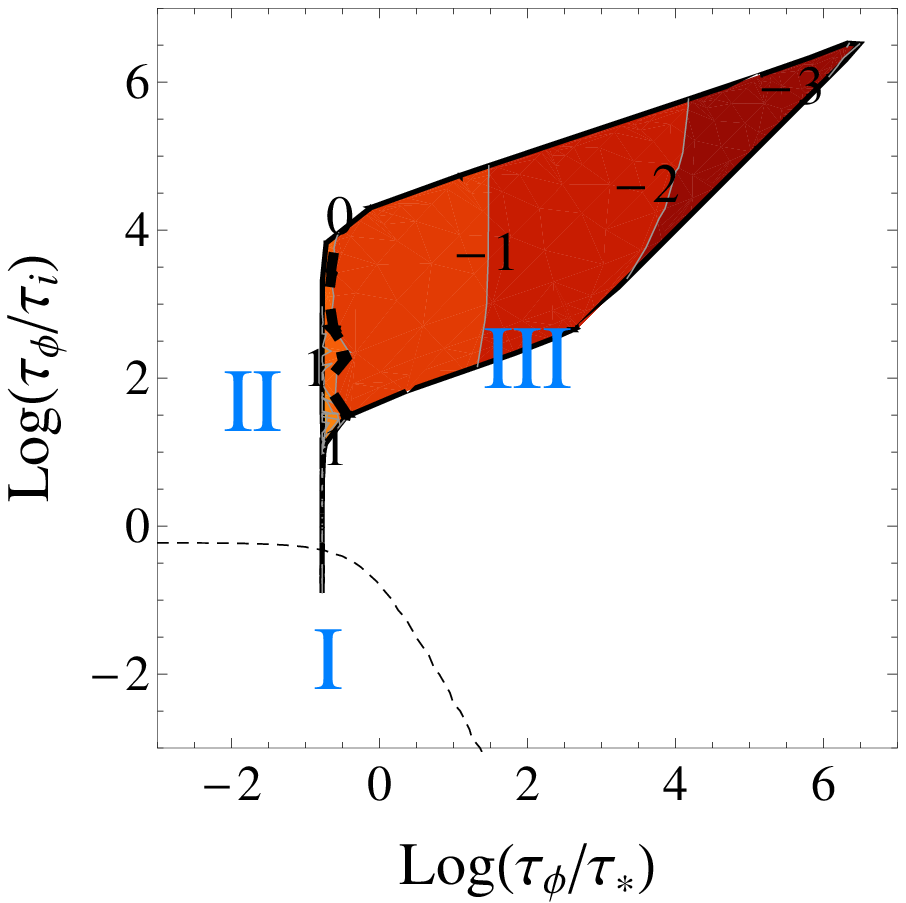}}
\caption[short caption]{\label{a_priori9}Analytically calculated
  quantum correction to the conductance at zero magnetic field,
  $\delta\sigma \left( B=0 \right)$ in units of $2 e^2/ h$ for two
  different Fermi energies, (a) and (b). Black dashed line:
  transition from positive to negative quantum correction,
  $\delta\sigma \left( B=0 \right)=0$.  Black line: Transition from
  PMC to NMC.  (c) Same as (a) but displayed as function of $\tau
  _{\phi }/\tau _i$ and $\tau _{\phi }/\tau _{\ast}$.}
\end{center}
\end{figure*}

The quantum correction to the conductance shown in Eq.
(\ref{deltasigma}) is plotted in Fig. \ref{a_priori9} as function of
impurity strength $V$ (and, correspondingly, $K_0$) and the
correlation length $\xi$.  We note that the transition line between a
positive and a negative quantum correction at $B=0$, as indicated by
the dashed line, does not coincide with the transition from negative
to positive magnetoconductance which is indicated by the thick black
line. Therefore, there is a region, denoted as II, where the quantum
correction to the conductance is positive but the conductance
increases with magnetic field, as one would expected for weak
localization.  Only in region I the positive quantum correction
coincides with the NMC expected for weak antilocalization. In region
III, the negative quantum corrections yields PMC as expected for weak
localization.  These regions can be related to the different types of
magnetoconductance sketched in Fig. \ref{draft}: applying a magnetic
field in the region II, we expect a nonmonotonic magnetoconductane,
where the conductance first increases, reaching a maximum at $B =
B_{ex}$, and then decays toward the classical conductance, as in case
(2) of Fig. \ref{draft}. In the other regions, I and III, the MC is
monotonic, with PMC in region III and NMC in region I.

In Fig. \ref{a_priori9a} the result for the quantum correction shown
in Fig. \ref{a_priori8a} is displayed as function of $\tau _{\phi
}/\tau _i$ and $\tau _{\phi }/\tau _{\ast}$.  Since the warping term
is dominant for large correlation lengths $\xi>a_0$, and $\tau _{\phi
}/\tau _{W}$ does not depend on $\xi$ and $V$, the regions I and II
collapse onto a line in Fig.  \ref{a_priori9a}.

\subsection{Quantum Corrections to Thermopower\label{kant_thermo}}

In this section we use the analytical theory to study the quantum
corrections to the thermopower and the resulting magnetic field
dependence.  In particular, we find out in which regime these
corrections are large and whether, in graphene, they are dominated by
the quantum corrections to the Fermi energy slope of the
conductance rather than by the weak localization corrections to the
conductance, as in standard metals.

\subsubsection{ Quantum Corrections at Zero Magnetic Field } \label{qcb0}

We first consider the amplitude of the quantum corrections to the
thermopower as defined by Eq. (\ref{thermo-sign}), namely, as the
difference between the value at zero magnetic field and the classical
thermopower.  Thus we need to use the weak localization correction to
the conductance at zero magnetic field, $\delta\sigma \left( B=0
\right)$, see Eqs. (\ref{deltasigma}) and (\ref{final}), and insert it
into the Mott formula.

Inserting the dependence of the scattering rates on the Fermi energy,
as given by Eqs. (\ref{90}), (\ref{91}), (\ref{92}), and
(\ref{93}), we obtain the slope of the Fermi energy dependence of
the conductance at $B=0$ as
\begin{eqnarray} \label{slope}
\frac{\partial \delta\sigma \left( B=0 \right)}{\partial
  E_F} &=& \frac{ e^2}{\pi h}
\frac{1}{E_F} \vartheta\left(t_{i},t_{z},t_w\right),
\end{eqnarray}
where
\begin{eqnarray}
\vartheta\left(t _{i},t_{z},t_w\right) &=& -2+ \frac{2}{1+2 t_i }+
\frac{4(1- t_w )}{1+t_i+ t_z +t_w} ,\label{thermo_ana1}
\end{eqnarray}
 which depends on the parameter ratios $t_i =\tau _{\phi }/\tau _i$
 and $t_z = \tau _{\phi }/\tau _{z}$, as well as explicitly on the
 warping rate ratio $t_w = \tau_{\phi}/\tau_w$.

\begin{figure}
\begin{center}
\subfigure[] {\label{tau_z_domain}
\includegraphics[width=0.24\textwidth]{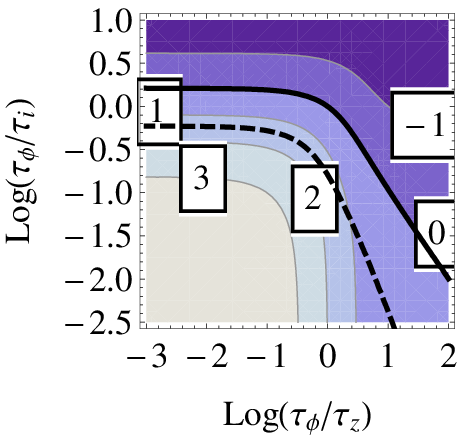}}
\subfigure[] {\label{tau_w_domain}
\includegraphics[width=0.22\textwidth]{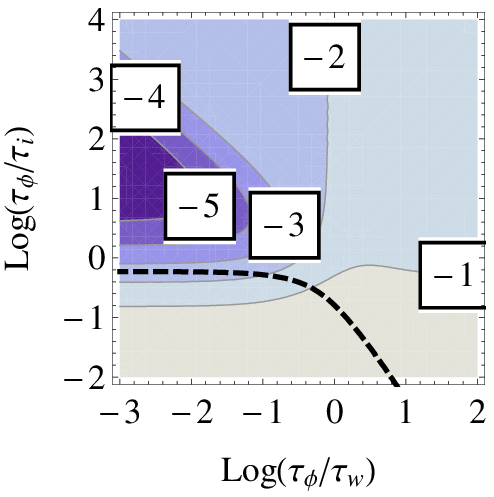}}
\caption[short caption]{ The function $\vartheta$,
  Eq. (\ref{thermo_ana1}) (a) for $\tau_{\phi}/\tau_w=0$ (b) for
  $\tau_{\phi}/\tau_z=0$.  The continuous black line indicates the
  sign change of the function $\vartheta$. Dashed line: transition
  between PMC and NMC.\label{startingpoints_fig}}
\end{center}
\end{figure}

We note that while the quantum corrections to the conductivity are
diverging, in the limit of $1/\tau_{\phi} \rightarrow 0$, which
corresponds to low temperatures and large system sizes, the quantum
corrections to the slope of the Fermi energy dependence of the
conductance, Eq. (\ref{slope}), converge to a finite value of order $
\frac{ e^2}{\pi h} \frac{1}{E_F}$, which depends on
the scatterings rates as follows:

1. When the warping term is negligible, $t_w \approx 0$, the function
$\vartheta$ converges to $-2$ for large $t_i$, the weak localization
regime, as seen in Fig. \ref{tau_z_domain}.  In the weak
antilocalization regime of small intervalley scattering ($t_i\ll1$),
$\vartheta$ turns positive.

2.  When the warping term dominates, and $t_z \rightarrow 0$, the
function $\vartheta$ is negative and converges for large intervalley
scattering $t_i \gg 1$, the weak localization regime, to $-6$ as seen
in Fig. \ref{tau_w_domain}.  Remarkably, $\vartheta$ (and thereby $d
\delta \sigma/dE_F$) remains for $t_z \rightarrow 0$ negative for
all values of $t_i$, even in the regime of weak antilocalization.

\begin{figure}
\begin{center}
\subfigure[$E_F = 0.01 t$] {\label{a_priori9b}
\includegraphics[width=0.23\textwidth]{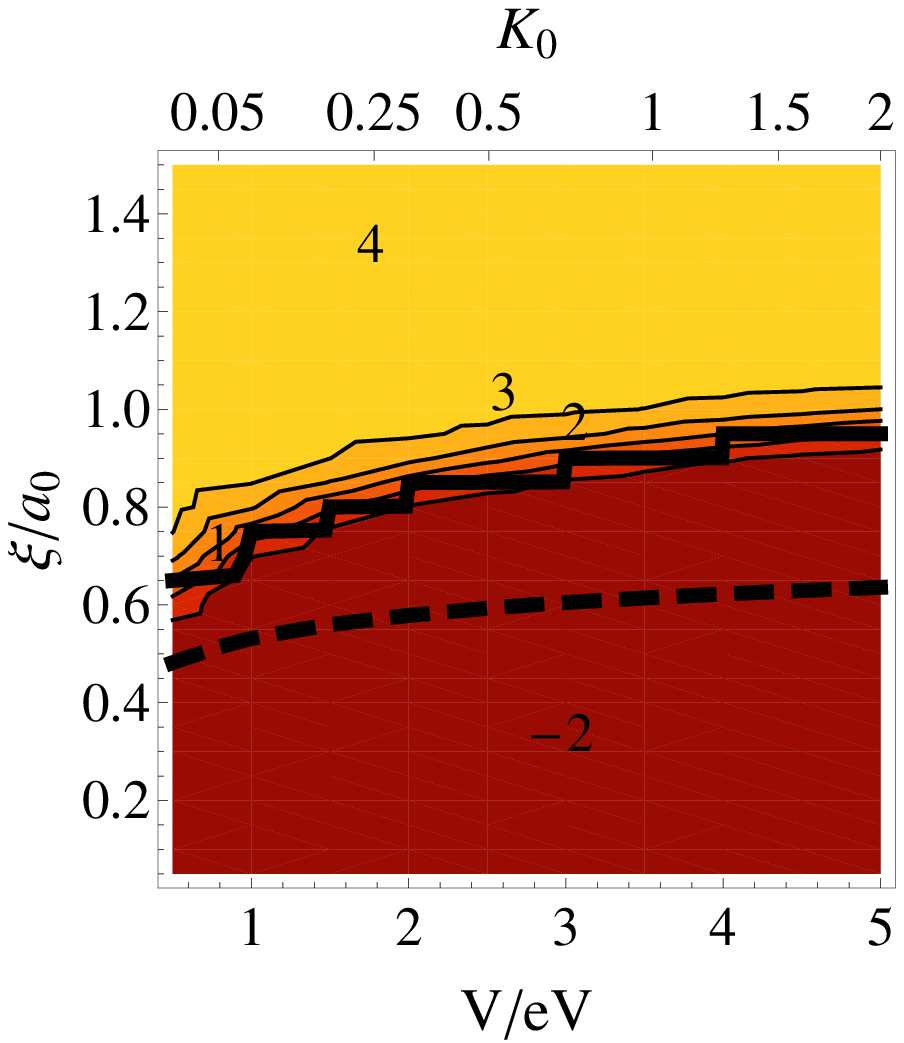}}
\subfigure[$E_F = 0.1 t$] {\label{a_priori9b2}
\includegraphics[width=0.23\textwidth]{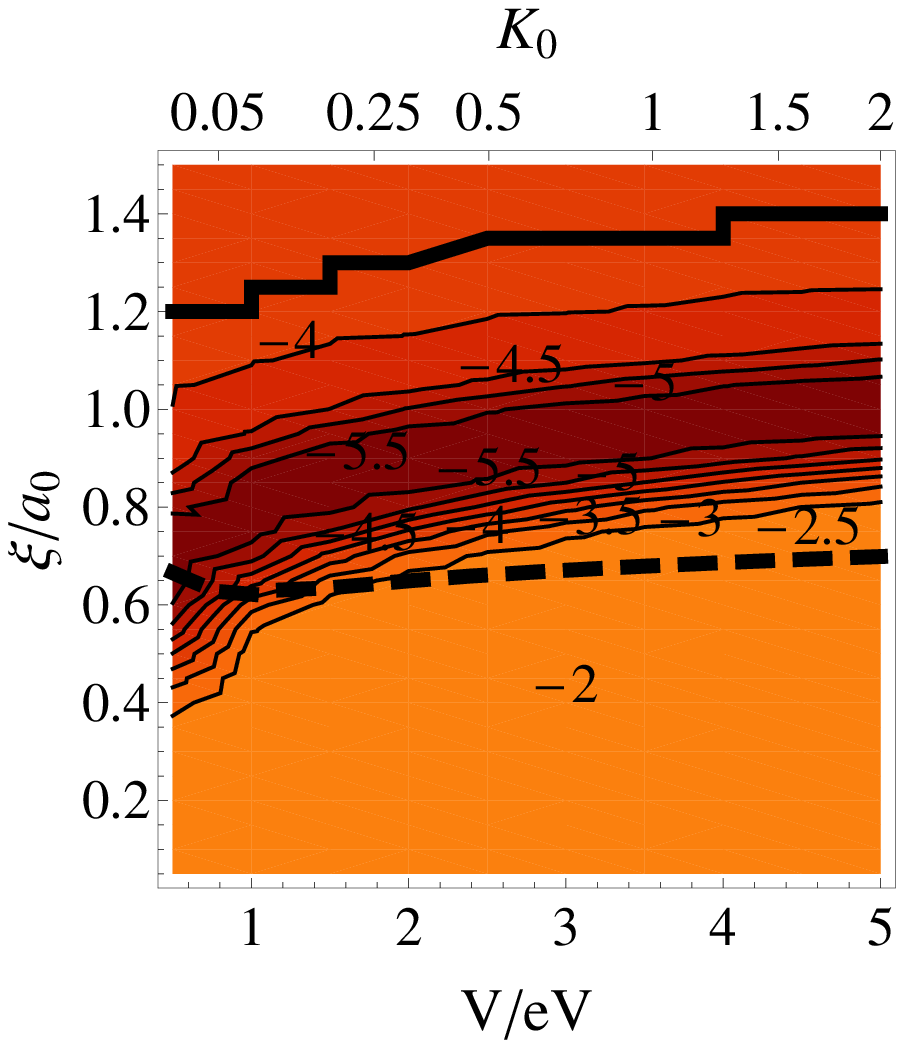}}
\caption[short caption]{\label{a_priori9.2} $\vartheta$,
  Eq. (\ref{thermo_ana1}), for two different Fermi energies as
  function of $\xi$ and $V$.  As in Figs. \ref{a_priori8a} and
  \ref{a_priori8b}, the black dashed line indicates the transition
  from positive to negative quantum correction to the conductance,
  $\delta\sigma \left( B=0 \right)=0$.  Black line: Transition from
  PMC to NMC.  }
\end{center}
\end{figure}

In Fig. \ref{a_priori9.2} $\vartheta$ is plotted as a function of
$\xi$ and $V$. For small Fermi energy, $E_F =0.01 t$,
$\vartheta$ is positive in the whole region of weak antilocalization
(corresponding to phase I in Figs. \ref{a_priori8a} and
\ref{a_priori8b}).  This positive enhancement of $\vartheta$ (and
therefore positive quantum correction to the slope of the Fermi energy
dependence of the conductance at $B=0$, Eq. (\ref{slope}))
matches well the numerical results of Sec. \ref{thermo_section}).  In
that regime, the intravalley scattering rate $1/\tau_z$ is expected to
dominate over the warping rate, and the values of $\vartheta$ indeed
agree with those obtained in Fig. \ref{tau_z_domain} where $t_w=0$.
At higher Fermi energies, Fig. \ref{a_priori9b2}, the warping rate
becomes more important, as $t_W \sim E_F^4$ increases faster than
$t_z$ with $E_F$. Indeed, $\vartheta$ remains negative for all
values of $\xi$ and $V$, in agreement with Fig. \ref{tau_w_domain}
where the intralayer scattering rate is set to zero, $t_z=0$. In the
regime I of weak antilocalization, only a slight increase of
$\vartheta$ is seen, while it remains negative.

\begin{figure}
\begin{center}
\subfigure[$E_F = 0.01 t$] {\label{final1}
\includegraphics[width=0.23\textwidth]{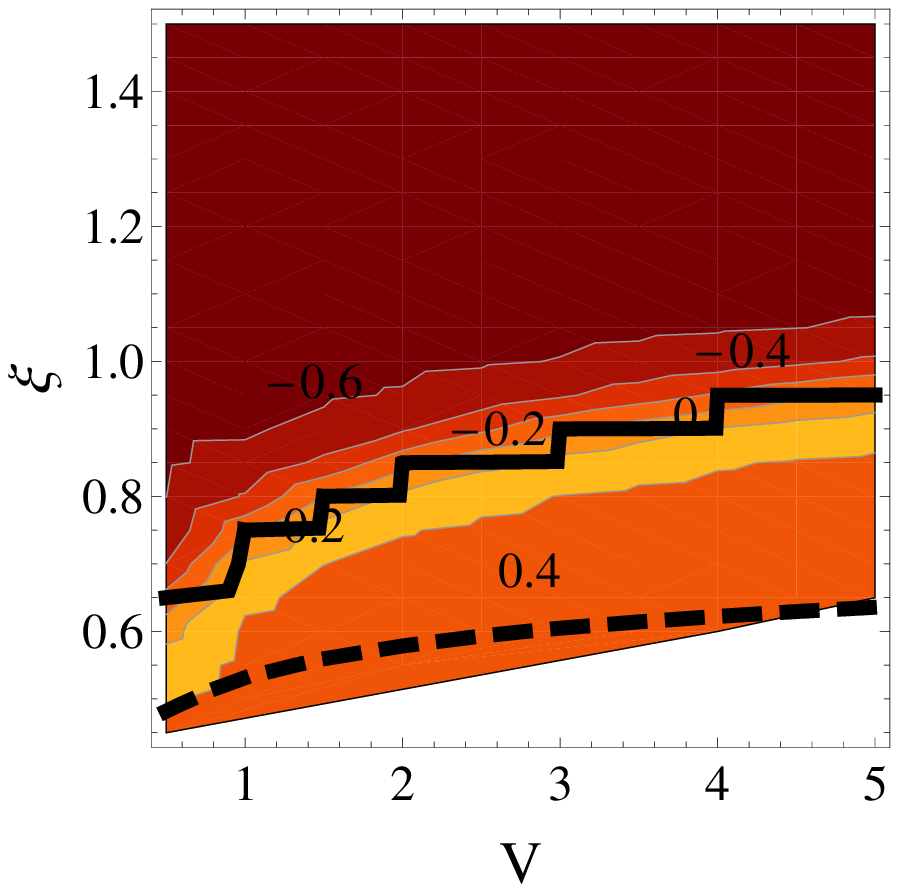}}
\subfigure[$E_F = 0.1 t$] {\label{final2}
\includegraphics[width=0.23\textwidth]{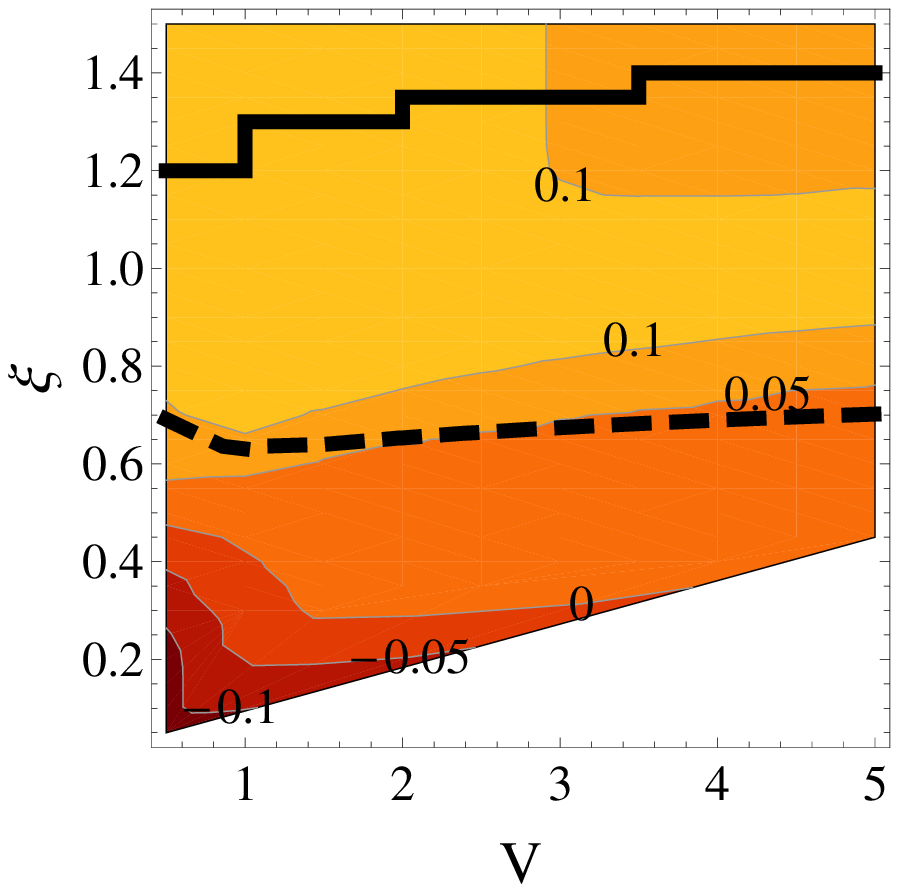}}
\caption[short caption]{\label{finalfigure} Analytically calculated
  $\delta S/T$ obtained from Eq. (\ref{deltaS}) in units of $\mu V /
  K^2$ for two different Fermi energies. Black line: transition
  from PMC to NMC. Dashed line: $\delta \sigma = 0$. White area: weak
  localization correction to conductance exceeds classical value.}
\end{center}
\end{figure}

Now, we are in a position to consider the quantum corrections to the
thermopower $\delta S$ as given in Eq. (\ref{squantum}).  For small
Fermi energies, these corrections are dominated by the second term in
Eq. (\ref{squantum}), resulting in the weak antilocalization regime
(I) in a negative correction $\delta S/T <0$ of order $\frac{\pi^2}{3}
\frac{k_B^2}{|e| eV} = 0.024 \mu V/K^2$.  In the regime of weak
localization, the quantum correction to the thermopower becomes
positive.  Since the classical magnetoconductance increases with gate
voltage, we find that the first term in Eq.  (\ref{squantum}) becomes
dominant at large Fermi energie, and one recovers $\delta S/S_{cl}
\approx - \delta G/G_{cl}$, which is characteristic of  standard
metals.  To study the competition between these two terms in more
detail, we need an expression for the classical conductivity.
 In the strong scattering limit of Gaussian impurities and for
  Coulomb scatterers one obtains
   a quadratic dependence on Fermi energy\cite{RevModPhys.81.109}
\begin{equation}
\sigma_{cl}=\frac{4e^2}{\pi h}+c_{\xi}\frac{2e^2 E_F^2}{h V^2},\label{classic}
\end{equation}
which is well justified in the limit of strong scatterers
\cite{PhysRevB.74.235443}.    The  prefactor
 $c_{\xi}$ is of the order of unity and increases from short range to long range scatterers by a factor of 2\cite{PhysRevB.74.235443}.
Note that we use here the notation introduced
 in Sec. \ref{connection}, where $V$ is a measure of the total impurity strength
   averaged over all impurities and increases with
    the density of impurities  $n_{\rm imp}$ as
     $
    V\sim \sqrt{n_{\rm imp}}$.
We then obtain $S_{cl}$ by  inserting
Eq. (\ref{classic}) into the Mott formula, Eq. (\ref{mott}).
we obtain $S_{cl}$ by inserting Eq. (\ref{classic}) into the Mott
formula, Eq. (\ref{mott}).
With Eq. (\ref{theta}) for $\delta\sigma$
and Eq. (\ref{slope}) for $\partial \delta\sigma/\partial
E_F$, we can use Eq. (\ref{squantum}) to find the dependence
of $\delta S/T$ on the impurity parameter and the Fermi energy. We obtain
\begin{equation}
\frac{\delta S}{T}=\frac{\pi^2 k_B^2 V^2}{3|e|(2V^2+\pi
  E_F^2)}\left(\frac{\pi E_F}{2V^2+\pi
  E_F^2}\theta-\frac{1}{2 E_F}\vartheta\right)\label{deltaS}.
\end{equation}
This result is displayed in Fig. \ref{finalfigure}, where we set $c_{\xi}=1$. The white area
corresponds to the regime where $\sigma_{\text{Diff}} \ll \Delta
\sigma (B=0)$, where Eq. (\ref{squantum}) is no longer valid.
(We note that, more generally one needs to  take into account that the classical
 conductivity Eq. (\ref{classic}) depends also on the range of impurities,  and  for weak scatterers may attain a weaker logarithmic energy dependence\cite{PhysRevB.74.235443}. This  will change these results quantitatively,
  but is not expected to change them  qualitatively, which is why
   we choose Eq. (\ref{classic}) and  leave  the inclusion of a more
    consistent  quantitative analysis for the classical conductivity for  future studies).

\begin{figure*}
\subfigure[$E_F = 0.01 t$] {\label{final_f}
  \includegraphics[width=0.6\textwidth]{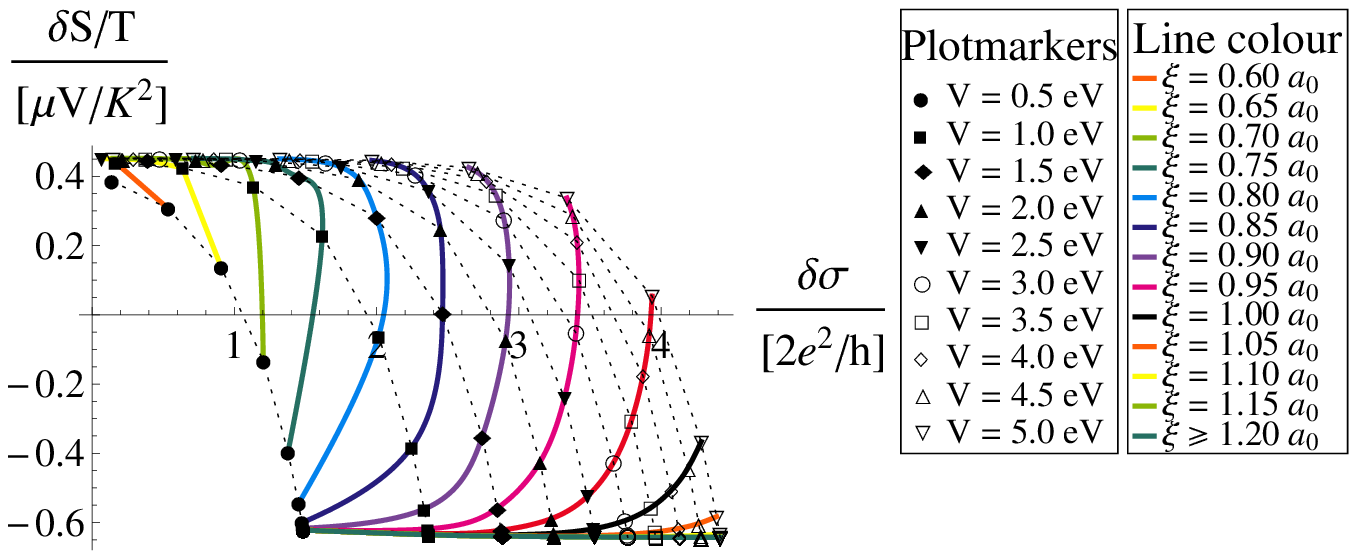}}
\subfigure[$E_F = 0.1 t$] {\label{final_g}
  \includegraphics[width=0.37\textwidth]{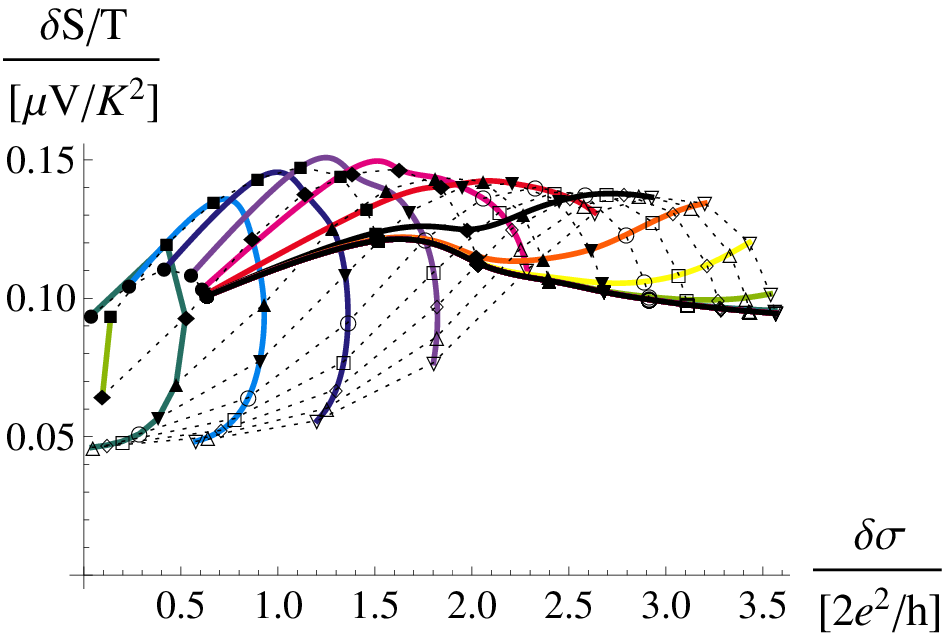}}
\caption[short caption]{\label{last_one} Analytically calculated
  $\delta S/T$ - $\delta \sigma$ diagram at zero magnetic field
  displaying the relation between the quantum corrections to
  thermopower and conductivity corrections for the phases of WAL, I
  and II, for various impurity parameters $\xi$ and $V$, for two
  different Fermi energies. Dotted black lines: same impurity
  strength for different $\xi$.}
\end{figure*}

Comparing the results with Fig. \ref{a_priori9.2} we can see that the
term due to the quantum corrections to the Fermi energy slope of the
conductance, the $\vartheta$ term in Eq. (\ref{deltaS}) is dominant in
our parameter range.  The quantum corrections of the conductance, the
$\theta$ term, plays a minor role here.  However, that term gains
importance with higher Fermi energy. For small Fermi energy
the intravalley scattering $1/\tau_z$ dominance is visible in phase I
(WAL), where we detect a negative amplitude of the correction $\delta
S/T$, while in phase II and III the correction is positive. For higher
Fermi energy, the warping term $1/\tau_w$ is becoming dominant
and we can observe a positive correction for the complete phase I and
II and even a small correction for parts of the phase III.

To see the connection between the electrical conductivity correction
and the thermopower correction, we plot $\delta \sigma$ versus $\delta
S/T$ in Fig. \ref{last_one}. We focus on the range of weak
antilocalization (phases I and II).  For small Fermi energy,
Fig. \ref{final_f}, we can see the transition from positive to
negative thermopower correction $\delta S$ when increasing the
impurity size $\xi$. We find negative $\delta S<0$ in phase I while
positive $\delta S>0$ is seen in phase II. An increase of the impurity
strength is moving the system toward positive $\delta S>0$. This
behavior changes at higher Fermi energy, see Fig. \ref{final_g},
since the warping term becomes more important. For short-range impurities,
the system in phase II does not show a clear relation between $\delta
\sigma$ and $\delta S/T$.  An increase of the impurity strength tends
to lower the thermopower correction.  With increasing impurity range,
when $\xi \approx a_0$, we observe an increase of the thermopower
correction with an increase of the conductivity correction near the
transition from NMC to PMC. For longer ranged impurities, the system
is in phase I of NMC. In that regime we find good agreement with the
numerical results and observe a clear relation between an increase of
$\delta S/T$ and $\delta \sigma$. We note that the detailed parameter
dependence may vary, depending on the value of the classical
conductance $\sigma_{\text{Diff}}$ and the low-energy cutoff
$1/\tau_{\phi}$.

\subsubsection{Magnetothermopower\label{kant_magneto_thermo2}}

\begin{figure}
\begin{center}
\subfigure[$\kappa (t_i,t_z,t_w)$ in a $\tau_\phi/\tau_z$ -
  $\tau_\phi/\tau_i$ diagram for $1/\tau_w=0$.] {\label{entwz}
  \includegraphics[width=0.23\textwidth]{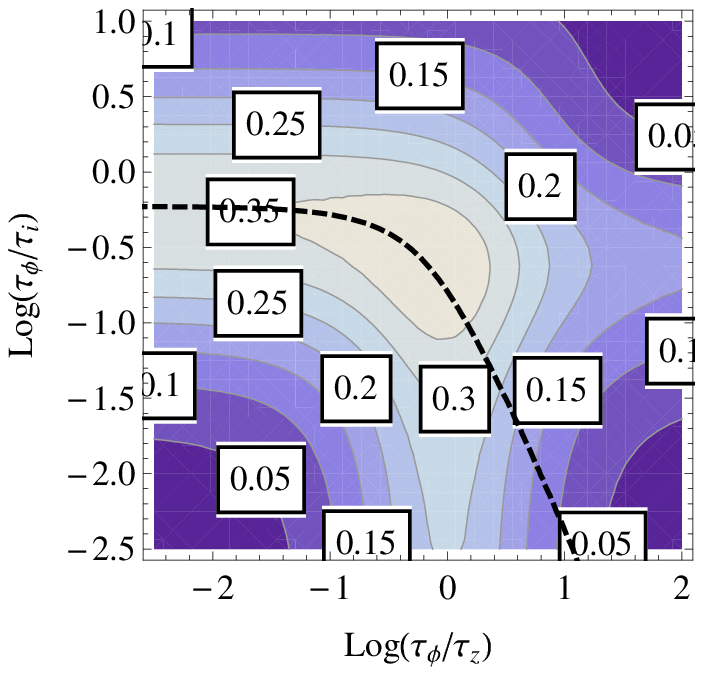}}
\subfigure[$\kappa (t_i,t_z,t_w)$ in a $\tau_\phi/\tau_w$ -
  $\tau_\phi/\tau_i$ diagram for $1/\tau_z=0$.] {\label{entww}
  \includegraphics[width=0.23\textwidth]{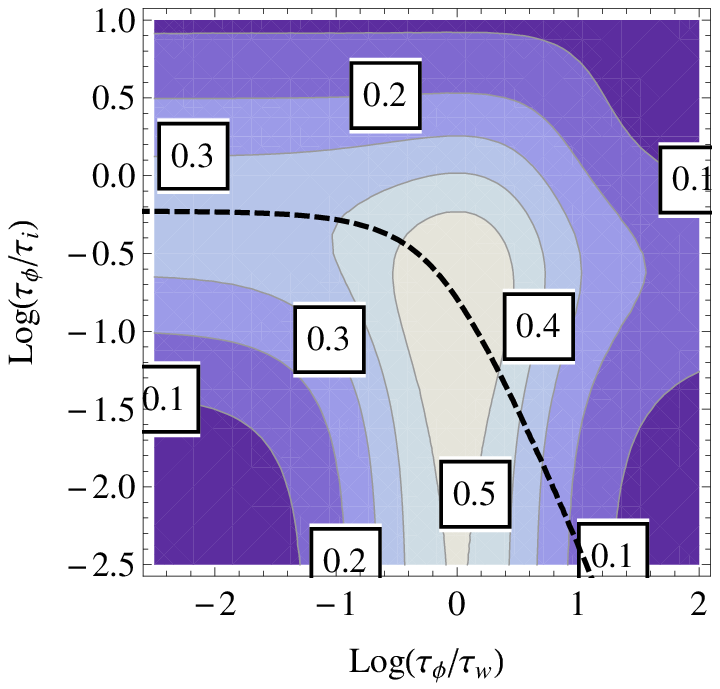}}
\caption[short caption]{\label{entw} $\kappa (t_i,t_z,t_w)$ displayed
  for the two limiting cases $1/\tau_w=0$ and $1/\tau_z=0$. Dashed
  lines: NMC to PMC transition from Fig. \ref{res2}. Positive numbers
  indicate positive magnetothermopower (PMT), an increase of $\delta
  S/T$ for small magnetic fields.}
\end{center}
\end{figure}

Next, we consider how these quantum corrections change when applying a
magnetic field.  We focus first on the derivative with respect to the
Fermi energy. To this end, we can use the expansion at weak
magnetic fields for $\Delta \sigma(B)$, Eq. (\ref{res}), and take its
derivative with respect to the Fermi energy $E_F$.  For the
finite-size samples used in the numerical calculations, we can
substitute $1/\tau_{\phi} = E_c = D/\Lambda^2$ and find that the
prefactor in Eq. (\ref{res}) does not depend on $E_F$.  Thus, we
find
\begin{eqnarray}
\Delta \frac{d\sigma (B)}{d E_F} & \thickapprox & \frac{e^2}{24\pi
  h}{\left(\frac{4 e B \Lambda^2 }{\hbar }\right)}^2 \frac{4}{E_F}
\kappa (t_i,t_z,t_w)\nonumber\\ \phantom{a}\label{res1}
\end{eqnarray}
where
\begin{eqnarray}
 \kappa (t_i,t_z,t_w)&=&\left[\frac{2
     t_i}{{\left(1+2t_i\right)}^3}+\frac{t_i +t_z+2 t_w}{{\left(1+ t_i
       +t_z+t_w\right)}^2}\right].
\label{ress}
\end{eqnarray}
This is  a purely positive magnetothermopower whose amplitude increases
with system size.  We display $\kappa (t_i,t_z,t_w)$ in the limits of
$1/\tau_w=0$, Fig. \ref{entwz}, and $1/\tau_z=0$, Fig. \ref{entww}.

For higher magnetic fields we can use the full expression and find,
\begin{equation}
\frac{\partial \Delta\sigma \left( B \right)}{\partial E_F}
= \frac{e^2}{\pi h} \frac{4}{E_F} \varsigma \left( \tau
_i,\tau _z,\tau_w,\tau _{\phi },\tau_B\right),
\end{equation}
where
\begin{widetext}
\begin{equation}
\begin{split}
\varsigma \left( \tau _i,\tau _z,\tau_w,\tau _{\phi },\tau_B\right) =
& \left(\frac{\tau_B }{2\tau_0}\Psi_1\left[\frac{1}{2}+\frac{
    \tau_B}{2\tau_0}\right]-\frac{\tau_B}{\tau_i}
\Psi_1\left[\frac{1}{2}+\tau_B
  \left(\frac{2}{\tau_i}+\frac{1}{\tau_\phi
  }\right)\right]\right.\\ &\left.-\tau_B
\left(\frac{1}{\tau_i}+\frac{2}{\tau_w}+\frac{1}{\tau_z}\right)\Psi_1\left[\frac{1}{2}+\tau_B
  \left(\frac{1}{\tau_w}+
  \frac{1}{\tau_i}+\frac{1}{\tau_z}+\frac{1}{\tau_\phi
  }\right)\right]\right),\label{magneto}
\end{split}
\end{equation}
\end{widetext}
and $\Psi_1$ is the polygamma function $\Psi_n$ with $n=1$, also known
as trigamma function. Performing the derivative shows that the
pseudospin-singlet isospin-singlet term does not contribute.  When the
magnetic field reaches $B_{\text{max}}$ the magnetothermopower
amplitude vanishes as does also the magnetoconductivity.  We find that
the thermopower correction $\delta S/T$ has a peak for higher magnetic
fields when the value for $B=0$ is positive. This is in accordance
with our numerical finding, see Fig. \ref{delle}.

\section{\label{compare} Analysis of Experimental Results}

There have been several reports on magnetoconductance experiments on
single layer graphene.  The weak localization amplitude is typically
of the order of $2e^2/h$.
\cite{PhysRevB.78.125409,PhysRevLett.103.226801,1367-2630-11-9-095021,PhysRevLett.98.136801,PhysRevLett.100.056802,A366}
Tikhonenko and coworkers showed that it is possible to observe weak
localization up to a temperature of 200
K.\cite{PhysRevLett.103.226801} In their work, they found a clear
transition between PMC and NMC as the temperature is changed.  An
increase of the temperature decreases the MC amplitude due to the
increasing of the dephasing rate $1/\tau_{\phi} (T)$. For example, a
change from 5 Kelvin to 14 Kelvin in
Ref. \onlinecite{PhysRevLett.103.226801} reduces the amplitude at
gate voltage $V_{G} = 40 V$ by 70\%.  For this reason, to maximize
the amplitude, most experiments are done at temperatures below $10 K$.
The MC effect is visible in all experiments at magnetic fields of up
to $B=0.1T$.  In addition, in all experiments there are observable
magnetoconductance fluctuations, whose amplitudes are however much
smaller than the total MC amplitude.

We focus on Ref. \onlinecite{PhysRevLett.103.226801} at the lowest
temperature $T=5K$ since it has the largest amplitude.  The scattering
rates reported by the authors are obtained by fitting the analytical
theory of Ref. \cite{PhysRevLett.97.146805}. They are found to be
$\log[\tau_{\phi}/\tau_{\ast}]= 1.3, 2, 2.4$ and
$\log[\tau_{\phi}/\tau_{i}]= -0.2, 1.17, 2$ for the three gate
voltages $V_G= 7,20,40 V$.

The Mott formula, Eq.(\ref{mott}), can also be written in terms of the back gate voltage $V_{\rm BG}$ as
\begin{equation}
S =\frac{\pi ^2}{3} \frac{k_B^2T}{ e}\frac{1}{\sigma}\frac{d \sigma}{d
  V_{\text{BG}}}{ \left.\frac{d V_{\text{BG}}}{d
    E}\right|}_{E=E_F},\label{tatr}
\end{equation}
with
\begin{equation}
E_F = \hbar v_F \sqrt{\pi n_{2D}}\propto \pm\sqrt{|V_{BG}|},\label{zusatz}
\end{equation}
where $n_{2D}$ is the two dimensional carrier density.\\
Using the dephasing rate $1/\tau_{\phi} =0.1$ ps$^{-1}$ as measured in
Ref. \onlinecite{PhysRevLett.103.226801} at $T=5K$, identifying $V_G
= V_{BG}$ and taking for the impurity concentration $3\%$ we find a
very good agreement between experiments and our results shown in
Fig. \ref{a_priori1}. We find the same values for the scattering rate
ratios $\tau_{\phi}/\tau_{\ast}$ and $\tau_{\phi}/\tau_{i}$ if we set
the impurity parameters to $\xi \lesssim 0.5\, a_0$ and $V \approx
0.5$ eV in the ab initio calculations outlined above. Thus, we
conclude that the analysis of the experimental results with the
theory allows us to make detailed predications about the potential
amplitude and range of the impurities in the graphene samples.  For
the sample of Ref. \onlinecite{PhysRevLett.103.226801} which was
produced by mechanical exfoliation of graphite and deposition on an
oxidized Si wafer, we can conclude that the typical range of the
impurities in that sample is $\xi \lesssim 0.5 a_0$.  If the impurity
concentration were known from an independent measurement, their
average strength $V$ could be inferred from these magnetoconductance
measurements as well.  Using the measured mobility of $ \mu_e = 12 000
cm^2/(Vs)$ one could in principle estimate one of these parameters.

Comparing with the analytical results, Fig. \ref{a_priori1}, we see
that the experimental results for gate voltages $V_G= 7V$ and $V_G= 20
V$ are in the regime where the intravalley scattering rate $1/\tau_z$
is dominant while at a gate voltage of $40V$ this rate is superceded
by the warping rate $1/\tau_w$.  A higher accuracy of  the data
analysis can be achieved by fitting the experimental data at different
gate voltages to the same parameters and making use of the analytical
gate voltage dependence of $\tau _{\phi }/\tau _i$ and $ \tau _{\phi
}/\tau _{\ast}$, as in Eq. (\ref{prop2}).

There have been several reports on the measurements of thermopower in
single layer
graphene.\cite{PhysRevB.82.245416,PhysRevB.80.081413,PhysRevLett.102.096807,PhysRevLett.102.166808}. We
found above that the magnitude of the quantum corrections to
thermopower is of the order of $1 \mu V / T^2$, see Figs. \ref{05.3d}
and \ref{last_one}, which is not much smaller than the measured values
for the classical thermopower of single-layer
graphene.\cite{PhysRevB.80.081413,PhysRevLett.102.096807,PhysRevLett.102.166808}
We note that the dependence of $\delta \sigma $ on $E_F$ strongly
depends on the Fermi energy dependence of the dephasing rate
$1/\tau_{\phi}$.  At low temperatures the dephasing is dominated by
the electron-electron scattering, yielding\cite{0022-3719-15-36-018}
\begin{equation}
           \frac{1}{\tau_{\phi}} = \alpha \frac{ k_B T}{2 E_F \tau_0}
           \ln ( 2 E_F \tau_0/\hbar).
\end{equation}
Since $1/\tau_0 \sim E_F$, Eq.(\ref{92}), close to the Dirac
point, one finds that $\tau_{\phi}$ is independent of the Fermi energy
$E_F$, in good agreement with the experimental results.
\cite{PhysRevLett.103.226801}

High magnetic field thermopower measurements at room temperatures have
been performed,\cite{PhysRevB.86.155414}
and a theory based on the SCBA has been given in Ref. \cite{PhysRevB.76.035402}.
The temperature scale of
these experiments ranges from room temperature down to several Kelvin,
which would  be also  the right
 temperature  range to observe the magnetothermopower due to
quantum corrections studied here. To this end the
 thermopower  measurements
would have to be  performed at magnetic fields below $ B_{max} \sim 0.1$ Tesla.  We are not
aware that such measurements have been performed to date.

  As mentioned before,  the weak localisation
   corrections to  the conductance and thermopower become
    suppressed, when the magnetic length $l_{B}$ is on the order of
     the elastic mean free path $l_e = v \tau$
      or when $1/\tau \sim v/l_B$, which defines the maximal
       magnetic field $B_{\rm max}$  at which  weak localisation
        corrections can be expected.
       The spacing between the Landau levels in graphene
        is known to be anomalously large, $\hbar \omega_c = \sqrt{2}v/l_B$. The
         condition that Landau levels become smeared out by
          disorder yields therefore $1/\tau \sim v/l_{B}$,
          and in the weak localisation regime, $B< B_{max}$ effects of Landau bands
            can indeed be disregarded.  For $B> B_{max}$
             the magnetoconductance is classical but can also be sensitive to the range of impurities, and acquire a $\sqrt{B}$ dependence \cite{12106081,vasileva}. Thus,
              the magnetothermopower accordingly can be  expected to become
              stronger than previously known\cite{PhysRevB.76.035402}.

There is increasing evidence that ad-atoms and vacancies are important
to understand the transport properties of graphene. The transport theory
of graphene with such strong impurities forming resonances has recently
been studied in Ref.\onlinecite{PhysRevLett.111.146601}, who found that there is a
regime close to the Dirac point where the transport  differs from
the predictions based on the transport theory with nonresonant impurities
used here. It will therefore be an important project to study the thermopower
in the presence of such impurities. However, Ref.\onlinecite{PhysRevLett.111.146601} confirms
that there is a regime away from the Dirac point where the effect of strong
impurities can be described by the model used here.

\begin{figure}
\begin{center}
\includegraphics [width=8cm] {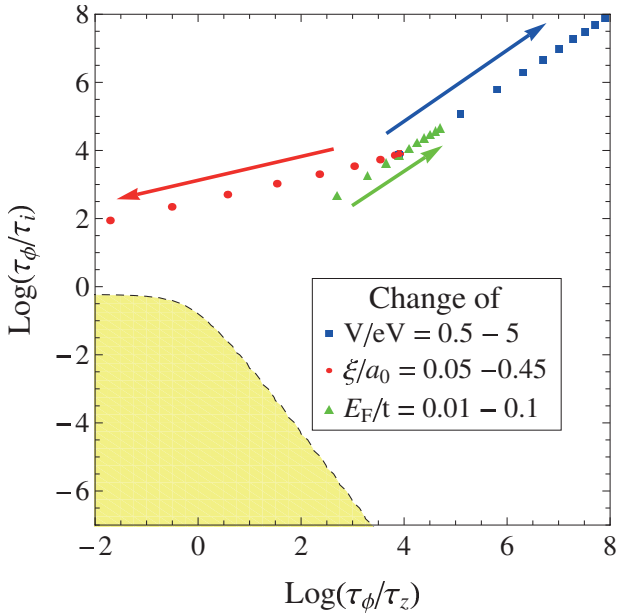}
\caption[]{\label{old_a_priori1} Magnetoconductance in the $\tau
  _{\phi }/\tau _i$ - $\tau _{\phi }/\tau _z $ diagram for $1/\tau_w
  =0$.  The ratios of scattering rates and dephasing rates
  $\tau_{\phi}/\tau_i$, $\tau_{\phi}/\tau _z $ are calculated from the
  parameters $\xi$ (red dots), $V$ (blue squares) and $E_F$
  (green triangles) which are varied in the range given in the inset.
  Arrows indicate the direction of increasing parameter.  The
  variation with $V$ and $E_F$ is shown for $\xi/a_0 =.05$,
  $\xi$ is varied at fixed $V/eV =0.5$ and $E_F/t=.04$.
  Dashed line: localization transition. White area: PMC. Yellow area:
  NMC.}
\end{center}
\end{figure}

\section{\label{conclusion2}Conclusions and Discussion}

We have studied the quantum corrections to the conductivity and the
thermopower in monolayer graphene by analyzing numerical results in
conjunction with the analytical theory.  The quantum corrections to
the thermopower result in large magnetothermopower, which we
demonstrate to be a very sensitive measure of the impurities in
graphene.  While there are experimental measurements of
magnetoconductance of single-layer graphene which could be used to
determine the average range and strength of the impurities, future
measurements of the magnetothermopower could provide additional
information about the graphene samples. The strong magnetic field
dependence of the thermopower, the {\it magnetothermopower}, is a
direct measure of the quantum corrections to the thermopower. In
contrast to usual metals, its amplitude in graphene is not simply
related to the amplitude of the electrical conductivity, but is also
governed by the quantum correction to the gate voltage slope of the
conductivity.

We demonstrate that the numerical calculations not only can be fitted
with the analytical theory, but we also find even good agreement when
the scattering rates entering the analytical theory are calculated
directly from the numerical input parameters. This allows a more
detailed understanding, since the dependence on gate voltage and other
variable parameters can be studied, revealing more accurate
information on the impurity parameters in graphene.

Besides the sign of the magnetoconductance, showing a transition
between PMC and NMC, we also analyzed the amplitude of the quantum
corrections.  We found that the transition from PMC to NMC does not
coincide with the transition from positive to negative quantum
corrections at vanishing magnetic field $B=0$.  We analyzed the
relation between the amplitude of the quantum corrections to the
conductance and the magnetothermopower. It appeared strongest close to
the Dirac point and for long-range impurities.

Here, we studied the thermopower using the Mott formula, which is only
valid for small ratios $T / E_F$.  Still, experiments show
that it remains valid close to the Dirac
point.\cite{PhysRevB.82.245416,PhysRevB.76.193401} We intend to study
its validity in a future publication by using directly the Kubo
formula for the thermopower.  For the aspect ratio 1 studied in this
paper, we did not find a strong sensitivity to the form of the
boundary conditions, armchair or zigzag.  The quantum corrections are
expected to increase for larger aspect ratios. Also the warping rate
$1/\tau_{w}$ is expected to be sensitive to the aspect ratio, being
suppressed when the width of the graphene samples is reduced.  This
effect is similar to the reduction of the Dyakonov-Perel spin
scattering rate in quantum wires.\cite{PhysRevLett.98.176808} The
other rates $1/\tau_i$ and $\tau_z$ are expected to depend only weakly
on the aspect ratio.  We note that we assumed in this paper that the
dominant impurity scattering processes are elastic.  The effect of
magnetic scatterers on the weak localisation corrections in graphene
will be studied in a separate work.

\begin{acknowledgments}
The authors would like to thank Hyunyong Lee, Paul Wenk, Vincent
Sacksteder, Seung-Geol Nam and Prof. Hu Jong Lee for useful
discussions.
We gratefully acknowledge that this research was
supported by World Class University (WCU) program, Division of
Advanced Materials Science as well as the Asian Pacific Center of
Theoretical Physics (APCTP) at POSTECH university. We also thank the
Jacobs University Foundation for supporting this research.
\end{acknowledgments}

\appendix

\section{Matrix notation of the Hamiltonian\label{matrix}}
For a better understanding of the structure of our formulation, it
useful to write the Hamiltonian in matrix notation. We present here
$H_1$, $H_2$ and $H_{imp}$ in this way. In particular, the $H_{imp}$
matrix makes it easier to see the connection of each part of the
impurity potential to the isospin-pseudospin description and the
matrix notation of the Gaussian potential which is shown in section
(\ref{connection}). We have
\begin{eqnarray}
H_1 &=& v_F \zeta_3 \otimes \vv{\sigma } \cdot \vv{k}\\\label{bigH2}
&=&v_F
\begin{pmatrix}
 0 &  k_x-ik_y & 0 & 0 \\
 k_x+ik_y & 0 & 0 & 0 \\
 0 & 0 & 0 & k_x+ik_y \\
 0 & 0 & k_x-ik_y & 0
\end{pmatrix},\nonumber\\
\phantom{a}
\end{eqnarray}
and
\begin{equation}
H_2 = \mu \left(\sigma
_1\left(k_x^2-k_y^2\right)-2\sigma_2\left(k_xk_y\right)\right)
\end{equation}
\begin{equation}
=\mu\left(\begin{matrix}
 0 & {(k_x+i k_y)}^2 & 0 & 0 \\
 {(k_x-i k_y)}^2 & 0 & 0 & 0 \\
 0 & 0 & 0 & {(k_x+i k_y)}^2 \\
 0 & 0 & {(k_x-i k_y)}^2 & 0
\end{matrix}\right).
\end{equation}

The second term of the impurity Hamiltonian, as given in
Eq. (\ref{Hnew}), can be rewritten in matrix form as
\begin{widetext}
\begin{equation}
\left(
\begin{array}{cc|cc}
 V_{3,3} & V_{1,3}-i V_{2,3} & V_{3,1}-i V_{3,2} & -V_{1,1}+i V_{1,2}+i V_{2,1}+V_{2,2} \\
 V_{1,3}+i V_{2,3} & -V_{3,3} & V_{1,1}-i V_{1,2}+i V_{2,1}+V_{2,2} & V_{3,1}-i V_{3,2} \\
 \hline
 V_{3,1}+i V_{3,1} & V_{1,1}+i V_{1,2}-i V_{2,1}+V_{2,2} & -V_{3,3} & V_{1,3}-i V_{2,3} \\
 -V_{1,1}-i V_{1,2}-i V_{2,1}+V_{2,2} & V_{3,1}+i V_{3,1} & V_{1,3}+i V_{2,3} & V_{3,3}
\end{array}
\right).
%\label{bigM}
\end{equation}\\
\end{widetext}

\section{\label{appendix_two} Magnetoconductance for $1/\tau_w \rightarrow 0$}
In Fig. \ref{old_a_priori1} we present the results of the ab initio calculations for the MC as a function of $\tau _{\phi }/\tau _i$ and $\tau _{\phi }/\tau _z$.\\

\bibliographystyle{apsrev}
%\bibliography{ref}

\end{document}